\documentclass[a4paper,useAMS,usenatbib]{mn2e}
\usepackage{epsfig}
\usepackage{multirow}

\def\lsim{ \lower .75ex \hbox{$\sim$} \llap{\raise .27ex \hbox{$<$}} }
\def\gsim{ \lower .75ex \hbox{$\sim$} \llap{\raise .27ex \hbox{$>$}} }
\def\g{{\tt GALFORM}}
\def\gr{{\tt GRASIL}}

\begin{document}

\title[Modelling the dusty universe]
{Modelling the dusty universe I: Introducing the artificial neural 
network and first applications to luminosity and colour distributions}

\author[Almeida et~al.]{
\parbox[t]{\textwidth}{
\vspace{-1.0cm}
C.\,Almeida$^{1}$,
C.\,M.\,Baugh$^{1}$,
C.\,G.\,Lacey$^{1}$,
C.\,S.\,Frenk$^{1}$,
G.\,L.\,Granato$^{2}$,
L.\,Silva$^{3}$,
A.\,Bressan$^{4,2,3}$,
}
\\
$^{1}$Institute for Computational Cosmology, Department of Physics, 
University of Durham, South Road, Durham, DH1 3LE, UK.\\
$^{2}$INAF - Osservatorio Astronomico di Padova, Vicolo Osservatorio 5, 35122 Padova, Italy.\\
$^{3}$INAF - Osservatorio Astronomico di Trieste, via Tiepolo 11, 34131 Trieste, Italy\\ 
$^{4}$INOE, Lusi Enrique Erro 1, 72840 Tonantzintla, Puebla, Mexico.\\
}
 
\maketitle
\begin{abstract}
We introduce a new technique based on artificial neural networks which
allows us to make accurate predictions for the spectral energy
distributions (SEDs) of large samples of galaxies, at wavelengths
ranging from the far-ultra-violet to the sub-millimetre and radio.
The neural net is trained to reproduce the SEDs predicted by a hybrid
code comprised of the \g~semi-analytical model of galaxy formation,
which predicts the full star formation and galaxy merger histories,
and the \gr~spectro-photometric code, which carries out a
self-consistent calculation of the SED, including absorption and
emission of radiation by dust. Using a small number of galaxy
properties predicted by \g, the method reproduces the luminosities of
galaxies in the majority of cases to within 10\% of those computed
directly using \gr. The method performs best in the sub-mm and
reasonably well in the mid-infrared and the far-ultra-violet. The
luminosity error introduced by the method has negligible impact on
predicted statistical distributions, such as luminosity functions or
colour distributions of galaxies.  We use the neural net to predict
the overlap between galaxies selected in the rest-frame UV and in the
observer-frame sub-mm at $z=2$. We find that around half of the
galaxies with a $850\mu$m flux above 5 mJy should have optical
magnitudes brighter than $R_{\rm AB} < 25$ mag. However, only 1\% of
the galaxies selected in the rest-frame UV down to $R_{\rm AB} < 25$
mag should have $850\mu$m fluxes brighter than 5 mJy. Our technique
will allow the generation of wide-angle mock catalogues of galaxies
selected at rest-frame UV or mid- and far-infrared wavelengths.

\end{abstract}

\section{Introduction}

Starting with IRAS, and continuing with ISO, the Spitzer Space
Telescope and AKARI, surveys using space-based infrared telescopes
have revealed that many galaxies emit a significant fraction of their
total luminosity at mid- and far-infrared (IR) wavelengths, this
emission coming from dust grains which have been heated by absorbing
optical or ultraviolet (UV) light from stars or AGN.  Measurements of
the integrated extragalactic background light reveal that the mid- and
far-IR wavelength range contain as much energy as the ultraviolet and
optical parts \citep{hauser}. Dust therefore plays a key role in
shaping the observational signature of the overall global star
formation history.  Building on the success of space-based infrared
telescopes as well as ground-based sub-mm instruments like the
Submillimetre Common User Bolometric Array (SCUBA) on the James Clerk
Maxwell Telescope, a number of new instruments and space missions are
planned which will map the universe at wavelengths sensitive to
emission from dust (e.g. Herschel, SCUBA-2, LMT, ALMA).  Some of these
new instruments will allow wide field surveys to be carried out, such
as the Herschel ATLAS survey which will cover 600 square degrees and
will provide accurate measurements of the clustering of galaxies
selected by their far IR emission. These new surveys will be targetted
by other telescopes, building up multi-wavelength coverage.  It is
therefore essential to develop theoretical tools which can take
advantage of these new data and which make predictions of galaxy
spectral energy distributions (SEDs) over a wide range of wavelengths.

In this paper we build on a hybrid model introduced by Granato
et~al. (2000) which combines the semi-analytical galaxy formation code
\g~(Cole et~al. 2000) with a spectro-photometric code \gr~(Silva
et~al. 1998). The semi-analytical model uses simple physically
motivated recipes and prescriptions to follow the baryonic process
believed to be important for galaxy formation (see Baugh 2006 for an
overview). \gr~takes the star formation history predicted for each
model galaxy by \g~and makes an accurate calculation of the SED from
the far-UV to the radio. \gr~calculates the absorption of starlight by
dust self-consistently by radiative transfer, using the dust mass
obtained from the gas mass and metallicity, and the disk and bulge
scalelengths predicted by the model.  The spectrum of dust emission is
calculated by solving for the radiative equilibrium temperatures of
individual dust grains. The hybrid \g~ plus \gr~ model successfully
reproduces the abundance of Lyman-break galaxies at high redshift
(detected through their emission in the rest frame UV) and the number
counts and redshifts of submillimetre selected galaxies (Baugh
et~al. 2005). This model has also been used to predict the number
counts and redshift distributions of galaxies as measured in the mid
and far IR by the Spitzer Space Telescope (Lacey et~al. 2008).

To build mock catalogues with information about the spatial
distribution of galaxies for wide field surveys like the Herschel
ATLAS, we need to use the hybrid \g~plus \gr~model to populate large
volume N-body simulations.  In this paper we use the Millennium
Simulation of the evolution of structure in a cold dark matter
universe (Springel et~al. 2005). The simulation volume is $500 h^{-1}$
Mpc on a side and contains around 20 million dark matter haloes at the
present day. To build a mock catalogue for the Herschel ATLAS, which
extends to $z \approx 2 $, would require us to populate around 30
snapshots from the Millennium, which would run to around 500 million
dark matter haloes.  The \gr~code takes several minutes to run for
each galaxy, so to process on the order of one billion galaxies would
take around 100 years on current large computers.

In this paper we explore an alternative approach in which we train an
artificial neural network to mimic the calculation of SEDs by \gr.  We
show that it is possible to construct a neural net which, starting
from a small number of galaxy properties which can be readily
predicted by \g, can produce reasonably accurate predictions of the
luminosity which would result from a direct calculation with \gr. We
note that a complementary approach in which an artificial neural
network is trained to speed up part of the calculation carried out by
\gr~has been developed by Silva et~al. (2009, in preparation).

Here we introduce the neural net technique and apply it to study the
overlap between Lyman-break galaxies and submillimetre selected
galaxies.  We give a brief overview of \g~and \gr~in Section 2 and
explain how they are combined into a hybrid code to predict the
spectral energy distributions of galaxies. In Section 3, we give some
theoretical background to artificial neural networks. Section 4 is
devoted to an investigation of the accuracy of the neural net in
predicting galaxy luminosities for different choices for the set-up of
the net.  We apply the new technique to the prediction of the
luminosity functions of Lyman-break galaxies at $z=3$, mid-IR selected
galaxies at $z=0.5$ and submillimetre galaxies at $z=2$ in Section 5,
where we compare the results from the neural net against the direct
calculations from \gr. We show how well the model can predict colour
distributions in Section 6. In Section 7, we examine the overlap
between galaxies selected in the rest-frame UV and in the observer
frame sub-millimeter.  Finally, in Section 8, we present our
conclusions.  Throughout we assume the cosmology of the Millennium
simulation with a present-day matter density of $\Omega_{\rm M}=0.25$
and a cosmological constant of $\Omega_{\lambda}=0.75$.

\section{Theoretical background I: Modelling the galaxy population}

In this section, for completeness, we give a brief overview of the 
semi-analytical galaxy formation model {\tt GALFORM} and the 
spectro-photometric code {\tt GRASIL}. We also explain how these codes can 
be used in combination to predict the full spectral energy distributions 
of a population of galaxies.   

\subsection{The galaxy formation model: {\tt GALFORM}}
\label{section:model}

The fate of baryons in a universe in which structure in the dark matter 
forms hierarchically depends on a range of often complex and nonlinear 
physical phenomena. The {\tt GALFORM} code models these processes using 
physically motivated recipes. Some parts of the model are better understood 
than others. For example, the merger history of dark matter haloes has 
been modelled extensively using N-body simulations of gravitational 
instability and accurate Monte Carlo techniques have been developed to 
replicate the merger histories (e.g. Parkinson, Cole \& Helly 2008). 
On the other-hand, the rate at which stars form from a 
reservoir of cold gas is not well understood theoretically 
and is modelled by adopting  
a prescription which contains parameters. The values of the parameters are 
fixed by requiring that the model reproduces a subset of observations of 
the galaxy population. The philosophy behind semi-analytical modelling is 
set out in the review by Baugh (2006). Full details of the {\tt GALFORM} 
model are given by Cole et~al. (2000) and in subsequent papers 
which have presented developments of the original model (Benson et~al. 
2003; Baugh et~al. 2005; Bower et~al. 2006; Font et~al. 2008). 
A useful summary of the model used in this paper, that of Baugh 
et~al. (2005), is given by Lacey et~al. (2008). The Baugh et~al. model 
reproduces the observed abundance of Lyman-break galaxies and galaxies 
detected with the SCUBA instrument. We use the ANN model to investigate 
the overlap between these populations in Section 7.

The key point to have clear is that {\tt GALFORM} predicts the full
star formation and chemical enrichment history of galaxies. The
starting point is the merger history of a dark matter halo. The rules
describing the baryonic physics are applied to gas in the merger tree,
starting from the branches which are in place at the earliest time.
The code then follows the gas cooling, star formation, feedback
processes and galaxy mergers. The star formation history, which
includes the metallicity of the stars made at each timestep, is the
primary ingredient required to compute the spectral energy
distribution of a galaxy. In its standard mode of operation, {\tt
GALFORM} uses a stellar population synthesis model (such as the one
devised by Bruzual \& Charlot 2003) to construct a composite stellar
population for each galaxy. Extinction of starlight by dust is
calculated by assuming that the dust and stars are mixed together,
rather than by treating the dust as a foreground slab. {\tt GALFORM}
predicts the half-mass radius of the disk and bulge components of each
galaxy (see Cole et~al. 2000; tests of the model for calculating sizes
are presented in Cole et~al. and also in Almeida et~al. 2007 and
Gonzalez et~al. 2008). Assuming a random inclination angle at which to
view the galactic disk, the attenuation of starlight is computed using
the tabulated results of radiative transfer calculations carried out
by Ferrara et~al. (1999).

\subsection{The spectro-photometric model: {\tt GRASIL}}

The {\tt GRASIL} code (Silva et~al. 1998) can be used to accurately
model the observed SEDs of galaxies over a wide range of wavelengths
-- from the far-UV to radio. In the standard application of {\tt
GRASIL}, a parameterized star formation history is tuned until a given
observed SED is reproduced \citep[e.g][]{bressan}.  The unique selling
point of {\tt GRASIL} is its sophisticated handling of the extinction
and reprocessing of starlight by dust. The galaxy as an axially
symmetric system with a disk and bulge component. The dust is assumed
to be divided into two phases: a diffuse component and dense,
star-forming molecular clouds, with the mass fraction between the two
being a model parameter. Stars are born within molecular clouds and
then escape after a few Myr. The extinction of the light from a set of
stars depends on their age relative to this escape time. High mass
stars, which typically dominate the emission at ultraviolet
wavelengths, spend a significant fraction of their short lifetimes
within the optically thick molecular clouds. Consequently, the
emission at these wavelengths is heavily extincted. The time for a
star to escape from a molecular cloud is a model parameter. {\tt
GRASIL} calculates the radiative transfer of starlight through this
dust distribution (molecular clouds and cirrus), and then solves for
the temperature distribution of the dust grains at each point in the
galaxy self-consistently based on the local stellar radiation field.
This temperature distribution is then used to calculate the dust
emission.  Effects of very small grains, subject to temperature
fluctuations, as well as polycyclic aromatic hydrocarbons (PAHs) are
included.  The model is calibrated against available data of normal
and starburst galaxies in the local universe \citep[][]{bressan, vega,
panuzzo, schurer}. The self-consistent calculation of dust
temperatures by {\tt GRASIL} avoids the need to impose a dust
temperature by hand, as is common in other models.

\subsection{A hybrid model: {\tt GALFORM} plus {\tt GRASIL}}

Granato et~al. (2000) described how the {\tt GRASIL} code can be used
to compute the SEDs of {\tt GALFORM} galaxies. The semi-analytical
code predicts the star formation history of each galaxy, outputting
the star formation rate in all the progenitors of the galaxy, stored
in bins of metallicity. The model also outputs the scale lengths of
the galaxy's disk and bulge, and the mass and metallicity of the cold
gas.  {\tt GRASIL} takes this information and produces an unextincted
and an extincted SED for the galaxy. The calculation carried out by
{\tt GRASIL} improves over the standard calculation made by {\tt
GALFORM} in two main areas: i) dust extinction at short wavelengths,
which is strongly affected by molecular clouds, is calculated more
accurately; and ii) the emission of radiation by dust is included.

\section{Artificial Neural Networks}
\label{section:ann}

Artificial neural networks (ANNs) are mathematical constructs  
designed to replicate the behaviour of the human brain. Given a 
training set of observations consisting of inputs with an 
associated set of outputs, the role of the ANN is to ``learn'' from 
these observations in order to be able to predict 
the output from a new set of inputs. The origins of the technique 
date back to \citet{first}, who developed a simple network using 
{\itshape artificial neurons} to perform logical operations. 
However, the concept of learning was only introduced a 
few years after this by \citet{hebb} and implemented by 
\citet{rosenblatt58,rosenblatt62}. Nowadays ANNs are widely 
used in computer science, finance, physics, mathematics, astronomy 
and many other areas. Typical applications include pattern 
recognition, function approximation, prediction and forecasting, 
and categorization. Even though neural networks have traditionally 
been viewed as black boxes, for which the user has little knowledge  
of their internal workings, they offer a number of advantages 
over other data mining and analysis tools, such as the ability to 
learn and applicability to a wide of problems. Also, ANNs can 
be readily parallelized. 

\subsection{Basic concepts}

In simple terms, the brain can be thought of as a collection of 
billions of special cells called neurons, which process 
information and are interconnected through synapses 
in a complex net. Neurons work by receiving electrochemical 
signals from other neurons, some of which will excite the cell 
whereas others will inhibit it. The neuron adds up these 
inputs and if the sum exceeds a certain threshold, it will 
transmit the same signal to other neurons. In this case 
the neuron is said to be activated.

ANNs are similar to their biological 
counterparts: they consist of simple computational units 
(also called neurons or nodes), which are connected in 
a network. For every neuron we need to specify the input 
connections and their associated weight, $w_j$. The neuron 
multiplies the input by its weight and adds the contributions 
from the interconnected units. The sum is then mapped by 
the activation function, $f$, to the output value, which, 
in turn, will become an input to the next group of adjacent 
neurons. If we define $i_{k}$ as the input signal coming from 
neuron $k$, and $w_{jk}$ as the weight between the input 
$k$ and neuron $j$, then the output, $o_ j$, from 
the neuron is given by:
\begin{equation}
 o_j = f \left(\sum_{k=1}^n w_{jk}\, i_k\right).
\end{equation}

There are numerous types of ANNs which differ in the way 
the neurons are organized and exchange information. It 
is common to group the neurons into 
layers. In general, there is an input layer, an output 
layer and some number of hidden layers in between. The input 
layer is responsible for handling the input data. It is clear that 
there is no activation function associated with this layer, 
because the output values of their neurons are simply set 
to be equal to their input values. The output of the 
network is recovered from the output layer. Using only one 
input layer and one output layer, it is possible to construct 
a very simple network called a perceptron.
The perceptron can recognize simple patterns in data. 
For more difficult tasks, we need hidden layers between the 
input and output layers. The term ``hidden'' is used because 
the user does not have direct access to the inputs and outputs 
dealt with by these layers.

Networks with more than just an input and output layer 
are called multilayer networks or multilayer perceptrons. 
They are the most widely used due to their ability to learn 
nonlinear functions.

The three most popular network configurations are: the perceptron 
(no hidden layers), the feed-forward and the recurrent network (the  
latter two cases both use hidden layers). Feed-forward nets 
are the most widely employed due to their simplicity. These ANNs 
pass information from the input layer, through the hidden layers 
to the output neurons. In recurrent networks, on the other hand, 
the output from the neurons can be fed backwards, through feedback 
connections, and act as input. Such behaviour is similar to that 
found in the biological brain. Even though recurrent networks can 
perform better than the feed-forward nets, they suffer from a 
major drawback: training is more difficult due to their 
oscillatory, even chaotic behaviour, resulting in longer computing times.

\subsection{Training}

There are two types of learning: supervised and unsupervised. 
In the first case, the network is presented with a target 
consisting of a set of inputs with associated outputs. The ANN 
adapts its weights in order to reproduce the desired output. 
In unsupervised learning, the network does not have a target output. 
In this case, the aim is to find patterns and to group the data. 
In this paper we focus on supervised learning.

There are several different approaches to supervised learning. 
Most share the common feature that the ANN learns by comparing the 
predicted output to the target output. The algorithm of this  
process is simple: (i) start with an untrained net; (ii) determine 
the output from a given input; (iii) compare the output to the target 
output and compute an error; (iv) adjust the weights in order to 
reduce the error.

The most widely used learning algorithm is the backpropagation 
algorithm, which was introduced by \citet{backprop}. This 
algorithm finds the 
local minimum of the error function:  
\begin{equation}
 \varepsilon = \frac{1}{2} \sum_k (t_k - o_k)^2,
\end{equation}
where $t_k$ represents the desired or target output values, and $o_k$ 
is the predicted output from the neuron.
Using the gradient descent method, it can be shown that the 
update to the weights from the hidden layer to the output layer 
is given by:
\begin{equation}
 \Delta w_{jk} = -\eta\,\Big(\frac{\partial \varepsilon}{\partial w_{jk}}\Big) = \eta\,\delta_j h_k,
\label{eq:ann.deltaw}
\end{equation}
with $\eta$ known as the learning rate, 
$\delta_j = (t_j-o_j)\,f^\prime(i_j)$ (where the activation function, 
$f$, is differentiable) and $h_k$ is the output from the preceding 
hidden neuron. A similar expression can be found for the variation 
of the weights between the input and hidden layers.

It is clear that if the surface corresponding to the error function 
has multiple local minima then this method, as originally defined, 
will only guarantee convergence towards one of the minima but not 
necessarily to the global minimum. However, this is not an insurmountable 
problem since, during the first steps of the gradient descent, the 
weights will gradually move towards the global minimum. Moreover, in 
the worse case scenario, the weights will converge to a local minimum 
in the vicinity. There are several methods to avoid this behaviour: we 
could add an extra factor to Eq.~\ref{eq:ann.deltaw}, 
$\beta\Delta w_{jk}^{0}$, called the momentum, which 
has the same direction as the previous step change, 
$\Delta w_{jk}^{0}$, and is controlled by the coefficient $\beta$.
Alternatively, we can train the network several times using the 
same training sample but with different initial random weights. 
The latter approach is the one we will follow in this paper. 
A further refinement is the resilient backpropagation 
algorithm \citep{rprop}, in which instead of adopting the full change 
in the weights specified by Eq.~\ref{eq:ann.deltaw}, we only use 
the sign of the derivative multiplied by a constant. We also adopt this 
method in our network. The resilient backpropagation algorithm has 
the advantage of being one of the fastest learning algorithms.

In an ideal situation, the ANN would, of course, find the optimal 
set of weights such that the error function is minimized. However, 
there is an important aspect that we have to bear in mind. One 
of the reasons why we use ANNs is that we need 
to achieve generalization; i.e. it is more important to find the 
network that best fits the testing or validation set, than it is to find 
the minimum of the error function for the training set. 
In fact, if the net is overtrained it will start to fit the noise associated 
with the data instead of the underlying signal. This leads to overfitting 
and consequently may affect the performance of the ANN on the validation 
sample. One of the procedures deployed to avoid overfitting is to use a 
so-called early stop. In this case, the training process is terminated 
when either the error function reaches a pre-defined threshold or a 
maximum number of iterations is reached. The latter choice is adopted 
in this paper. 

Finally, a brief word about the form of the activation function. As we saw, 
the activation function plays an important role in the neural network. 
It allow us to activate or deactivate neurons and adds the nonlinearity 
needed to solve complex problems. Evidently, activation functions only 
make sense for the hidden and output layers, not for input layers. 
There are many activation functions; in fact, any nonlinear function 
would fit the bill. The most common are: the sigmoid function, 
$f(x) = 1/\left(1+e^{-\alpha x}\right)$, where $\alpha$ is 
the {\itshape steepness}; the Gaussian, $f(x) = e^{-(\alpha x)^2}$; 
and the Elliot, $f(x) = \alpha x / \left(1 + |\alpha x|\right)$.
Our default choice is the sigmoid function; we contrast the performance of 
the ANN with this activation function against some of the others listed 
above in Section~4.5.3.

\section{Application of ANN to GALFORM plus GRASIL}
\label{section:performance}

Our main objective is to predict a galaxy's spectral energy distribution 
using a small set of its physical properties as predicted by \g~. 
This is far from a simple 
proposition, due to the complexity of the individual spectra and the 
wide range of spectral energy distributions found in a population of 
galaxies. In this section we explain how we use the ANN to predict 
spectra or luminosities, showing the first results and discussing some 
performance issues.

\subsection{Training and testing samples}

The training process is crucial for ANNs. The better the network 
learns about the characteristics of the training set, the better it 
will perform when predicting the spectra of a new set of galaxies.

\begin{figure}
{\epsfxsize=8.truecm
\epsfbox[18 144 592 718]{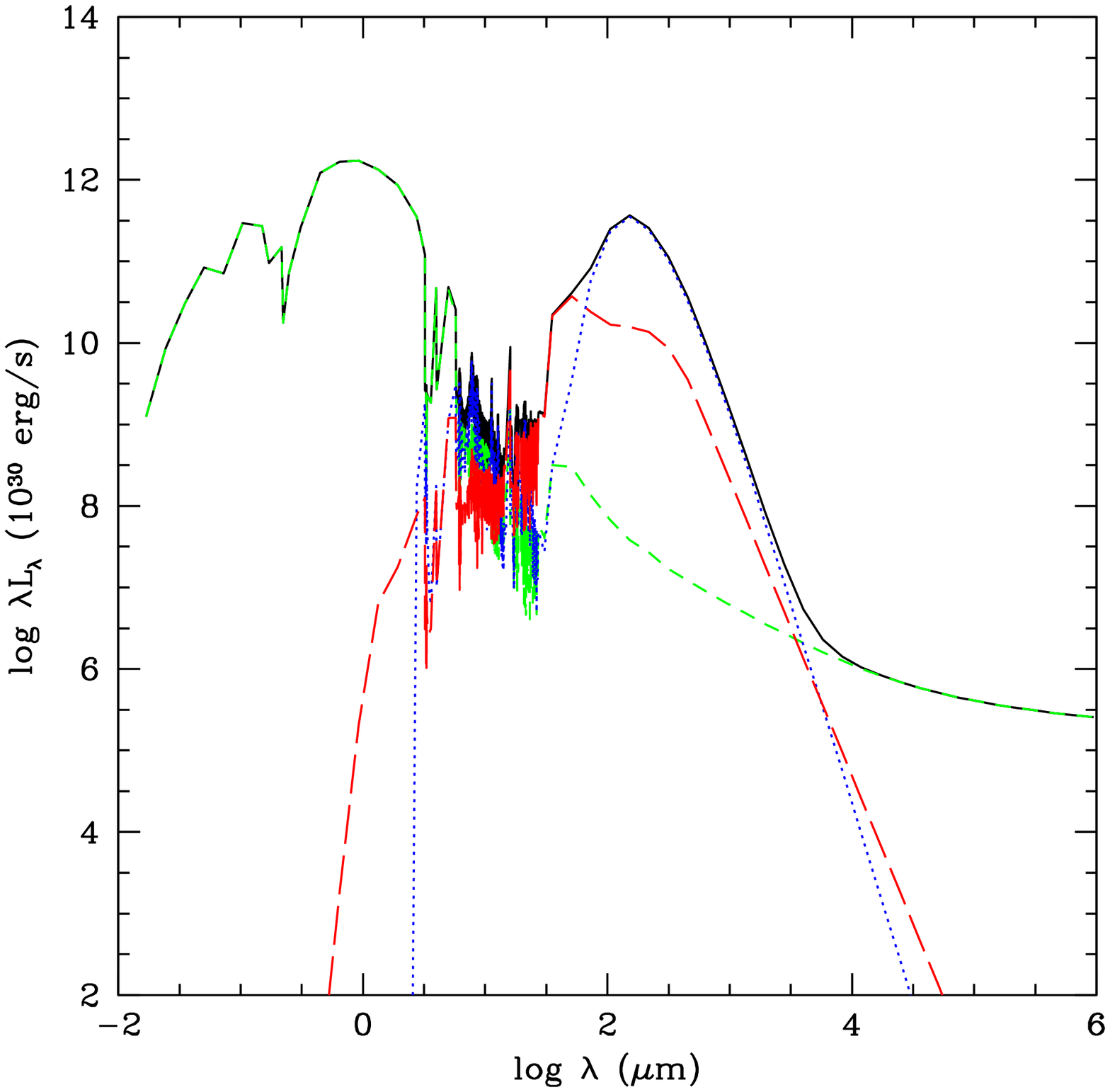}}
{\epsfxsize=8.truecm
\epsfbox[18 144 592 718]{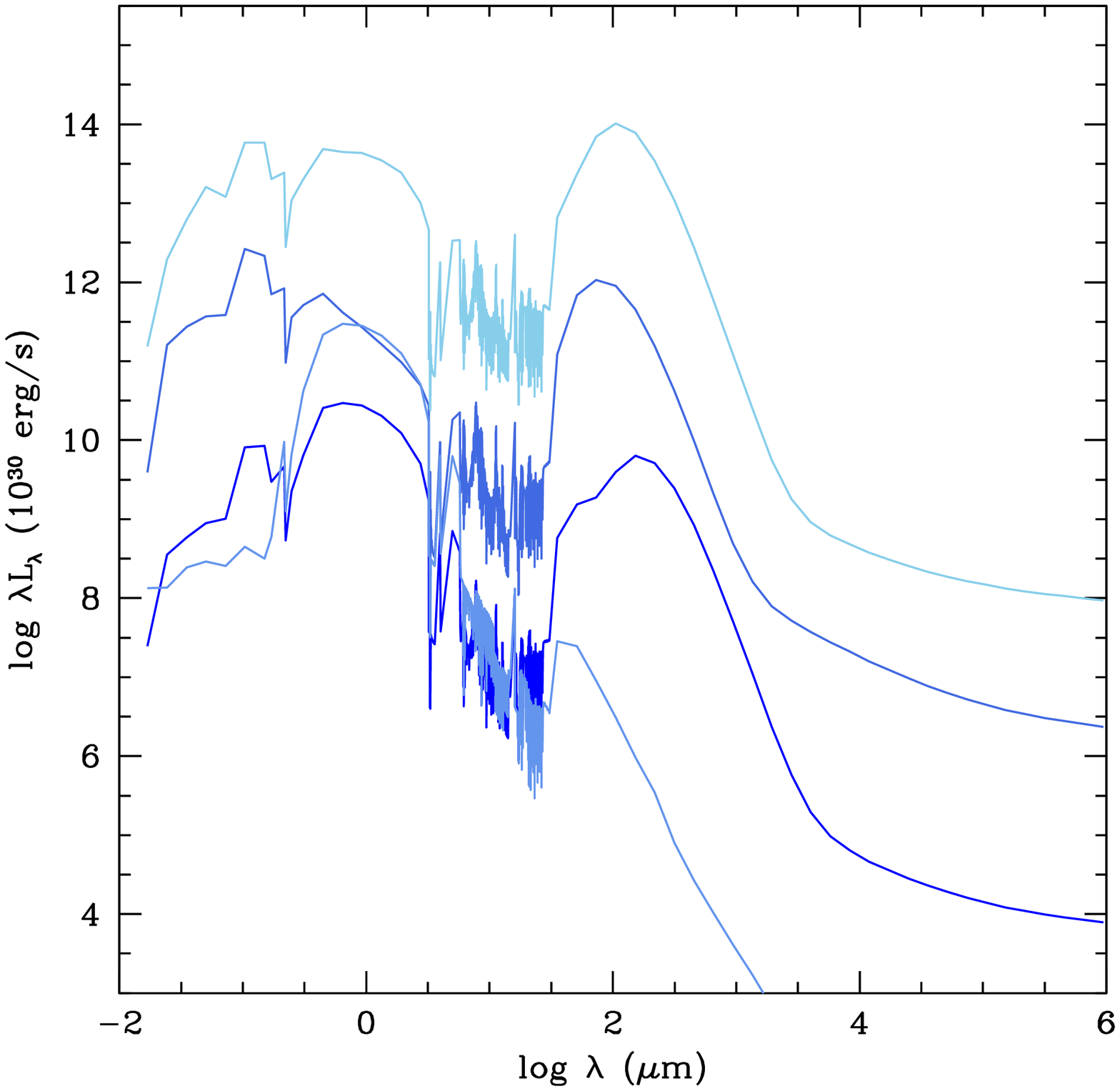}}
\caption{ Example galaxy SEDs, as computed by \gr~using star formation
histories predicted by {\tt GALFORM}. In the top panel, we show the
different components of a galaxy spectrum: the black line shows the
total SED, which is the result of adding the extincted star light
(green short-dashed line), and the emission from diffuse cirrus (blue
dotted line) and molecular clouds (red long-dashed line) emission. The
stellar contribution plotted here includes emission from dust in the
envelopes of AGB stars, and also thermal and synchrotron radio
emission at long wavelengths.  Selected examples of total spectra are
shown in the bottom panel.}
\label{fig:ann.seds}
\end{figure}

The galaxy spectra calculated by \gr~are far from simple. As noted in
section~\ref{section:model}, \gr~calculates the stellar emission, dust
extinction and dust emission, using the star formation and metal
enrichment histories predicted by {\tt GALFORM}. As a consequence,
\gr~spectra are complex and varied. The SEDs we compute from
\gr~comprise, in our application, 456 wavelength bins (this number can
be varied in \gr), so the output has a high dimensionality.
Fig.~\ref{fig:ann.seds} shows some examples of spectra produced by
\gr. In the top panel, we plot the spectral energy distribution of a
randomly selected galaxy (black line), showing the different
contributions (extincted starlight, molecular dust clouds and diffuse
dust). The mid-infrared emission in this particular galaxy is
dominated by PAH molecular bands and the far-infrared by cirrus
(diffuse dust) emission. Further examples of total galaxy SEDs are
shown in the bottom panel of this plot.

\begin{figure}
{\epsfxsize=8.truecm
\epsfbox[18 144 592 718]{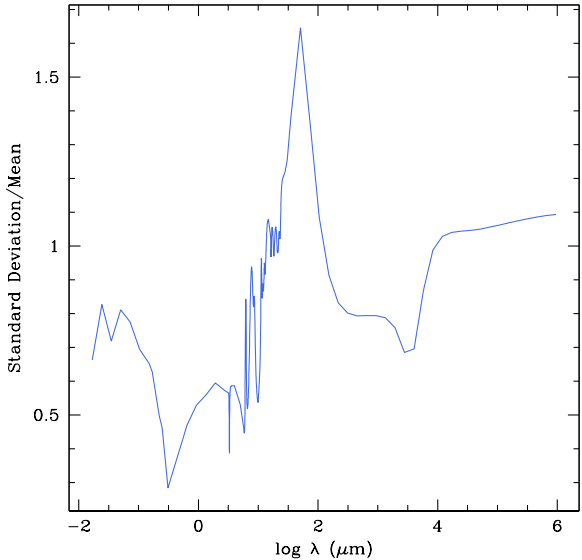}}
\caption{ The ratio between the standard deviation and the mean of
\gr~spectra for a representative sample of \g~galaxies. The spectra
were all normalized by dividing by the bolometric luminosity.  }
\label{fig:ann.sd}
\end{figure}

Fig.~\ref{fig:ann.sd} shows a quantitative view of the complexity 
of the spectra output by \gr~for a population of galaxies. We plot 
the ratio between the standard deviation and the mean of the normalized 
spectra for a representative sample of galaxies. This plot shows that the 
ultraviolet, mid infrared, microwave and the radio regions of the 
spectrum show the most variety in galaxy SEDs. The visible and 
far-infrared parts of the model spectra show, by comparison, 
less variance.

Each spectrum is composed of 456 flux bins, so our first approach will
be to set the number of output neurons in our net to be 456, one for
each flux bin. Later on we will try different methods in order to
reduce the dimensionality and variance of the output space.  For use
in the ANN, we first normalize the total luminosity in each SED to
unity. We then assign the logarithm of the flux at each wavelength to
these outputs in order to reduce the dynamic range of the training
data.

The selection of the input for the ANN is less straightforward.  The
natural choice would be to adopt the same input as used directly by
\gr~to create the spectra, i.e. the star formation and metal
enrichment histories along with the gas mass and metallicity, and the
scale-lengths of the disk and bulge. However, this is hard to
implement due to the enormous number of input variables implied (more
than 3000 taking into account the different timesteps and bins of
metallicity in which the star formation histories are stored). This,
in turn, would represent a substantial amount of computing time and
complexity for the learning process.  To keep things simple, we
decided to use a small set of galaxy properties, measured at the
output redshift at which the galaxy's SED is required. After some
investigation, we found that a useful set of galaxy properties to
serve as input to the ANN is: total stellar mass, stellar metallicity,
bolometric luminosity, circular velocity of the disc measured at the
half-mass radius, the effective circular velocity of the bulge, disc
and bulge half-mass radii, V-band luminosity weighted age, V-band dust
extinction optical depth, metallicity of the cold gas, the mass of
stars formed in the last burst and the time since the start of the
last burst of star formation. (Recall bursts are triggered by galaxy
mergers or by disks becoming dynamically unstable; the latter process
does not operate in the Baugh et~al. model which is used as an example
in this paper.) We therefore construct an input layer with 12 galaxy
properties. It is important to note that this set of input galaxy
properties has been tuned for the \citet{baugh05} model at $z=0$. For
a different model, the the ANN might perform better with a different
set of galaxy properties as inputs. We also note that the above list
of input galaxy properties includes the circular velocities of the
disk and bulge - these properties affect the SED through their effect
on the efficiency of supernova feedback and on the star formation
timescale.

In this section, the training and testing samples were extracted from 
a large catalogue of galaxies from the Baugh et~al. model at $z=0$, 
following a similar 
procedure to that used by \citet{grasil}. The \g~catalogue is sampled  
to give equal numbers of galaxies in logarithmic bins of total stellar 
mass.\footnote{In a later section, we will use also galaxies 
which have an ongoing burst or which recently experienced a burst. 
In this case, the sample is constructed using logarithmic bins of 
burst mass instead.} 
This strategy yields 1945 galaxy spectra for the training set, each 
of which is composed of 456 flux bins, i.e. a $1945\times456$ data array. 
A further set of 1898 galaxies were used as a validation sample.

\subsection{Predicting spectra}
\label{subsection:ann.predicted.spectra}

To predict luminosities, we use a supervised feed-forward 
neural network composed of 12 neurons in the input layer (which 
correspond to the 12 galaxy properties listed above), 60 neurons 
in one hidden layer and 456 neurons in the output layer (which 
are set to be equal to the logarithm of the flux in each of the 
spectrum bins). The ANN architecture is therefore 12:60:456.

Unless otherwise specified, the following procedures and parameters 
were chosen: (i) in order to deal with the different ranges of the 
input and output properties, we subtract the mean (computed over the 
training sample) from each input 
and output and divide by the respective standard deviation; (ii) we 
adopt a sigmoid activation function; (iii) the maximum number of 
training epochs is set to 5000 (i.e. this is the criteria used to stop the 
training process); (iv) in order to guarantee convergence towards the 
global minimum of the error function (see previous section), we train 
the ANN ten times using different initial random weights, and select 
the one that gives the smallest root mean square logarithmic error 
(see definition below) for the validation sample. Later in this section, 
we will show how the results change on modifying the ANN parameters.

\begin{figure}
{\epsfxsize=8.5truecm
\epsfbox[18 144 592 718]{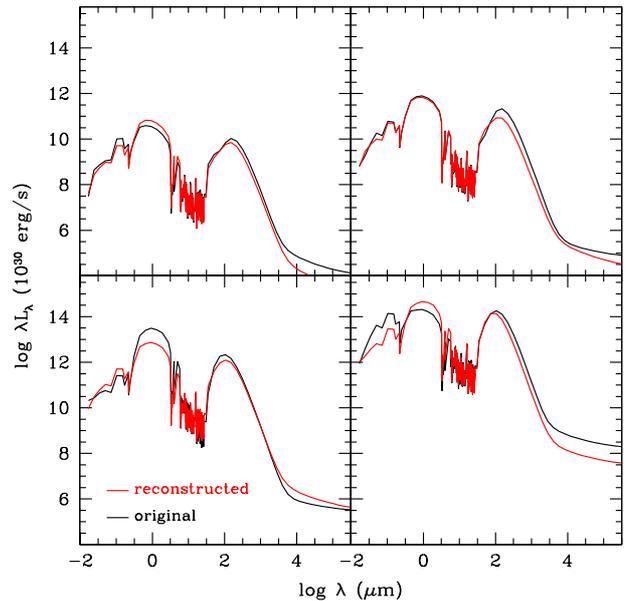}}
\caption{Four randomly selected examples of galaxy SEDs predicted by the 
ANN (red) compared with the SEDs calculated  directly by {\tt GRASIL} (black). 
}
\label{fig:ann.predicted.spectra}
\end{figure}

In Fig.~\ref{fig:ann.predicted.spectra} we plot four randomly 
selected examples of the spectra predicted by the ANN and compare 
these with the original spectra. Fig.~\ref{fig:ann.predicted.spectra} 
shows that even without further optimization, the spectra predicted 
using the ANN agree, on the whole, reasonably well with the original 
spectra, particularly at  visible and near-infrared wavelengths. 
However, in certain wavelength ranges, some galaxies exhibit predicted 
luminosities which differ by more than an order of magnitude from their true 
values.

\begin{figure*}
{\epsfxsize=8.truecm
\epsfbox[18 144 592 718]{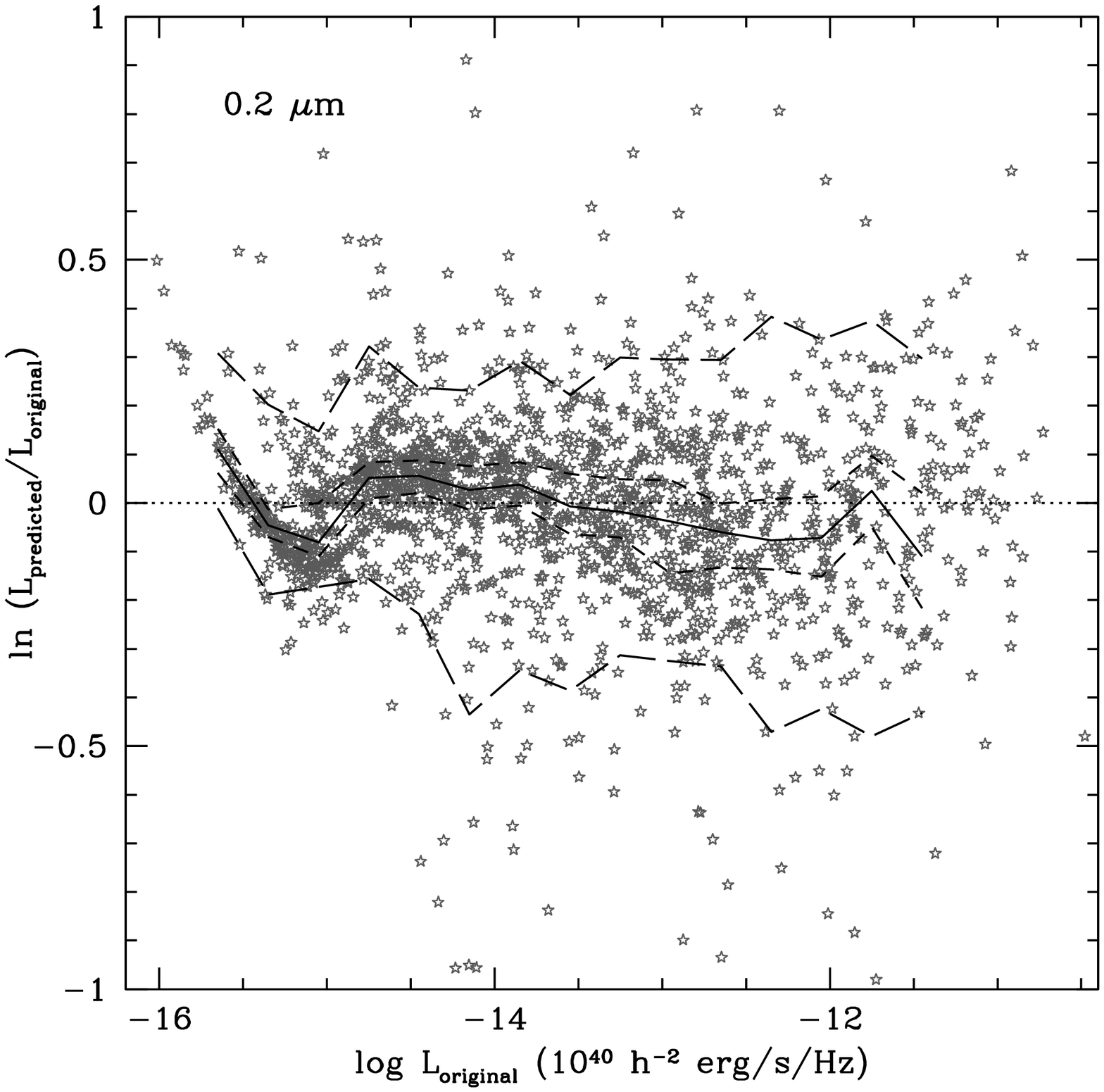}}
{\epsfxsize=8.truecm
\epsfbox[18 144 592 718]{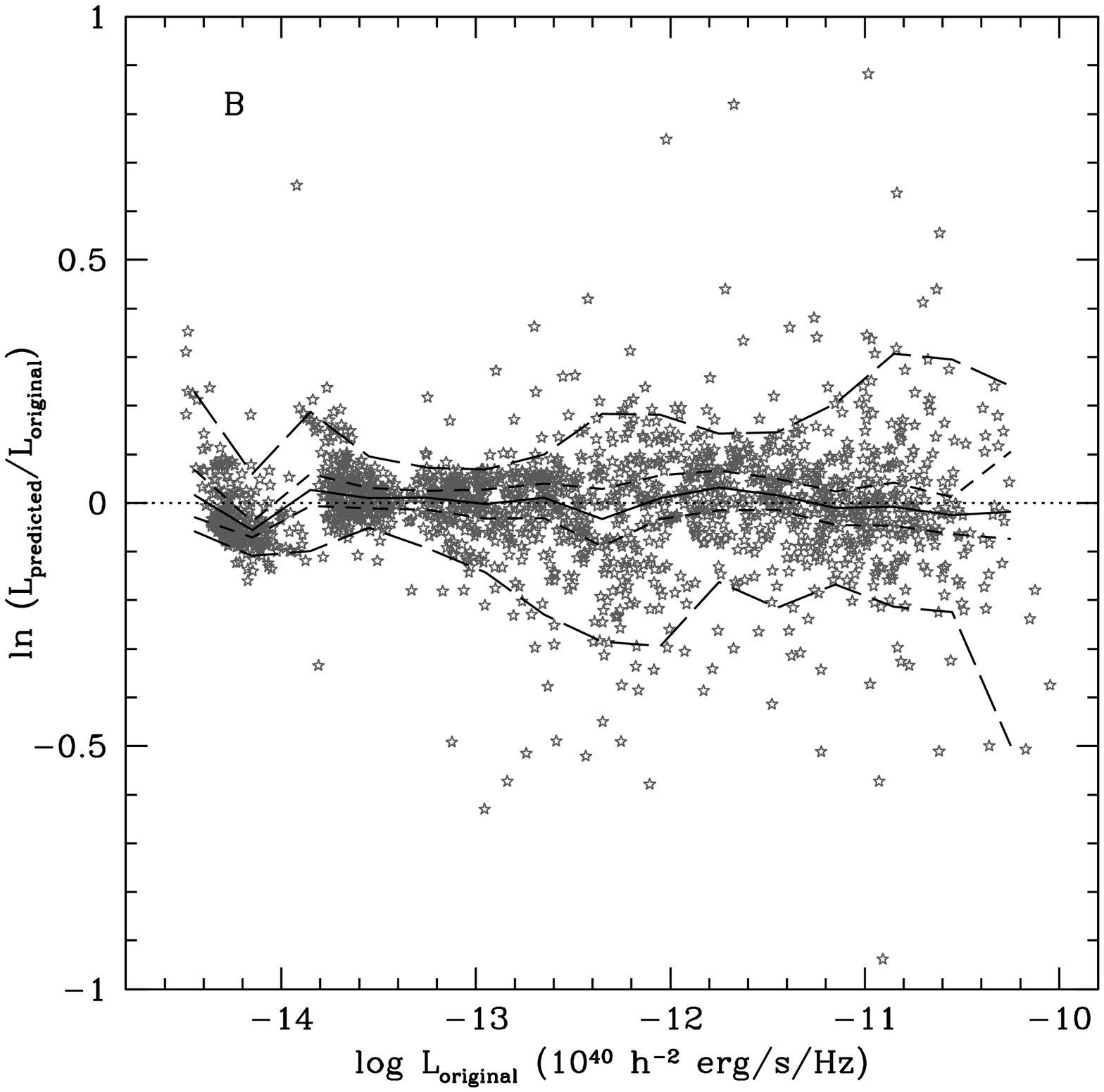}}
{\epsfxsize=8.truecm
\epsfbox[18 144 592 718]{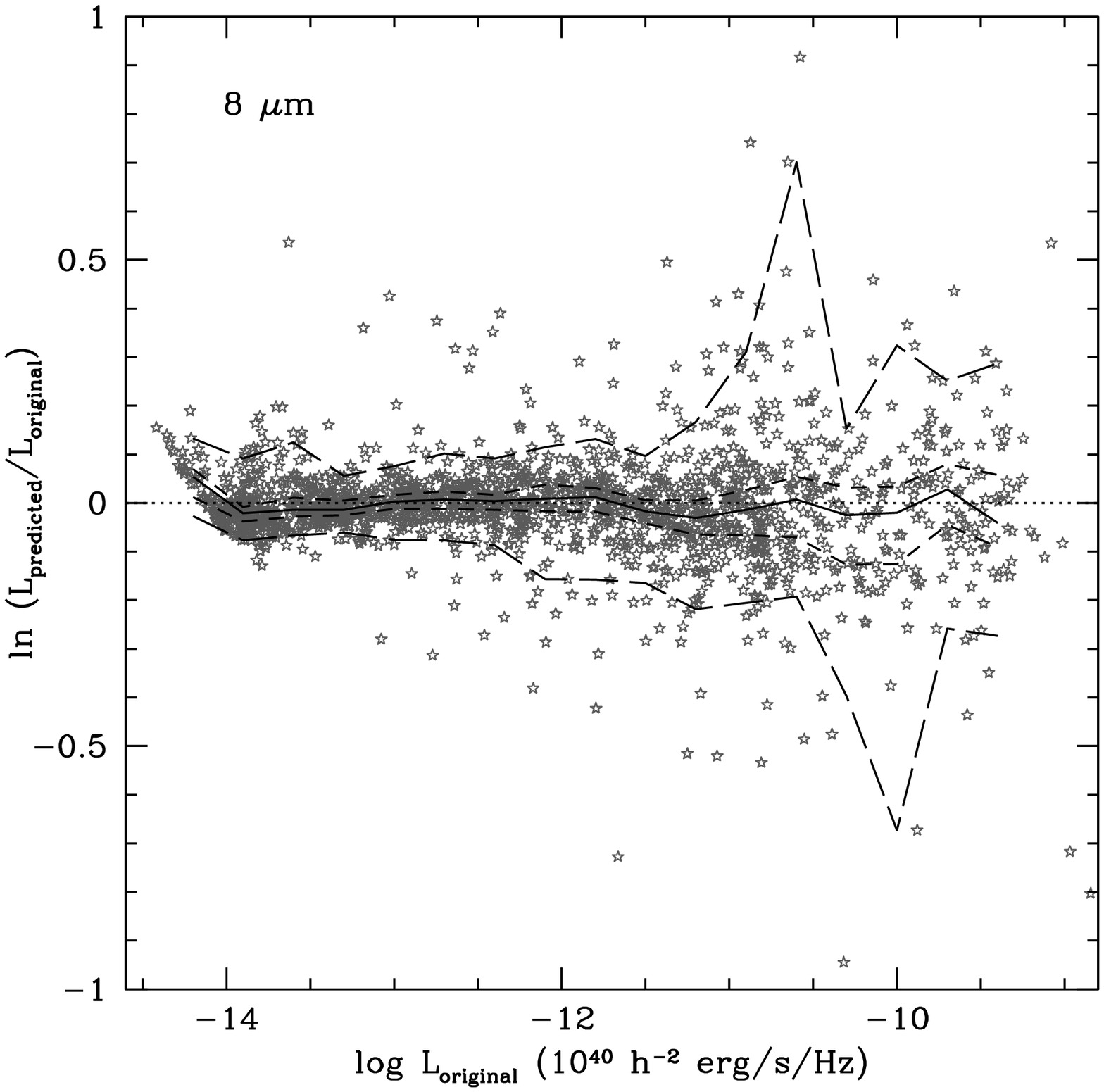}}
{\epsfxsize=8.truecm
\epsfbox[18 144 592 718]{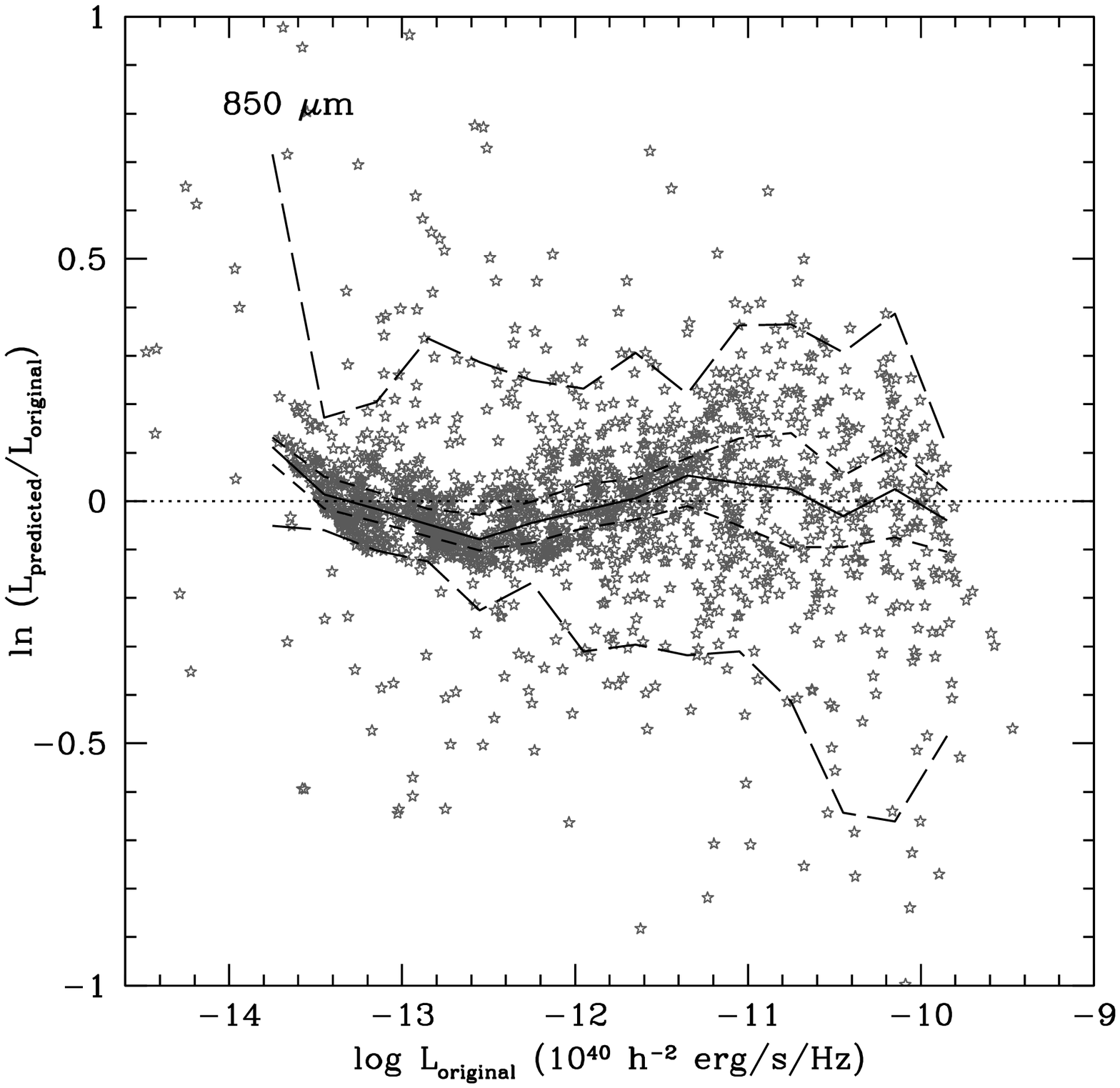}}
\caption{The logarithm of the ratio of the luminosity predicted by the ANN 
to the true luminosity, at selected wavelengths, for the case in which the 
ANN predicts the full SED. From top to bottom, left to right the panels 
show this ratio for the FOCA $0.2\,\mu$m, B ($0.44\,\mu$m), IRAC $8\,\mu$m 
and SCUBA $850\,\mu$m bands. The solid, short and long dashed lines 
show the median, the $33^{\rm rd}-66^{\rm th}$ and $5^{\rm th}-95^{\rm th}$ 
percentiles of the distribution, respectively.}
\label{fig:ann.compare.predicted.spectra}
\end{figure*}

To gain a more quantitative feel for the performance of the ANN, we 
plot in Fig.~\ref{fig:ann.compare.predicted.spectra} the ratio of the 
predicted to original luminosity for selected, representative 
wavelength bins: the 
FOCA (the Focal Corrector Anastigmat balloon borne camera) $0.2\,\mu$m, 
B ($0.44\,\mu$m), IRAC (the Infra Red Array Camera on Spitzer) $8\,\mu$m 
and SCUBA (Submillimetre Common User Bolometric Array) $850\,\mu$m bands. 
The statistics of the distributions are also summarized in 
Table~\ref{tab:ann.predicted.spectra}. Here the root mean squared logarithmic 
error is defined as:
\begin{equation}  
\varepsilon_L = \sqrt{1/n \sum^n [\ln (L_{\rm predicted}/
L_{\rm original})]^2} \, ,
\end{equation}  
and $P_{|e|<10\%}$ gives the percentage of galaxies with predicted 
luminosities which lie within 10\% of the true values. Note that 
$\varepsilon_L$ has a similar form to the error function which the 
ANN attempts to minimize, as given in Eq.~2. 

\begin{table*}
\begin{center}
  \begin{tabular}{cccccccc}
  \hline
 Band & $\varepsilon_L$ & $P_{|e|<10\%}$ & $p_{1}$ & Q$_1$ & Q$_2$ & Q$_3$ & $p_{99}$ \\
 \hline
 Bolometric		& 0.19     & 73.8 & -33.2   & -4.2  & 0.2  & 5.9 & 28.4 \\
 FOCA $0.2\,\mu$m	& 0.22     & 50.7 & -71.3   & -9.8  & -0.5 & 9.9 & 53.4 \\
 B ($0.44\,\mu$m)	& 0.12     & 75.8 & -38.3   & -5.1  & 0.2  & 5.6 & 31.6 \\
 IRAC $8\,\mu$m 	& 0.15     & 79.7 & -48.7   & -3.7  & 0.7  & 4.5 & 34.1 \\
 SCUBA $850\,\mu$m 	& 0.32     & 59.8 & -177.1  & -7.5  & 1.6  & 8.6 & 54.1 \\
  \hline
  \end{tabular}
   \caption{
Summary statistics for the distribution of the error on the spectra 
predicted by the ANN, when using the entire spectrum as the ANN output 
layer. We give errors on the predicted bolometric luminosity and for 
four different bands: FOCA $0.2\,\mu$m, B ($0.44\,\mu$m), IRAC $8\,\mu$m and SCUBA $850\,\mu$m. 
$\varepsilon_L$ is the root mean square error given by Eq.~4. 
$P_{|e|<10\%}$ shows the percentage of galaxies with predicted luminosities 
within 10\% of the true values. $p_1$, Q$_1$, Q$_2$, Q$_3$, $p_{99}$, give 
the $1^{\rm st}$ percentile, $1^{\rm st}$ quartile, median, $3^{\rm rd}$ 
quartile and $99^{\rm th}$ percentile of the error distribution, respectively.}
  \label{tab:ann.predicted.spectra}
  \end{center}
 \end{table*}

Fig.~\ref{fig:ann.compare.predicted.spectra} shows that overall there is 
reasonable agreement between the predicted and original luminosities 
in the analysed bands. For most galaxies, the predicted 
luminosity lies within 20\% of the original luminosity.
The performance of the ANN is wavelength dependent, with the results for 
the B and infrared bands being better than those for the ultraviolet and 
submillimetre bands. This is mainly due to the increased variance of 
the spectra in these later regions (see Fig.~\ref{fig:ann.sd}). 
We obtain a value of $\varepsilon_L=0.22$ in the UV $0.2\,\mu$m, 
$\varepsilon_L=0.12$ in the B band, $\varepsilon_L=0.15$ in the near-infrared 
and $\varepsilon_L=0.32$ in the SCUBA-$850\,\mu$m band. Also, the error has 
a tendency to increase with luminosity, as revealed by the broadening of  
the $5^{\rm th}-95^{\rm th}$ percentile range of the distribution. This 
suggests that the ANN has more difficulty dealing with bright galaxies. 
This could be due to the greater complexity of the star formation histories 
of these galaxies, with the mechanism invoked to suppress the formation of 
bright galaxies playing an increasingly important role for more luminous 
galaxies (i.e. superwind feedback in the case of Baugh et~al. 2005). This 
in turn will induce an increase in the variety of spectra produced by \gr.

In summary, this first attempt to predict spectra from a given set of 
galaxy properties, using artificial neural networks, has proven to work 
quite well. For most of the spectral range, we find that around 75\% of 
the predicted spectra deviate by 10\% or less from the true spectrum. 
However, there is a considerable error associated with this method, 
particularly for those wavelengths where the dust emission dominates 
over the stellar emission. In such cases, we found a high value of the 
statistic $\varepsilon_L$ and a reduction in the percentage of galaxies 
with predicted luminosities within 10\% of the true value. 

\subsection{Incorporating Principal Components Analysis of the spectra}
\label{subsection:ann.predicted.pca}

A simple way to speed up the ANN and to potentially boost its accuracy
is to reduce the dimensionality of the output, which in our case is
the number of bins used to describe the spectrum. The reason for this
is clear: the ANN should converge more rapidly to a trained network,
because there are fewer weights to be adjusted. The dimensionality of
the spectra can be reduced by using a principal components analysis
(PCA). The PCA works by finding patterns in the dataset, producing a
new set of linear, orthogonal basis vectors, which describe the
directions of maximum variance. Hence the spectra can be represented,
to a high level of accuracy, by a small number (e.g. around 10) of
basis vectors (each of which is 456 wavelength bins long).

The starting point of the PCA is the dataset of $n$ galaxy spectra
(again normalized to unit total luminosity and taken as logs), each of
which is described by a $m$-dimensional vector, $\vec{x}$ (in our case
$m = 456$). The data sample consists of $n\times m$ data points. The
first step of the PCA is to subtract the mean from all the data
dimensions, such that the mean of the $m$ data vectors is zero:
\begin{equation}
 \xi_{ij} = x_{ij} - \bar{x}_j ,
\end{equation}
where $i = 1, n$ and $j = 1, m$, and the mean is
\begin{equation}
 \bar{x}_j = \frac{1}{n}\sum_{i=1}^n x_{ij}.
\end{equation}
We then compute the covariance matrix
\begin{equation}
 C_{jk} = \frac{1}{n}\sum_{i=1}^n \frac{\xi_{ij}\,\xi_{ik}}{s_j\,s_k} ,
\end{equation}
where the variances of each variable are given by:
\begin{equation}
 s_j^2 = \frac{1}{n}\sum_{i=1}^n \xi_{ij}^2 .
\end{equation}
To find the axes of maximum variance we find the eigenvectors and 
eigenvalues of the covariance matrix:
\begin{equation}
 \vec{C}\,\vec{e}_j = \lambda\,\vec{e}_j .
\end{equation}
Note that we have assumed that the data set can be represented 
by a linear combination of the new eigenvectors.

The next step is to sort the eigenvectors in order of decreasing 
eigenvalue, which corresponds to decreasing variance. 
This is when the reduction of dimensionality is made. We can decide 
how many eigenvectors to retain, based on how many eigenvectors 
we think are sufficient to describe the original data to some desired 
level of accuracy. We keep the eigenvectors with the largest eigenvalues, 
which correspond to the axes along which the variance is highest. Once we 
have selected $p$ eigenvectors, we are ready to project our original 
data onto the new basis thereby retrieving the principal 
components (PCs) of the spectra:
\begin{equation}
 \vec{A} = \vec{\xi}\,\vec{E}_p .
\end{equation}
To go back to the original basis, from $p \leq m$ eigenvectors, 
and to (partially) reconstruct the original data, 
we use:
\begin{equation}
\label{eq:ann.reconstruct}
 \vec{x}_{\rm rec} = \vec{A}\,\vec{E}_{p}^{-1} + \bar{x}.
\end{equation}
If the entire set of principal components is used, no information is 
lost and the full spectrum can be recovered. With 20 principal 
components, however, we find that we can extract 99\% of spectral 
information.

The implementation of this method in the ANN framework is straightforward. 
We use the same input layer as in 
Section~\ref{subsection:ann.predicted.spectra}, 
i.e. 12 neurons corresponding to the selected galaxy properties, 
and 60 neurons will be used in the hidden layer. The number of output 
neurons is defined by the number of principal components we use. 
Once we establish this number, and having calculated the principal 
components and eigenvectors of the training sample, we use the ANN 
to predict the principal components instead of the full spectrum.
The network architecture in the case of $p$ PCs is 12:60:$p$. 
The final step is to reconstruct the full spectrum from the predicted 
principal components, using Eq.~\ref{eq:ann.reconstruct}, 
where $\vec{E}_p$ and $\bar{x}$ are the eigenvectors and 
mean of the training sample, and $\vec{A}$ is the predicted PCs. 
We use the same procedures and ANN parameters as defined in 
the previous subsection.

 \begin{table*}
\begin{center}
  \begin{tabular}{cccccccc}
  \hline
 Band & $\varepsilon_L$ & $P_{|e|<10\%}$ & $p_{1}$ & Q$_1$ & Q$_2$ & Q$_3$ & $p_{99}$ \\
 \hline
 Bolometric		& 0.19     & 78.3 & -33.8  & -4.2 & 0.5 & 5.4 & 27.7 \\
 $0.2\,\mu$m 		& 0.20     & 58.4 & -64.2  & -8.0 & 0.4 & 8.4 & 47.1 \\
 B (0.44 $\mu$ m)  			& 0.11     & 74.5 & -31.6  & -4.2 & 0.6 & 5.2 & 28.3 \\
 $8\,\mu$m 		& 0.16     & 78.7 & -80.2  & -3.9 & 0.5 & 4.3 & 30.7 \\
 $850\,\mu$m 		& 0.32     & 63.1 & -143.1 & -6.8 & 0.2 & 7.0 & 57.7 \\
  \hline
  \end{tabular}
   \caption{
Summary statistics for the error distribution of the spectra predicted 
by the ANN when using 20 principal components as the output layer.
Statistics are quoted for the bolometric luminosity and four  
different bands: FOCA $0.2\,\mu$m, B, IRAC $8\,\mu$m and SCUBA 
$850\,\mu$m. The description of the various quantities is given 
in Table~\ref{tab:ann.predicted.spectra}.}
  \label{tab:ann.predicted.pca}
  \end{center}
 \end{table*}

The statistics of the error distribution for this method are 
presented in Table~\ref{tab:ann.predicted.pca}. Using the principal 
component decomposition instead of the full spectrum does not lead 
to a dramatic improvement in the accuracy of the ANN. The results achieved 
with this technique are similar to those in the previous subsection, with, 
perhaps, a slight improvement: approximately 80\% of the predicted 
spectra now show bolometric luminosities which lie within 10\% of 
the original value, and $\varepsilon_L = 0.19$. Similar results are obtained 
for the different bands. Interestingly, both methods show similar 
difficulties when predicting the spectra, i.e. if a particular 
spectrum was not forecasted accurately by the method described 
in the previous subsection, then it is likely that it will differ 
substantially from its original value when the principal components 
are used to describe the spectrum. This is expected because with 
20 PCs we only lose 1\% of the spectral information.

The main advantage of compressing the spectra using PCA is the reduced 
computing time compared with using the full spectrum: the time 
required is reduced by a factor of $\sim 15$.

Table~\ref{tab:ann.different.pcs} shows the errors associated with the
bolometric luminosities of predicted spectra, using ANNs which
represent the spectra with different numbers of principal components.
The ANN predictions using PCA do not significantly improve when using
more than 10 PCs, as indicated by the percentage of spectra within
10\% of error, which is approximately constant at $\approx 80\%$.  We
remind the reader that most of the wavelength range of the spectrum
can be reconstructed with 10 or more principal components.  Hence we
expect the ANN to behave in a similar way as it does when using the
full spectrum.

\begin{table}
\begin{center}
  \begin{tabular}{ccc}
  \hline
 Number of 		& Bol. $\varepsilon_L$ & $P_{|e|<10\%}$ \\
Principal Components	&  & \\
 \hline
 1	& 0.79		& 16.5 \\
 3 	& 0.27 		& 27.8 \\
 5 	& 0.22		& 62.2 \\
 10	& 0.20		& 76.6 \\
 20	& 0.19		& 78.3 \\
 50	& 0.18		& 77.7 \\
 100 & 0.19		& 79.2 \\
 456 & 0.20		& 81.1 \\
  \hline
  \end{tabular}
   \caption
 {Summary statistics of the distribution of the predicted bolometric 
luminosity, for ANNs using different number of principal components 
(see Table~\ref{tab:ann.predicted.spectra} for a description of the quantities).}
  \label{tab:ann.different.pcs}
  \end{center}
 \end{table}

\subsection{Predicting the luminosity in a single band}
\label{subsection:ann.predicted.lums}

As we saw in the previous subsection, using just a few principal 
components of the spectrum, instead of the full 456 flux bins, 
facilitates the training process due to the reduced number of 
internal ANN weights which need to be adjusted. However, the gain 
in accuracy is marginal. In this section we explore another possible 
route to improve the accuracy of the ANN: the prediction of the 
luminosity in a single band instead of the full spectrum. The ANN 
becomes simpler in the sense that we only need to predict one variable, 
the band-pass luminosity. We would naturally expect the ANN to perform 
better for one band than in the case of trying to predict the full 
spectrum, as the power of the ANN is focused over a narrow range of 
wavelength. However, there is one drawback. If we require the luminosity 
in a range of bands, then with this approach we would need to train the 
ANN for each band in turn.\footnote{Note that later on we explore the 
performance of the ANN when predicting more than one band at a time.}

\begin{figure*}
{\epsfxsize=8.truecm
\epsfbox[18 144 592 718]{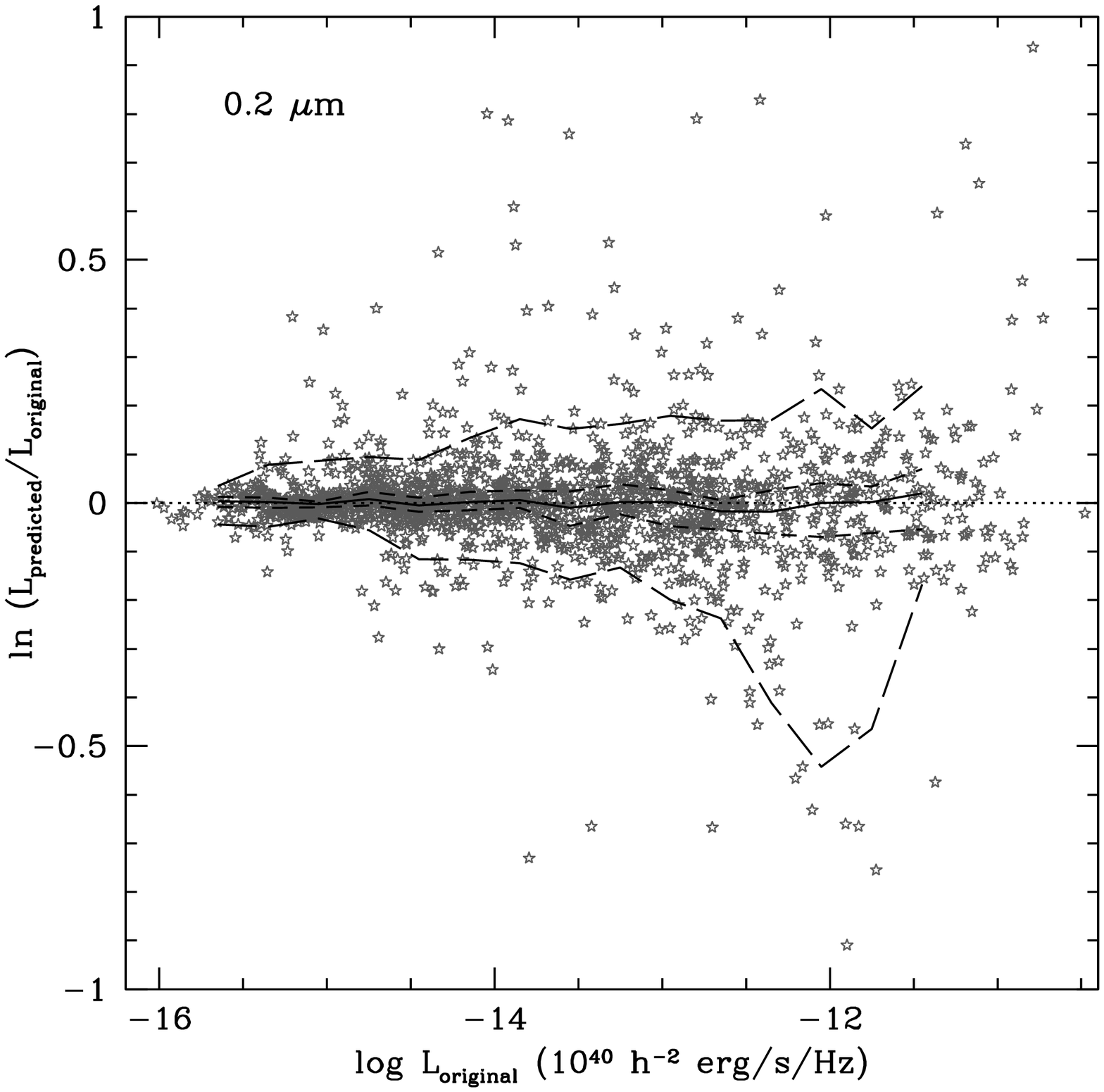}}
{\epsfxsize=8.truecm
\epsfbox[18 144 592 718]{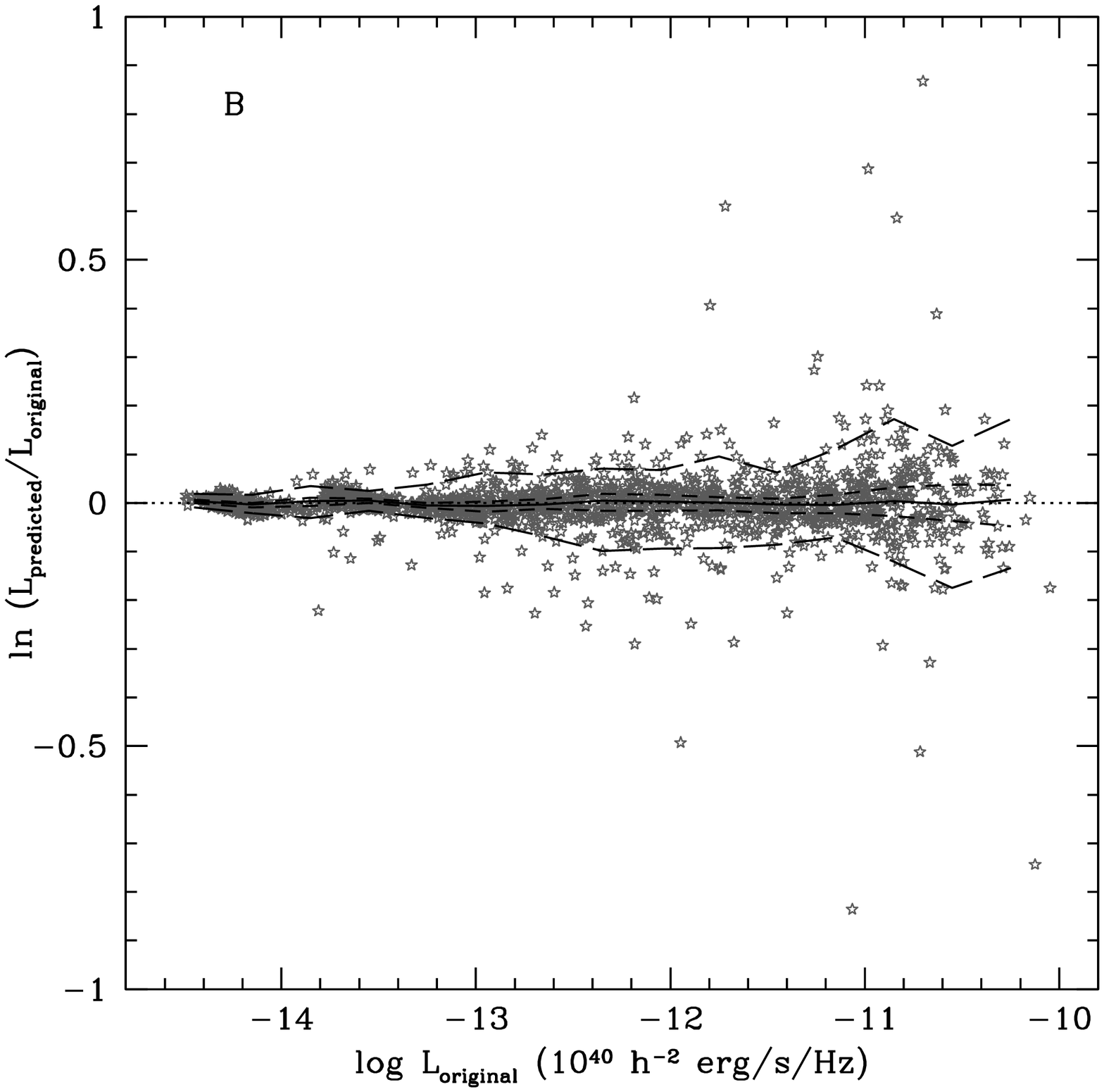}}
{\epsfxsize=8.truecm
\epsfbox[18 144 592 718]{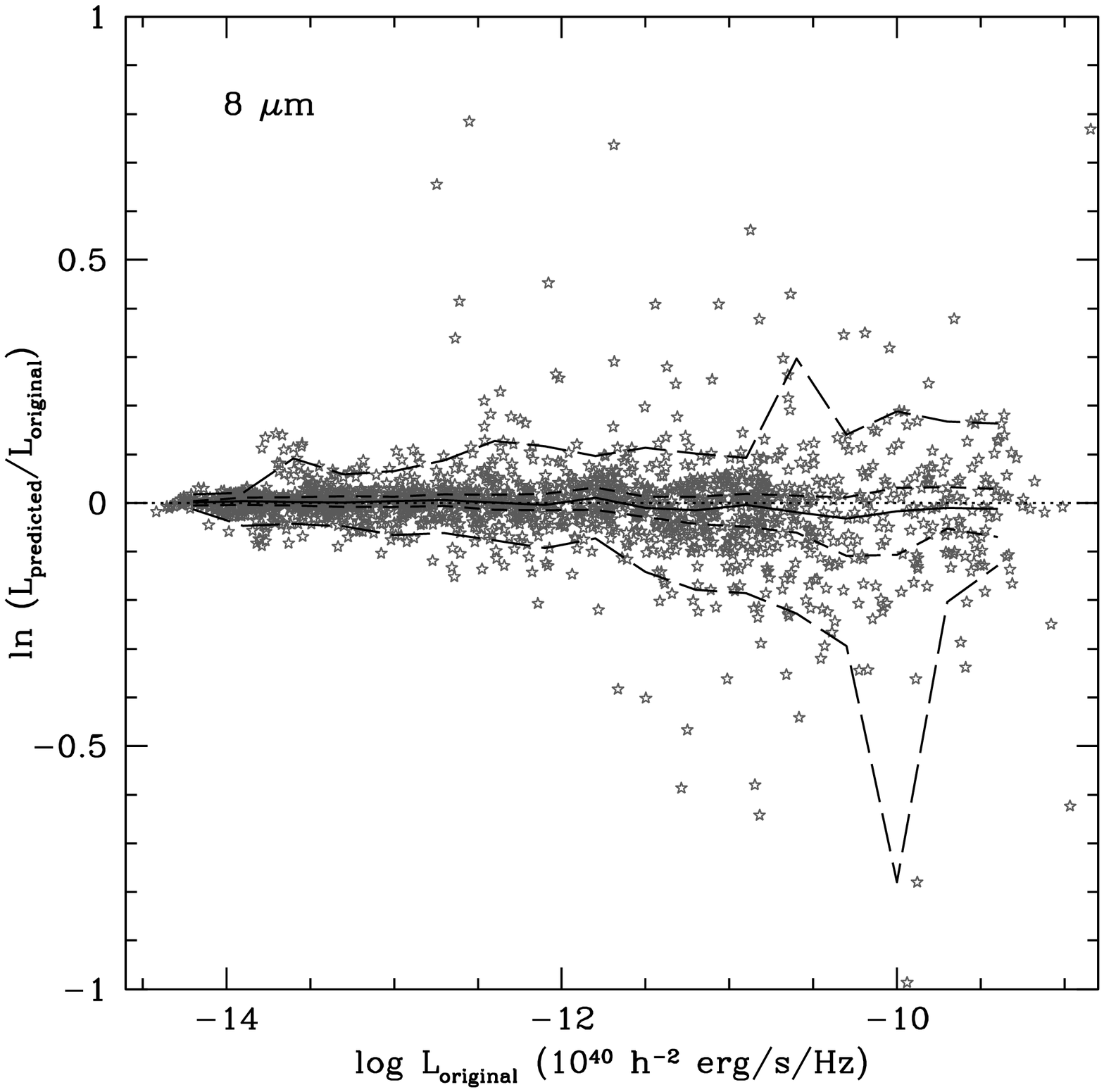}}
{\epsfxsize=8.truecm
\epsfbox[18 144 592 718]{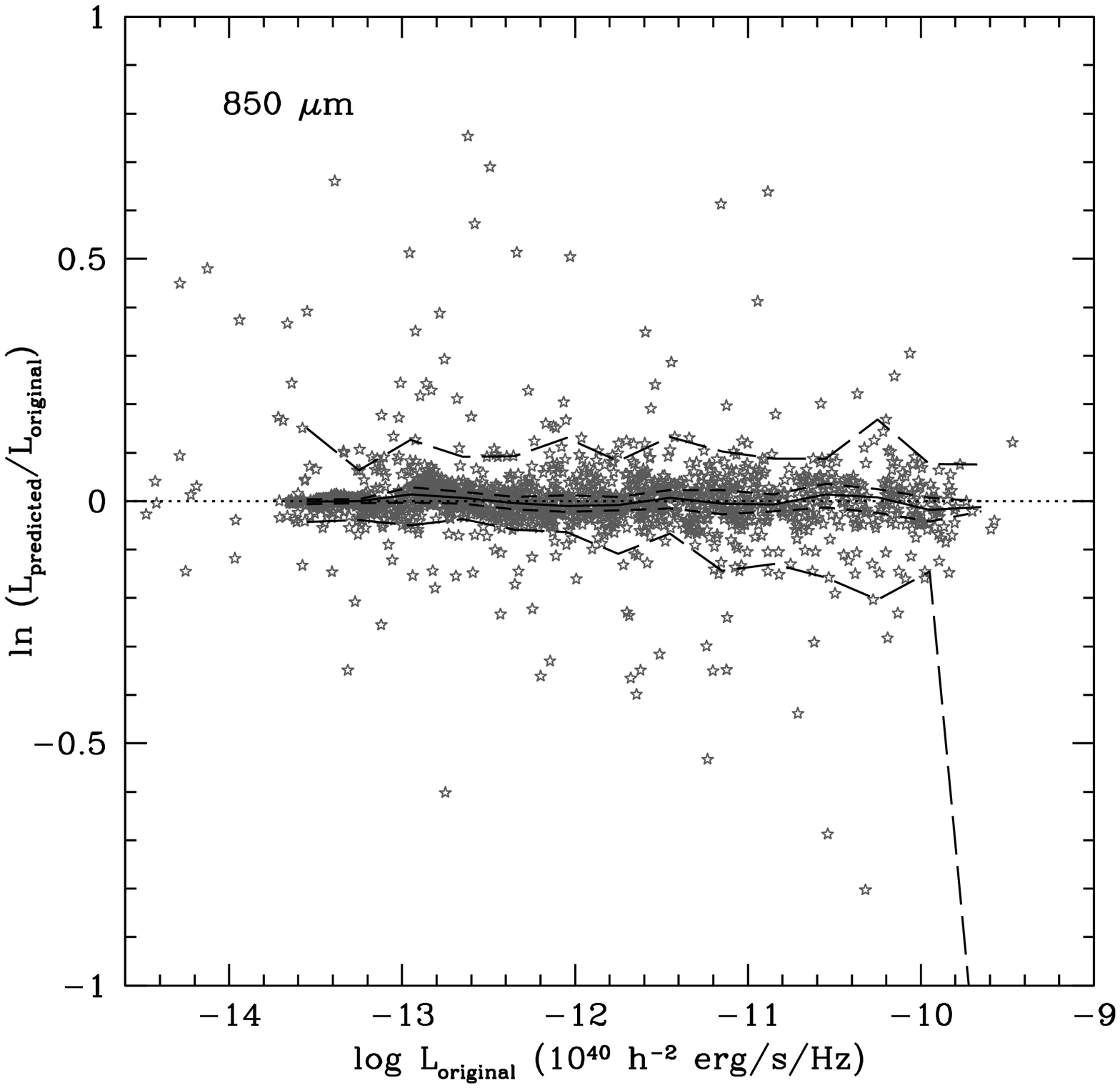}}
\caption{The logarithm of the ratio of predicted to true luminosity,
using the ANN applied to the prediction of a single band.  From
left to right, top to bottom we plot the ratios for the FOCA
$0.2\,\mu$m, B, IRAC $8\,\mu$m and SCUBA $850\,\mu$m bands. The solid
and dashed lines have the same meaning as in
Fig.~\ref{fig:ann.compare.predicted.spectra}.}
\label{fig:ann.compare.predicted.lums}
\end{figure*}
\begin{table*}
\begin{center}
  \begin{tabular}{cccccccc}
  \hline
 Band & $\varepsilon_L$ & $P_{|e|<10\%}$ & $p_{1}$ & Q$_1$ & Q$_2$ & Q$_3$ & $p_{99}$ \\
 \hline
 $0.2\,\mu$m 		& 0.14     & 80.1 & -49.4    & -3.4 & 0.0 & 3.6 & 36.6 \\
 B (0.44 $\mu$ m)			& 0.07     & 95.3 & -17.5   & -1.5 & 0.0 & 1.7 & 16.5 \\
 $8\,\mu$m 		& 0.16     & 88.5 & -43.4   & -2.2 & 0.0 & 2.9 & 28.8 \\
 $850\,\mu$m 		& 0.24     & 89.1 & -66.1 & -2.5 & 0.0 & 2.3 & 29.5 \\
  \hline
  \end{tabular}
   \caption {Summary statistics for the distribution of the error in the 
luminosities (FOCA $0.2\,\mu$m, B, IRAC $8\,\mu$m and SCUBA $850\,\mu$m) 
predicted by the ANN, when using one output neuron. See description of quantities 
in Table~\ref{tab:ann.predicted.spectra}.}
  \label{tab:ann.predicted.lums}
  \end{center}
 \end{table*}

To predict band luminosities we need to preprocess the training set spectra 
to calculate luminosities in a predefined set of bands (in this section 
we use the following bands: FOCA $0.2\,\mu$m, B ($0.44\,\mu$m), IRAC $8\,\mu$m 
and SCUBA $850\,\mu$m). We will start with a network configuration of 12:60:1, 
i.e. 12 neurons in the input layer, one hidden layer with 60 neurons each 
and 1 output neuron corresponding to the desired luminosity. The ANN is 
trained separately for each of the selected bands, using procedures and 
parameters similar to those used in Section~\ref{subsection:ann.predicted.spectra}. 
The results are shown in Fig.~\ref{fig:ann.compare.predicted.lums}, and the 
distribution of the predicted luminosity is given in Table~\ref{tab:ann.predicted.lums}.

The approach of training the ANN to predict one band at a time performs 
much better than the previous neural nets. The proportion of galaxies with 
predicted luminosities which are within 10\% of the true luminosity is 
significantly higher than before, 88\% compared with $\sim 68\%$, when the 
full spectra were used as the output. A similar improvement is seen in the 
root mean square logarithmic error. The results are particularly impressive 
for the B-band, where $\varepsilon_L = 0.07$ and more than 95\% of the population 
have predicted luminosities within 10\% of the original. 
As with the previous ANN, the results in the UV and infrared/submillimetre 
bands the ANN are not as good as in the B-band, due to the increased variety 
in the model spectra at these wavelengths. Nevertheless, the performance 
of the ANN is still markedly better than before even at these wavelengths 
(see Fig.~\ref{fig:ann.sd}). It is also apparent from this plot that the 
scatter around the desired relation increases slightly with the luminosity 
of the galaxy. No correlation was found between the errors and the other 
galaxy properties at the output redshift.

\begin{table*}
\begin{center}
  \begin{tabular}{cccccccc}
  \hline
 Band & $\varepsilon_L$ & $P_{|e|<10\%}$ & $p_{1}$ & Q$_1$ & Q$_2$ & Q$_3$ & $p_{99}$ \\
 \hline
$0.2\,\mu$m		& 0.15     & 79.5 & -34.5    & -3.8 & 0.1 & 3.5 & 42.4 \\
 B (0.44 $\mu$ m)			& 0.08     & 95.2 & -15.3    & -1.1 & 0.1 & 1.3 & 17.7 \\
 $8\,\mu$m 		& 0.13     & 88.9 & -28.8    & -2.6 & -0.2 & 2.4 & 24.3 \\
 $850\,\mu$m 		& 0.25     & 80.7 & -68.1   & -4.1 &  0.0 & 3.6 & 50.1 \\
  \hline
  \end{tabular}
   \caption {
Summary statistics for the distribution of the error on the 
luminosities predicted by the ANN, using four output neurons: 
FOCA $0.2\,\mu$m, B, IRAC $8\,\mu$m and SCUBA $850\,\mu$m. 
For a description of the quantities see 
Table~\ref{tab:ann.predicted.spectra}.
}
  \label{tab:ann.predicted.lums.3}
  \end{center}
 \end{table*}

As mentioned earlier, there is also the possibility of using the ANN to 
predict $n$ luminosities at a time. The advantage of this is that we 
would only need to train the network once, instead of having to train 
it $n$ times, once for each band. Table~\ref{tab:ann.predicted.lums.3} 
lists the statistics of the error distribution using this procedure. 
The ANN is trained using four neurons in the output layer, one for each 
band (FOCA $0.2\,\mu$m, B, IRAC 8 $\mu$m and SCUBA 850 $\mu$m).
This variant method performs slightly worse than the case in which 
the ANN is trained four times, once for each of the luminosity bands, 
with a percentage of 80\%, 95\%, 89\% and 81\% of galaxies with predicted 
luminosities within 10\% of the true luminosity, in the FOCA 
$0.2\,\mu$m, B, IRAC $8\,\mu$m and SCUBA $850\,\mu$m bands, respectively. 
Nevertheless, it still outperforms the first two methods that we explored 
in which the full spectrum was predicted. 

Unless otherwise stated, from now on we focus on the predictions of 
the best performing method as described at the beginning of this 
subsection, namely using the ANN to predict the luminosity in one band 
at a time.

\subsection{Performance for different ANN choices}

In this section we explore the architecture of the ANN, along with 
some of the parameter space of the ANN, the sample extraction and the 
effect of using different redshifts.

\subsubsection{Architectures}

 \begin{figure}
{\epsfxsize=8.truecm
\epsfbox[18 144 592 718]{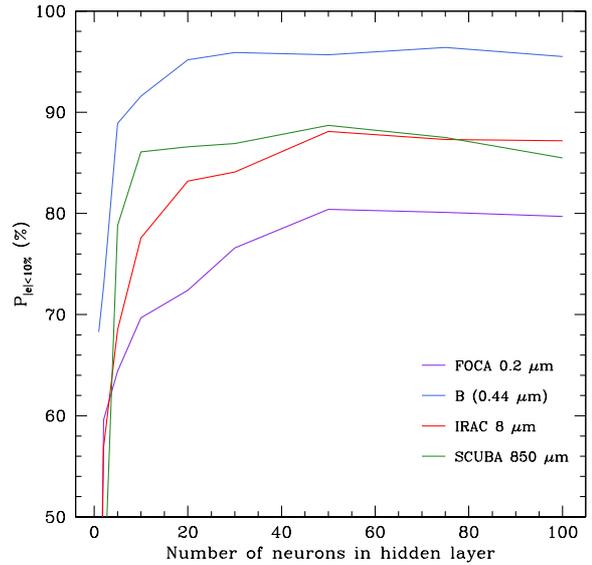}}
\caption{The evolution of the percentage of galaxies with predicted 
luminosities which lie within 10\% of the true luminosity, as a 
function of the number of neurons in the hidden layer. For all 
configurations tested, we stopped the training process after 5000 
epochs. The violet, blue, red and green lines show the results for 
the FOCA $0.2\,\mu$m, B, IRAC 8 $\mu$m and SCUBA 850 $\mu$m bands, 
respectively.}
\label{fig:ann.nhidden}
\end{figure}

So far, we have used a supervised feed-forward neural network 
with 12 input neurons, one hidden layer with 60 neurons and an 
output layer with one or more neurons, depending on the method 
under consideration. The number of input and output neurons are 
effectively determined by the setup of the problem. On the other 
hand, the number of hidden layers and the number of neurons each 
contains are parameters that are more subjective. Currently, there 
is no clear consensus on how many hidden units should be used 
\citep[see][]{scarselli}. Each application will have its optimal 
set of parameters, which can only be found by trial and error. 
In Fig~\ref{fig:ann.nhidden}, we show the evolution of the 
percentage of galaxies with predicted luminosities which lie 
within 10\% of the true luminosity, as a function of the number of 
neurons used in the hidden layer of the ANN. Each curve shows the 
performance for a different filter. The training process for all 
configurations was stopped after 5000 epochs. We see that there is 
little variation in the performance of the network for architectures 
with more than 20-30 neurons in the hidden layer. For two hidden layers, 
the results are similar to those found for one layer, for the same total 
number of neurons. However, in some cases, for example the FOCA 0.2 $\mu$m 
band, the use of two hidden layers seems to help slightly in terms of 
the $P_{|e|<10\%}$, which changes from 80.4\% to 82.5\%, for one and two 
layers respectively (there is also a reduction of the SCUBA 850 $\mu$m 
$\varepsilon_L$ by a factor of $\approx 2$, when two layers are used). The use 
of more than two layers does not improve the results further. Henceforth, 
we will use a configuration of 12:30:30:1 throughout this paper.

\subsubsection{Number of training epochs}

\begin{figure}
{\epsfxsize=8.truecm
\epsfbox[18 144 592 718]{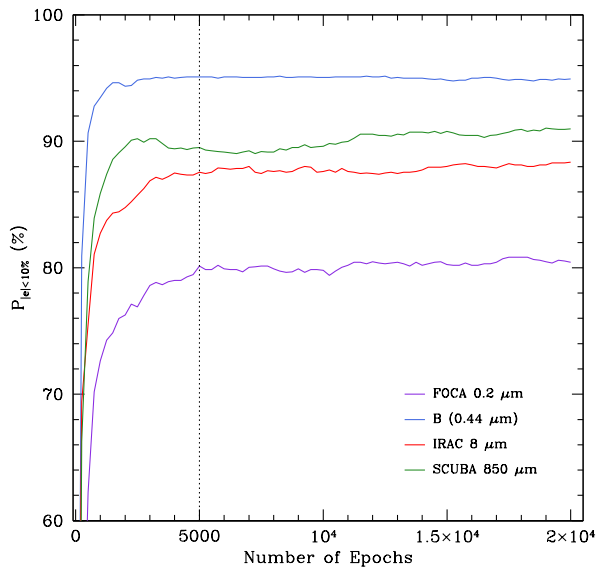}}
\caption {
The dependence of the percentage of galaxies with predicted luminosities 
within 10\% of the true luminosity, $P_{|e|<10\%}$, on the number of 
training epochs, for an ANN configuration of 12:30:30:1. The 
violet, blue, red and green lines show the results for the 
FOCA 0.2 $\mu$m, B, IRAC 8 $\mu$m and SCUBA 850 $\mu$m bands, respectively. 
The dotted line shows 5000 training epochs, for reference.
}
\label{fig:ann.epochs}
\end{figure}

As explained in Section~\ref{section:ann}, two methods can be used to 
stop the training process in order to avoid overfitting: applying 
a pre-defined error threshold or setting a maximum number of epochs. 
In this paper, we use a maximum number of epochs, 5000, as the 
criteria for early stopping. 
In Fig.~\ref{fig:ann.epochs} we show how the percentage of galaxies 
with predicted luminosities within 10\% of the true luminosity 
($P_{|e|<10\%}$) depends on the number of epochs, when using a network 
configuration of 12:30:30:1. All the four networks show rapid convergence: 
after 1000 iterations, $P_{|e|<10\%}$ in the FOCA 0.2 $\mu$m, B, IRAC 8 
$\mu$m and SCUBA 850 $\mu$m bands are 79\%, 94\%, 88\% and 87\%, respectively. 
The plot shows that, after 5000 epochs, the networks have already 
converged to their optimal states, after which there is no noticeable 
change in the $P_{|e|<10\%}$. The value of 5000 epochs was kept as the 
early stop parameter for the training of the ANN. 

\subsubsection{Choice of activation function}

\begin{table*}
\begin{center}
  \begin{tabular}{ccccccccc}
  \hline
 Activation Function 	& \multicolumn{2}{c}{0.2 $\mu$m} & \multicolumn{2}{c}{B} & \multicolumn{2}{c}{8 $\mu$m} 
	& \multicolumn{2}{c}{850 $\mu$m}\\
	&  $\varepsilon_L$ & $P_{|e|<10\%}$ & $\varepsilon_L$ & $P_{|e|<10\%}$ & $\varepsilon_L$ & 
	$P_{|e|<10\%}$ & $\varepsilon_L$ & $P_{|e|<10\%}$ \\
 \hline
 Elliot		& 0.14	& 78.6	& 0.07	& 94.4	& 0.13	& 87.5	& 0.23 	& 79.6 \\
 Sigmoid	& 0.14	& 80.1	& 0.07	& 95.3	& 0.16	& 88.5	& 0.24	& 89.1 \\
 Gaussian	& 0.16	& 77.9	& 0.07	& 94.6	& 0.14	& 87.7	& 0.27	& 74.6 \\
 Linear		& 0.55	& 20.5	& 0.22	& 49.7	& 0.39	& 22.0	& 1.08	& 11.0 \\
  \hline
  \end{tabular}
   \caption{The performance of neural nets with different hidden layer activation 
functions: elliot, sigmoid, gaussian and linear (see Table~\ref{tab:ann.predicted.spectra} 
for a description of the statistical quantities).}
  \label{tab:ann.activation.functions}
  \end{center}
 \end{table*}

Activation functions are needed in order to add nonlinearity to the 
training process. So far, for the hidden neurons, we have been using 
the sigmoid function, given by: $f(x) = 1/ \left(1+e^{-\alpha x} \right)$, 
where the coefficient $\alpha$ is commonly referred to as the steepness 
of the activation function; the steepness of choice is $\alpha = 0.02$.  
For the output neurons, the linear activation function, $f(x) = \alpha x$, 
was adopted.

In Table~\ref{tab:ann.activation.functions} we quote the performance of 
ANNs with different activation functions for the hidden layers for 
the FOCA 0.2 $\mu$m, B, IRAC 8 $\mu$m and SCUBA 850 $\mu$m bands.
As expected, the best results are achieved with the nonlinear functions. 
The sigmoid activation function seems to slightly outperform the elliot 
and gaussian ones. A linear activation function should have the effect 
of removing all nonlinearity from the neural network training, and 
consequently the ANN will be no more than a simple perceptron 
(see Section~\ref{section:ann}). As is clear from 
Table~\ref{tab:ann.activation.functions}, the capabilities of neural 
network are greatly reduced when we use only linear correlations. The 
errors associated with the predicted luminosities increase, which leads 
to merely 11\% of galaxies with errors smaller than 10\% in the 
SCUBA 850 $\mu$m band. Similarly poor results are obtained in the 
other bands.

In view of these results, we will use the sigmoid activation function 
throughout this paper.

\subsection{ANN Performance: Normal and burst galaxy samples}

\begin{table*}
\begin{center}
  \begin{tabular}{cccccccccc}
  \hline
 \multicolumn{2}{c}{Network } & \multicolumn{2}{c}{0.2 $\mu$m} & \multicolumn{2}{c}{B} 
	& \multicolumn{2}{c}{8 $\mu$m}  & \multicolumn{2}{c}{850 $\mu$m}\\
Train	& Test	& $\varepsilon_L$ & $P_{|e|<10\%}$ & $\varepsilon_L$ & $P_{|e|<10\%}$ & $\varepsilon_L$ 
& $P_{|e|<10\%}$ & $\varepsilon_L$ & $P_{|e|<10\%}$ \\
 \hline
 Normal	& Normal	& 0.14	& 80.1	& 0.07	& 95.3	& 0.16	& 88.5	& 0.24	& 89.1 \\
 Normal	& Burst		& 0.70	& 15.6	& 0.58	& 20.3	& 0.55	& 21.6	& 1.37	& 11.6 \\
 Quiescent & Quiescent	& 0.12	& 81.2	& 0.04	& 97.1	& 0.08	& 89.1	& 0.07	& 94.8 \\
 Burst	& Burst		& 0.22	& 59.9	& 0.07	& 88.7	& 0.15	& 63.9	& 0.47	& 79.1 \\
  \hline
  \end{tabular}
   \caption {The performance of several neural networks, trained with different 
galaxy samples: ``normal'', ``quiescent'' and ``bursts''. See text for further 
details, and also Table~\ref{tab:ann.predicted.spectra} for a description of the 
quantities listed).}
  \label{tab:ann.bursts}
  \end{center}
 \end{table*}

Until now, we have been using a sample extracted from the \g~catalogue 
following a similar procedure to that outlined in \citet{grasil} for 
``normal galaxies'' (see definition below). In this subsection, we 
distinguish between quiescent and burst galaxies, and analyse the 
performance of the ANN in both cases. Quiescent and burst galaxies 
are sampled differently in the \citet{baugh05} model at $z=0$. It is rare 
to catch a galaxy undergoing a starburst, so it is necessary to sample the 
bursts carefully to build up a statistical sample. In the present work, we 
do not calculate the burst spectrum for a fixed set of times after 
the start of 
the burst, as done in \citet{grasil}. Instead, we enlarge the burst 
sample by simply increasing the volume of the simulation run 
in {\tt GALFORM}. 

A galaxy is considered to be a burst galaxy, i.e. an ongoing burst or
a recent burst in which the stars formed in the burst still have an
impact on the spectral energy distribution, if $t_{\rm burst} \le 10
\, \tau_e$, where $t_{\rm burst}$ is the time since the start of the
most recent burst, and $\tau_e$ is the effective e-folding time for
the starburst (we assume that the burst terminates after 3
e-folds). In Table~\ref{tab:ann.bursts} we show the performance of the
ANN when applied to the quiescent and the burst samples
separately. Here, the ``normal sample'' represents the sample
extracted from the \g~catalogue using the procedure described
previously, selecting equal numbers of galaxies in logarithmic bins of
total stellar mass.  This selection also picks a small fraction of
galaxies that are undergoing a burst. The quiescent sample is a
``bursts-clean'' version of the normal sample, i.e. a normal sample
for which we selected galaxies with $t_{\rm burst} > 10\,\tau_e$ or
which had no burst in their history.  The burst sample was extracted
differently, selecting equal numbers of galaxies in logarithmic bins
of mass of stars formed in the most recent burst of star formation.
Table~\ref{tab:ann.bursts} clearly shows the difficulty experienced by
the network in predicting the spectra of burst galaxies, which is
mainly due to the large variety in their spectra.  When we train the
ANN using the normal sample and apply this to predict the burst
sample, the accuracy of the ANN is greatly reduced. The four-band
$P_{|e|<10\%}$ average, in this case, drops from 88\% to 17\%, mainly
due to submillimetre band. In order to improve these results, we
constructed separate quiescent and burst samples as described above.
Training and testing the network independently for these two samples
produces better results. For quiescent galaxies, $\varepsilon_L <
0.12$ for the four bands studied, with equally impressive results for
the distribution of errors ($\sim 90\%$ of the galaxies show predicted
luminosities with less than 10\% error). The results for bursts
galaxies are somewhat less impressive, particularly for the FOCA 0.2
$\mu$m and IRAC 8 $\mu$m bands, where $P_{|e|<10\%} \sim
60-70\%$. However, they clearly outperform the ANN trained using the
normal sample, with a four-band $P_{|e|<10\%}$ average of 73\%.
Henceforth, we shall train the ANN for burst and quiescent galaxy
samples separately.

\subsubsection{ANN Performance: different output redshifts}

\begin{table*}
\begin{center}
  \begin{tabular}{cccccccccc}
  \hline
 \multicolumn{2}{c}{Network }	& \multicolumn{2}{c}{0.2 $\mu$m} & \multicolumn{2}{c}{B} 
	& \multicolumn{2}{c}{8 $\mu$m} & \multicolumn{2}{c}{850 $\mu$m}\\
Train	& Test	& $\varepsilon_L$ & $P_{|e|<10\%}$ & $\varepsilon_L$ & $P_{|e|<10\%}$ & $\varepsilon_L$ 
	& $P_{|e|<10\%}$ & $\varepsilon_L$ & $P_{|e|<10\%}$ \\
 \hline
 $z=0$	& $z=0$		& 0.14	& 80.1	& 0.07	& 95.3	& 0.16	& 88.5	& 0.24	& 89.1 \\
 $z=0$	& $z=2$		& 0.62	& 58.5	& 0.36	& 63.4	& 0.50	& 53.1	& 0.56	& 34.1 \\
 $z=2$ 	& $z=2$		& 0.29	& 80.9	& 0.17	& 89.7	& 0.14	& 86.1	& 0.26	& 89.3 \\
  \hline
  \end{tabular}
   \caption {The performance of an ANN trained with galaxy samples 
extracted at different redshifts: $z=0$ and $z=2$. See 
Table~\ref{tab:ann.predicted.spectra} for a description of the quantities).}
  \label{tab:ann.z}
  \end{center}
 \end{table*}

It is necessary to analyse how the trained ANN performs at 
different redshifts. If it turned out to be the case that 
an ANN trained at one redshift performed equally well at other 
redshifts, then there would be no need to retrain the network 
to predict galaxy luminosities at different redshifts, thus saving 
computing time. So far in this paper, we have analysed the ANN predictions 
only at redshift $z=0$. Table~\ref{tab:ann.z} compares the root mean 
square logarithmic error and the percentage of galaxies with predicted 
luminosities (in the rest frame) that lie within 10\% of the original 
values, for normal samples (see definition above) at redshifts $z=0$ 
and $z=2$. The network trained at $z=0$ performs reasonably well when 
used to predict the luminosities at $z=2$: it is capable of reproducing 
$\approx 52\%$ of the luminosities within an error of 10\%, and $\varepsilon_L$ 
smaller than $\sim 0.6$ at wavelengths of  0.2 $\mu$m, 0.44$\mu$m (B-band), 8 $\mu$m and 850 $\mu$m. 
However, it is strongly advisable to train the ANN at the 
redshift of choice, as indicated by the table. If we train the net at $z=2$ 
instead and apply it at $z=2$, the forecasted luminosities are much 
more accurate (we achieve a four-band $P_{|e|<10\%}$ average of 87\%).
Our approach will be to train the ANN at each redshift for which 
it is applied. 

\subsection{Error analysis}

\begin{figure*}
{\epsfxsize=8.truecm
\epsfbox[18 144 592 718]{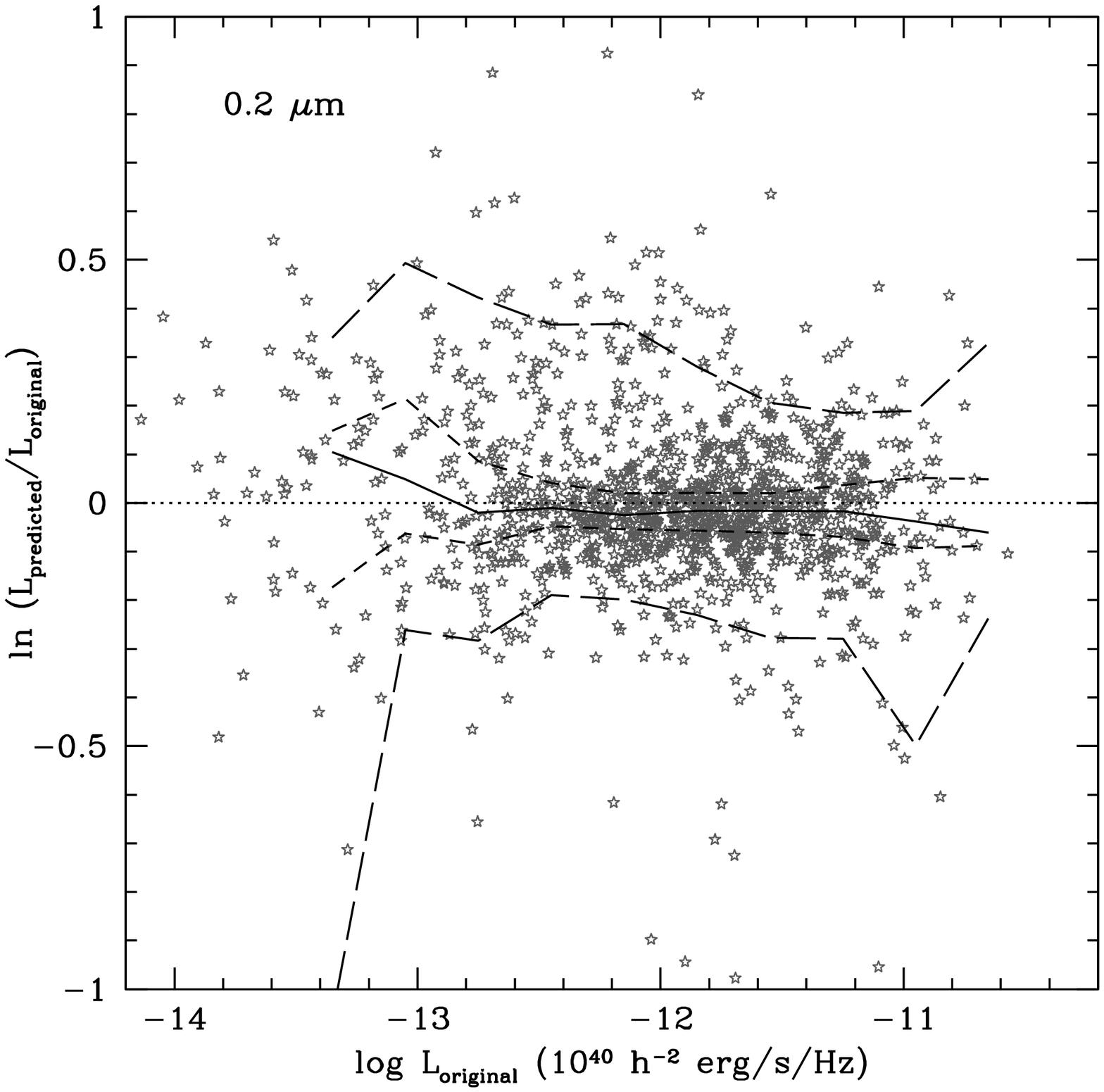}}
{\epsfxsize=8.truecm
\epsfbox[18 144 592 718]{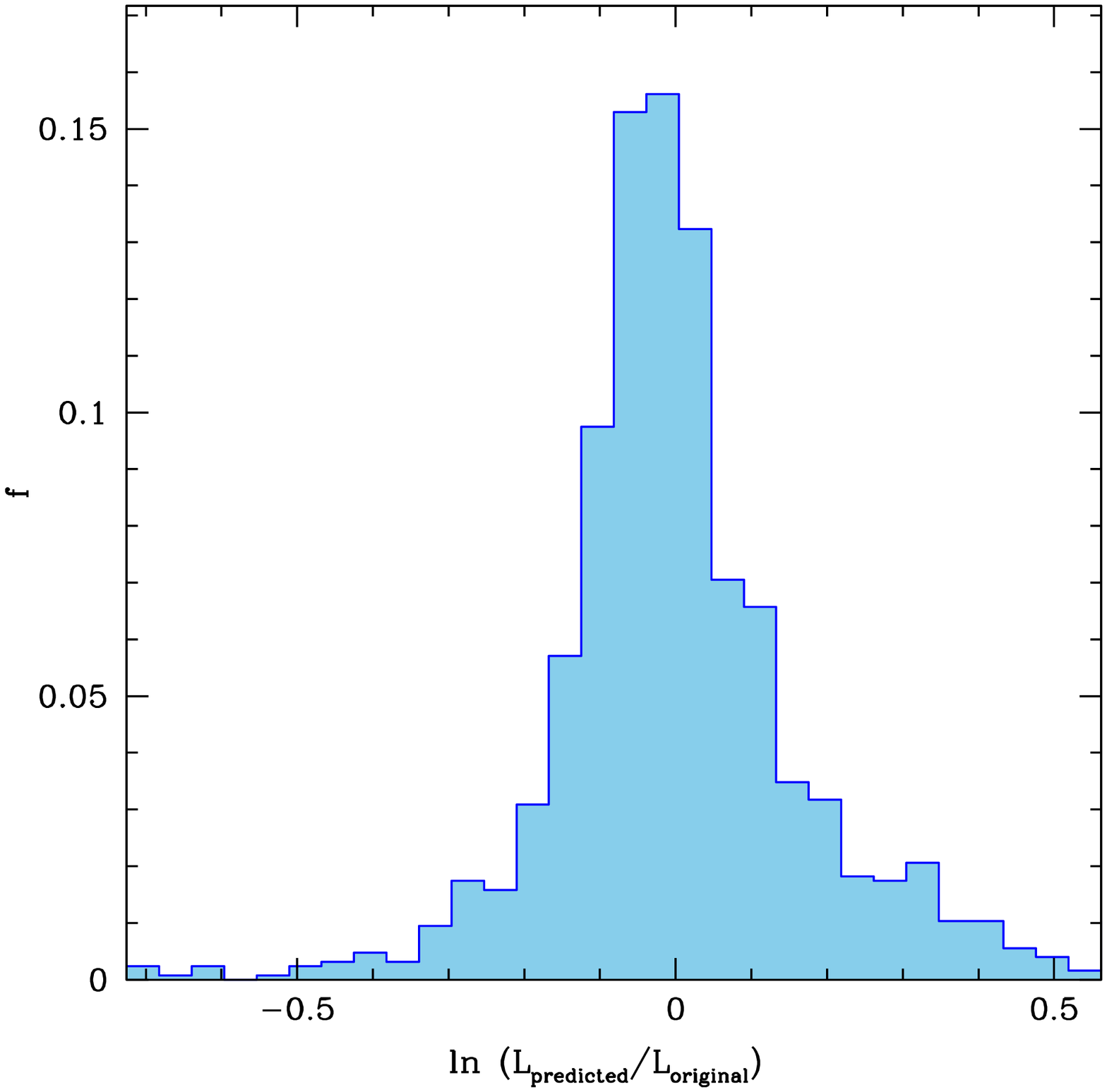}}
\caption{
The distribution of the error associated with the 0.2 $\mu$m 
luminosities predicted by ANN, for a sample of burst galaxies at $z=0$. 
In the left panel, the solid and dashed lines have the same meaning as 
in Fig.~\ref{fig:ann.compare.predicted.spectra}. In the right panel, 
the range of the x-axis is set to the 1st and 99th percentiles 
of the distribution, and the histogram is normalized to 
give $\sum_i n_i = 1$. }
\label{fig:ann.error.analysis}
\end{figure*}

\begin{figure*}
{\epsfxsize=14.truecm
\epsfbox[18 144 592 718]{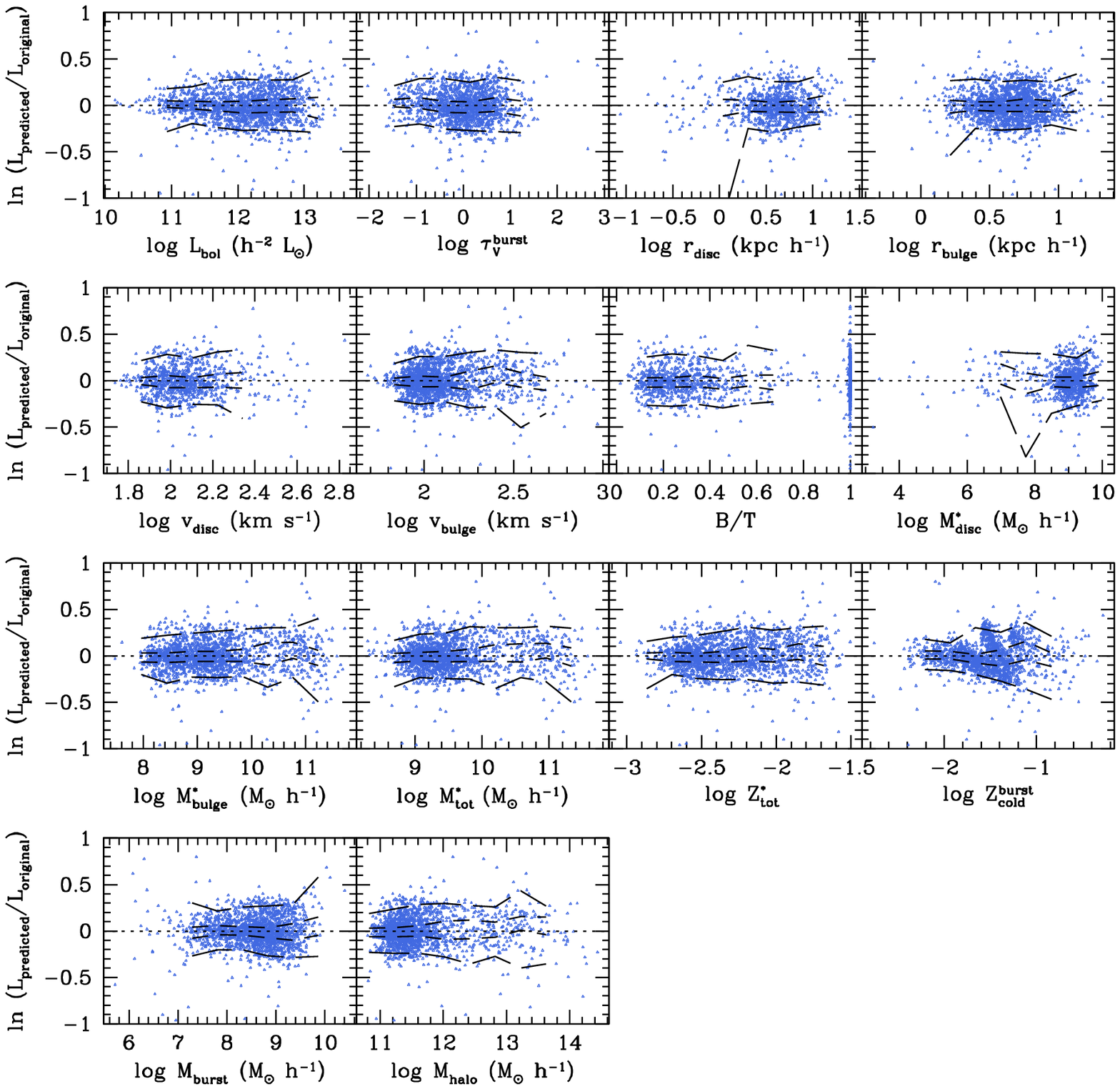}}
\caption{ The relation between the logarithmic error associated with
the $\lambda = 0.2\,\mu$m luminosities predicted by the ANN and
various galaxy properties, for the sample of burst galaxies. From left
to right and top to bottom we consider: bolometric luminosity, central
V-band extinction optical depth, disc and bulge size, disc and bulge
circular velocity, bulge-to-total mass ratio, disc, bulge and total
stellar mass, stellar metallicity and metallicity of the cold gas in
the burst, mass of stars formed in the last burst and mass of host
halo. The meaning of the dashed lines is the same as in
Fig.~\ref{fig:ann.compare.predicted.spectra}.  The zero error line is
plotted for reference (dotted line).  }
\label{fig:ann.uv.errors.props}
\end{figure*}

Previously, we tested the performance of the ANN when applied to 
predicting the quiescent and burst samples separately, and found 
that the errors associated with the prediction of the burst sample 
are larger than those for the quiescent galaxies.
The percentage of accurately predicted luminosities was particularly 
low at UV wavelengths (approximately 60\%). We now analyse further 
the error distribution associated with the predicted FOCA 0.2 $\mu$m 
band for the burst sample.

In Fig.~\ref{fig:ann.error.analysis} we plot the percentage error 
of the predicted luminosities as a function of their true, expected, 
values. As presented in Table.~\ref{tab:ann.bursts}, we find that 
$\sim 60\%$ have predicted luminosities within 10\% of the true values. 
The plot shows that the errors do not seem to be correlated with 
true UV luminosities. In fact, we find a weak correlation 
coefficient of 0.12. The independence of the errors from the luminosity 
will be an important factor when calculating luminosity-dependent 
quantities (e.g. luminosity functions) and sampling using luminosity. 

In an effort to further reduce the error associated with the predicted 
luminosities, we also investigated the relation between the errors and 
various galaxy properties for the burst sample. In 
Fig.~\ref{fig:ann.uv.errors.props} we plot the logarithm of the predicted 
to true luminosity against: bolometric luminosity ($\log L_{\rm bol}$), 
central V-band extinction optical depth for the burst component 
($\tau_{V}^{\rm burst}$), disc and bulge size ($r_{\rm disc/bulge}$) 
and circular velocities ($v_{\rm disc/bulge}$), bulge-to-total mass 
ratio ($B/T$), disc, bulge and total stellar mass 
($M_{\rm disc/bulge/tot}^*$), stellar metallicity ($Z_{\rm tot}^{*}$) 
and the metallicity of the cold gas in the burst ($Z_{\rm cold}^{\rm burst}$), 
the mass of stars formed in the last burst ($M_{\rm burst}$) and 
the mass of host halo ($M_{\rm halo}$). The plot reveals no clear 
correlation between the error associated with the predicted 0.2 $\mu$m 
luminosities and any galaxy property. The absolute value of the 
correlation coefficient is smaller than 0.05 for most of the properties, 
with slightly higher values ($\sim 0.12$) found for disc size and 
stellar mass. This implies that any sample built using the ANN method 
will have errors which are decoupled from the galaxy properties. 
Different network configurations were tried in order to improve 
the performance, but without any notable success.

\section{Predicting Luminosity Functions}
\label{section:ann.lf}

We are now ready to predict accurate luminosities for much larger 
\g~samples. 
In the previous section, we presented the first results from the new ANN 
technique. We calculated the luminosities for a sample of \g~galaxies 
in the FOCA 0.2 $\mu$m, B (0.44 $\mu$m), IRAC 8 $\mu$m and SCUBA 850 $\mu$m 
bands, and found that for more of than 80\% of the galaxies the predicted 
luminosity was within 10\% of the true luminosity. In this section, we show 
the impact of the error in the ANN predictions on the form of the luminosity 
function. It is important to closely reproduce the original model luminosity 
function, because this quantity is the most basic statistical description of 
a galaxy population. Here, we present predictions for the luminosity functions 
of Lyman-break galaxies at $z=3$, for galaxies selected in the mid infra-red at 
$z=0.5$, and SCUBA galaxies at $z=2$.

Throughout this section we use a 12:30:30:1 ANN architecture, which 
corresponds to 12 galaxy properties as the input layer, 30 neurons in each 
of the two hidden layers and one output neuron representing the galaxy 
luminosity in a given band (see the previous section for further details). 
The sigmoid function was adopted as the activation function of choice for 
the hidden neurons. Driven by the conclusions of the previous section, 
we split our sample into two: quiescent and burst galaxies, and train the 
network for each of these populations at the selected redshift.

\subsection{Lyman-break Galaxies at $z=3$}

Lyman-break galaxies (LBGs) were the first significant high redshift 
galaxy population to be isolated, and were used to measure the star formation 
history of the Universe at early epochs \citep{steidel}. These high 
redshift galaxies are selected from photometry in several optical 
bands which straddle the Lyman-break at $912\,\AA$ in a galaxy's rest 
frame for objects in the target redshift range ($z \ga 2$). 

In this section we use the ANN to predict the luminosities of $z=3$ 
galaxies at a rest frame UV wavelength of 0.17 $\mu$m, which corresponds 
to the observer frame R band at this redshift. We compare the 
luminosity function of the training set as computed using \gr~spectra 
with that obtained using the luminosities predicted by the ANN. 

\begin{figure*}
\centering
{\epsfxsize=8.truecm
\epsfbox[18 144 592 718]{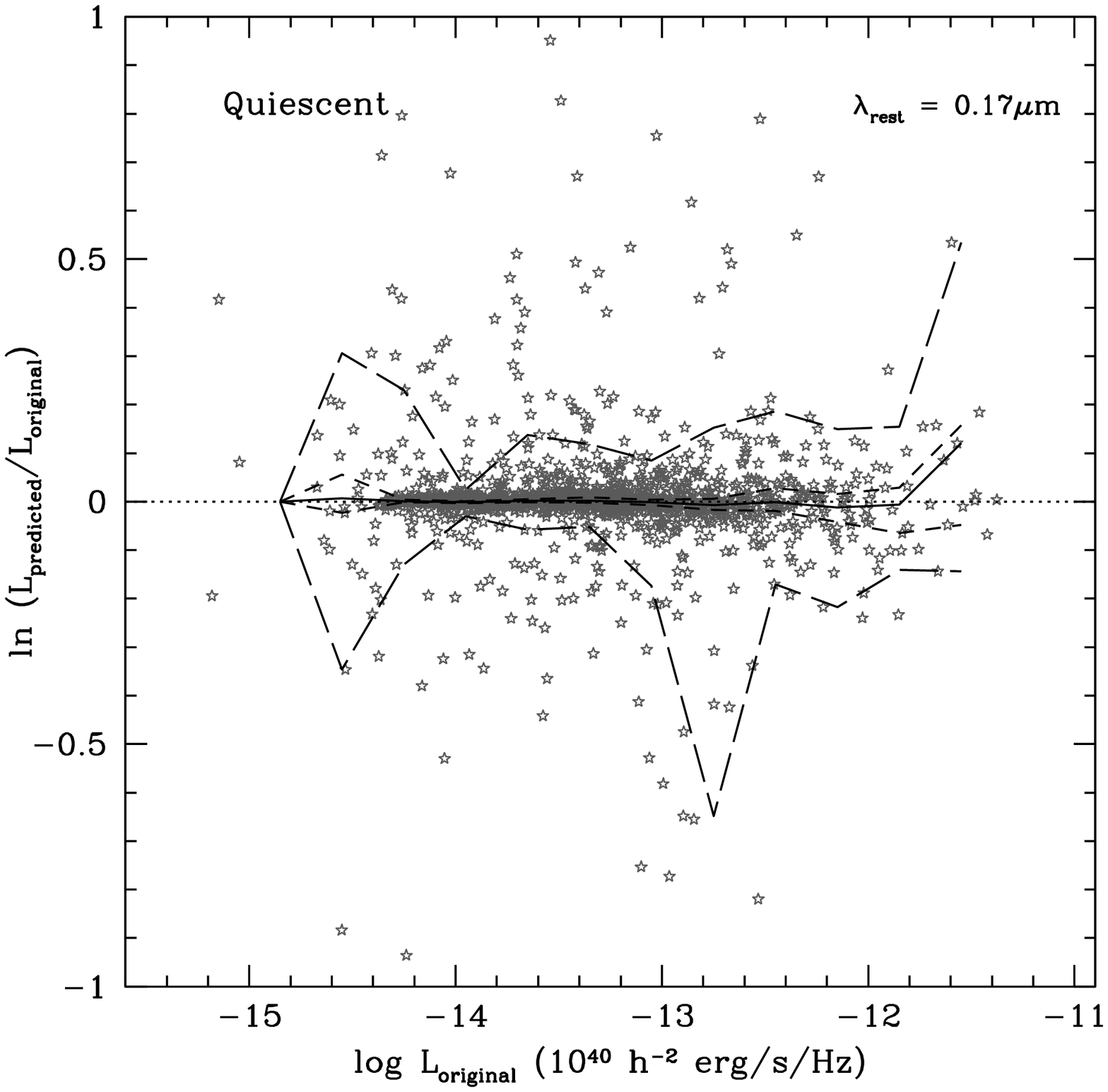}}
{\epsfxsize=8.truecm
\epsfbox[18 144 592 718]{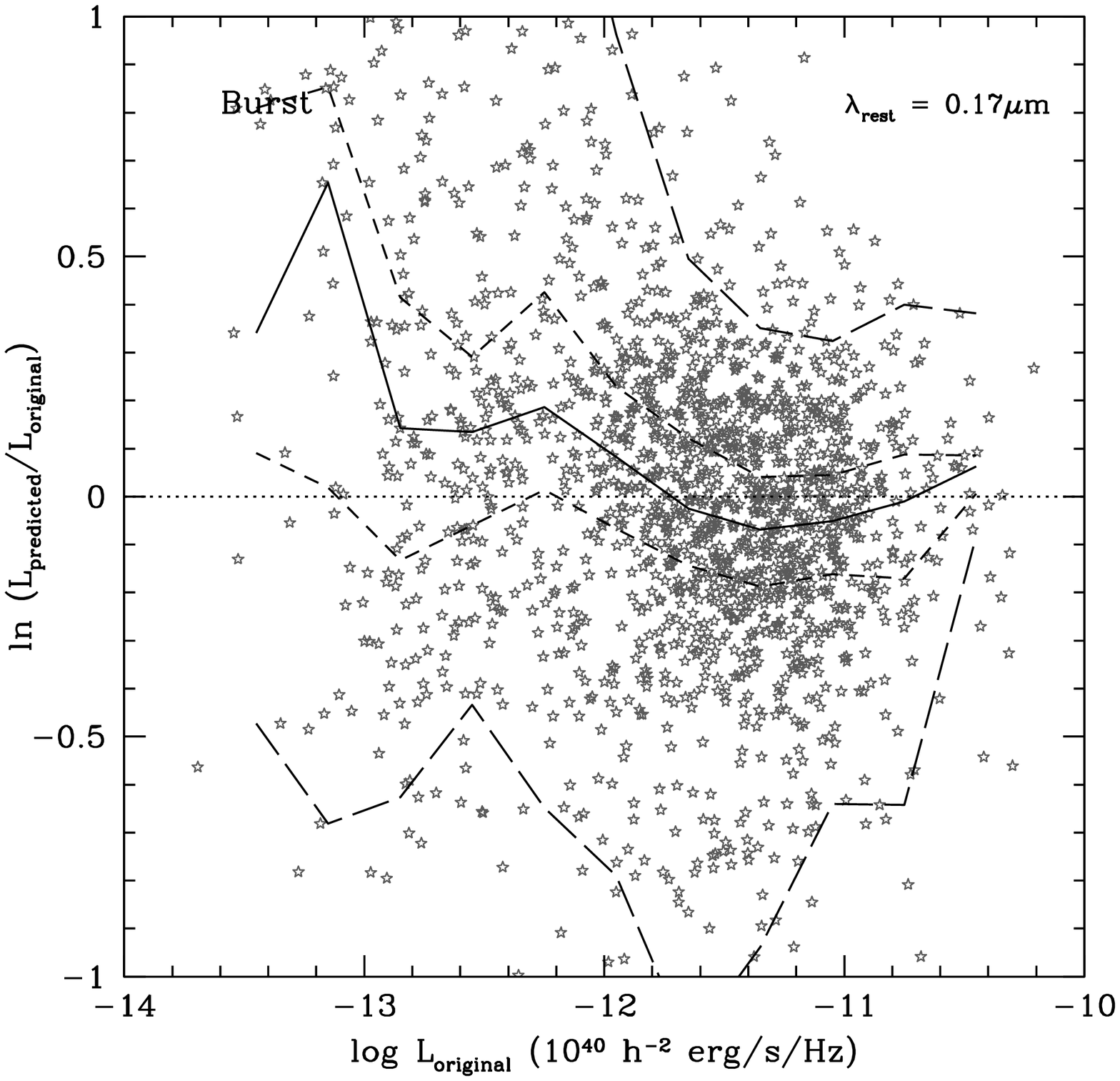}}
{\epsfxsize=8.truecm
\epsfbox[18 144 592 718]{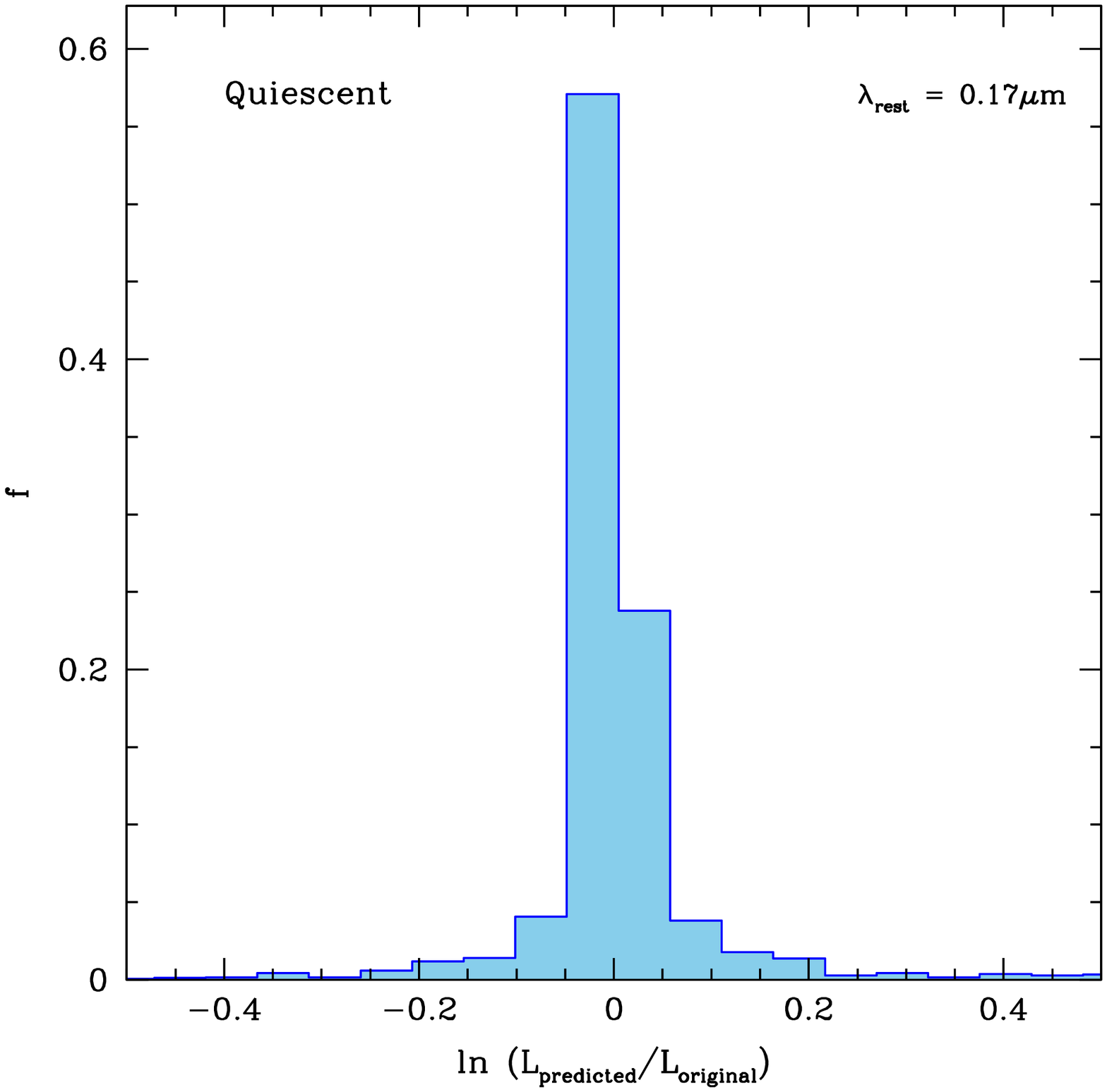}}
{\epsfxsize=8.truecm
\epsfbox[18 144 592 718]{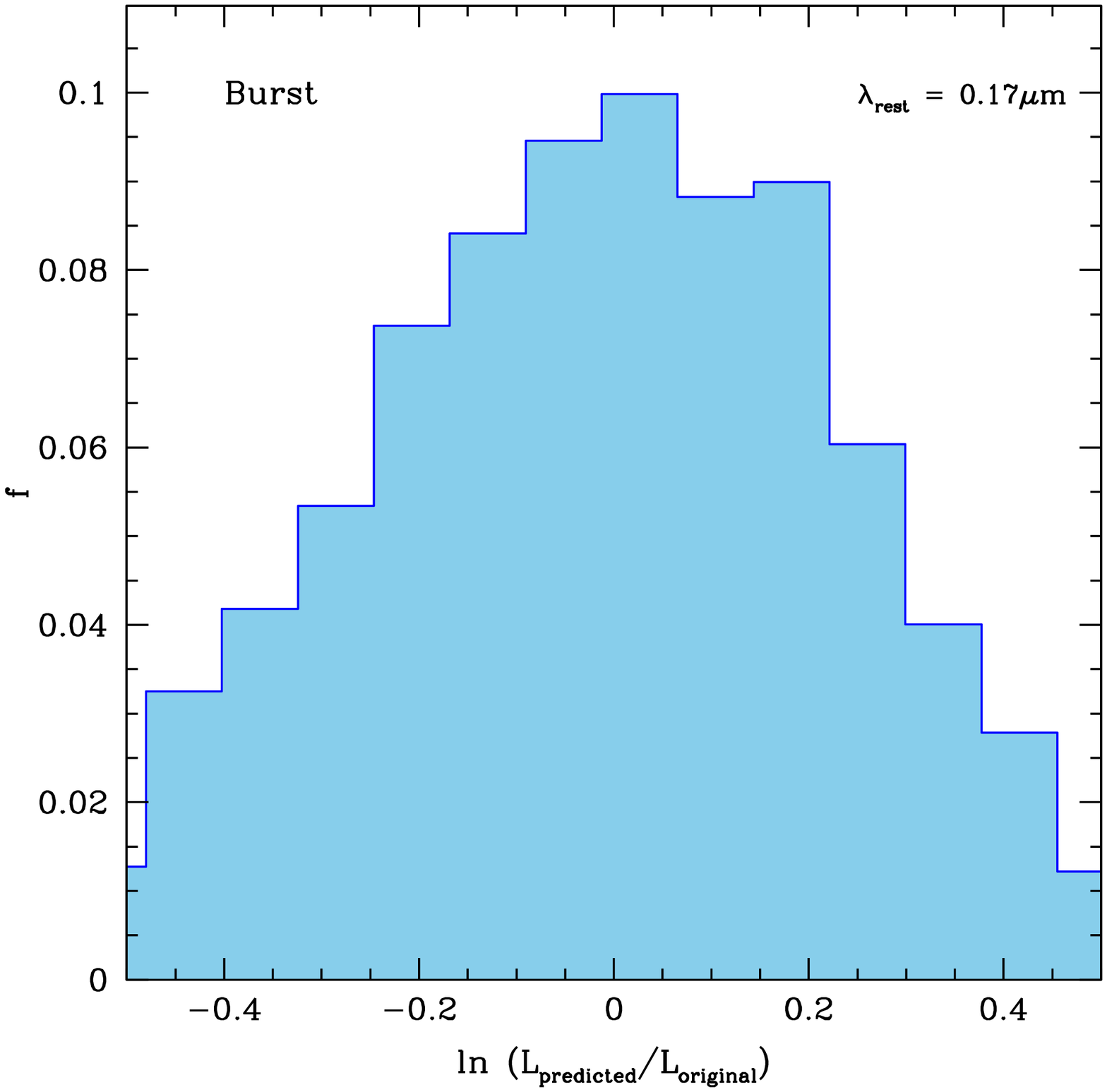}}
\caption{
A comparison between the ANN predicted rest frame UV ($\lambda = 0.17\,\mu$m) 
luminosities and the original luminosities, as extracted from \gr~model 
spectra. The left and right panels show the results for quiescent and 
burst galaxies, respectively. 
The solid and dashed lines in the upper panels have the same meaning as in 
Fig.~\ref{fig:ann.compare.predicted.spectra}. 
The distribution of the logarithm of the ratio 
of predicted to true luminosity is presented in the lower panels. 
The range of the 
x-axis for the lower panels is set to the $1^{\rm st}$ and $99^{\rm th}$ 
percentiles of the distribution for bursts, and the histograms are normalized 
to give $\sum_i n_i = 1$.}
\label{fig:ann.uv.errors}
\end{figure*}

\begin{table*}
\begin{center}
  \begin{tabular}{cccccccc}
  \hline
 Sample & $\varepsilon_L$ & $P_{|e|<10\%}$ & $p_1$ & Q$_1$ & Q$_2$ & Q$_3$ & $p_{99}$ \\
 \hline
 Quiescent	& 0.20     & 88.4 & -72.1    & -1.0 & 0.0 & 0.7 & 41.1 \\
 Burst 		& 0.40     & 54.2 & -329.7    & -9.9 & -0.1 & 9.5 & 67.6 \\
  \hline
  \end{tabular}
   \caption{Statistics for the error distribution of the rest frame UV 
luminosities predicted by the ANN, for quiescent and burst galaxies at $z=3$. 
The listed quantities are described in Table~\ref{tab:ann.predicted.spectra}.}
  \label{tab:ann.uv.errors}
  \end{center}
 \end{table*}

In Fig.~\ref{fig:ann.uv.errors} we plot the logarithmic error associated 
with the ANN predicted UV luminosities against the original luminosities 
(the target values), for quiescent (top-left panel) and burst (top-right 
panel) galaxies. In the lower panels we plot the error distribution for 
both samples. The statistics of these distributions are summarized in 
Table~\ref{tab:ann.uv.errors}. In the UV regime, the ANN works better for 
quiescent galaxies than for burst galaxies. For most of the quiescent 
galaxies, 88\% of the sample, the ANN predicts luminosities within 10\% 
of the true value, and only $\approx$4\% of galaxies have errors larger 
than 20\%. Burst galaxies show somewhat a bigger scatter around the 
expected values, which is driven by the higher intrinsic variety in their 
spectra at 0.17 $\mu$m. The value of $\varepsilon_L$ is higher for bursts, 
0.40. As noted in the previous section, Fig.~\ref{fig:ann.uv.errors} shows 
that the error distribution is approximately independence of the luminosity.

\begin{figure}
\centering
{\epsfxsize=8.truecm
\epsfbox[18 144 592 718]{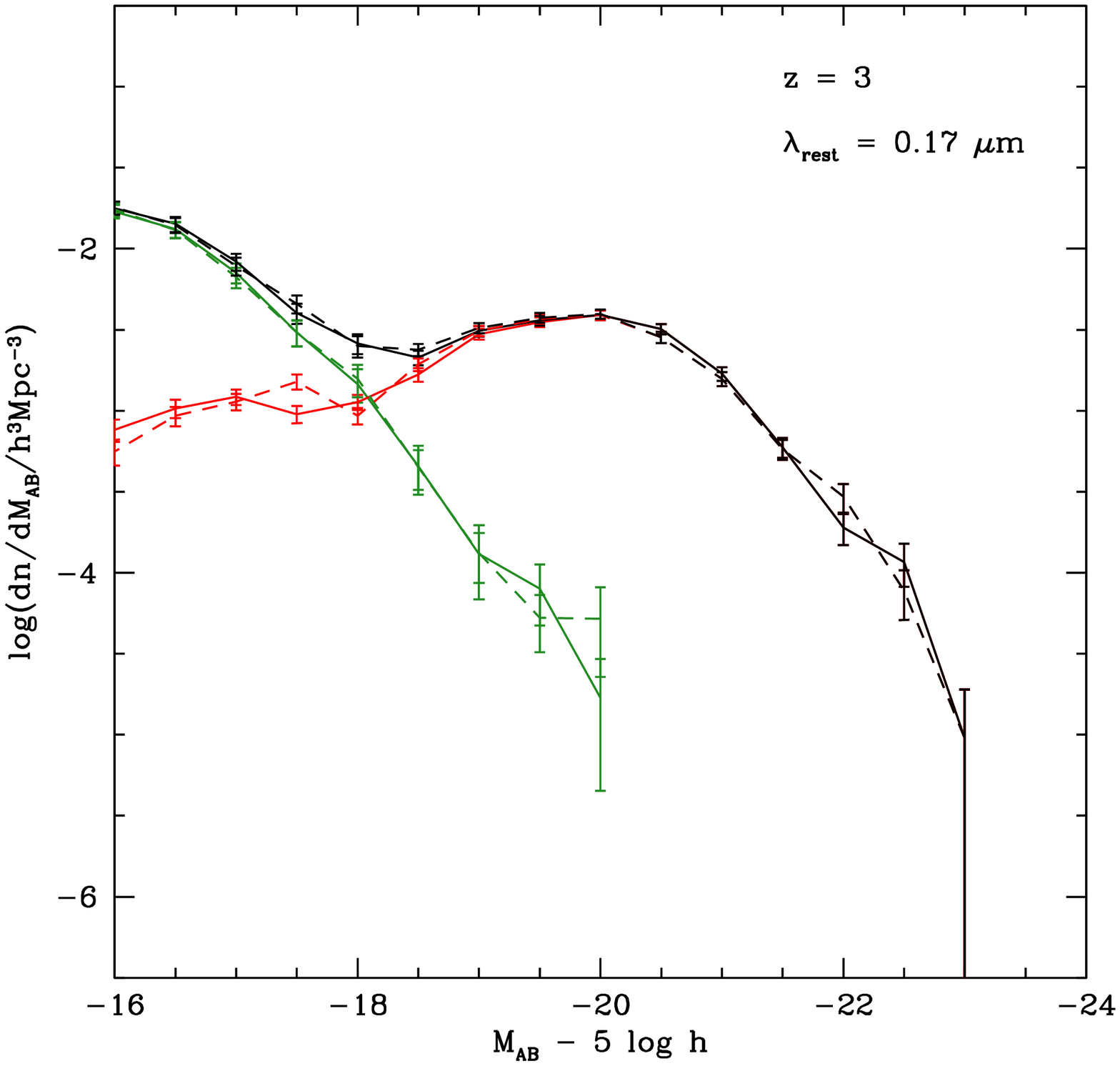}}
\caption{
The rest frame 0.17 $\mu$m luminosity function at $z=3$, calculated from the 
original luminosities (solid lines), as extracted from the \gr~spectra, and 
the luminosity function constructed from the ANN predicted luminosities 
(dashed lines). The black lines show the total luminosity function, whose 
components are quiescent galaxies (green lines) and burst galaxies 
(red lines). The error bars on the model luminosity function 
indicate the Poisson 
uncertainties due to the number of galaxies simulated.
}
\label{fig:ann.lf.uv}
\end{figure}

The rest frame 0.17 $\mu$m luminosity functions at $z=3$ are plotted 
in Fig.~\ref{fig:ann.lf.uv}. The luminosity function constructed using 
the luminosities predicted by the ANN (dashed lines) is in excellent 
agreement with the ``true'' luminosity function which uses the 
luminosities calculated 
directly with~\gr~(solid lines). This agreement holds for quiescent 
and burst galaxies separately, in spite of the somewhat larger errors 
in the case of burst luminosities. We carried out an independent 
test of this comparison, by perturbing the \gr~luminosities by the error 
distribution of the ANN predictions and reached the same conclusion. 
At bright luminosities, the true luminosity function is closer to a 
power-law than an exponential break, and so the errors would need to 
be much larger before they would lead to an appreciable difference in the 
luminosity function predicted by the ANN. The lack of a dependence 
of the size of the error on UV luminosity also helps to keep the 
shape of the luminosity function (see Fig.~\ref{fig:ann.uv.errors}) 
predicted by the ANN close to the true one. 

\subsection{Mid-Infrared Galaxies at $z=0.5$}

The development of new space-based infrared telescopes in the 1980s
opened up a new window in the electromagnetic spectrum, allowing us to
see galaxies which are heavily obscured in the optical and UV, but
which have substantial emission in the IR. This IR emission is the
result of the heating of the dust when it absorbs starlight and
consequently emits the energy at longer wavelengths. Further studies
made it clear that an important fraction of the star formation in the
Universe is obscured by dust \citep{smail, hauser, hughes}. These
discoveries made clear the importance of dusty galaxies for
understanding how galaxies are made. Any complete model of galaxy
formation must be able to make accurate predictions for the emission
from galaxies at IR wavelengths.

This section focuses on the mid-IR emission from galaxies. We predict 
the luminosities in the MIPS 24 $\mu$m band as used in recent 
observations in the infrared by the satellite \textit{Spitzer} 
\citep[see][for detailed comparisons between the \g~plus \gr~model 
predictions and observational data]{cedric}. We selected a galaxy 
population at $z=0.5$. 

\begin{figure*}
{\epsfxsize=8.truecm
\epsfbox[18 144 592 718]{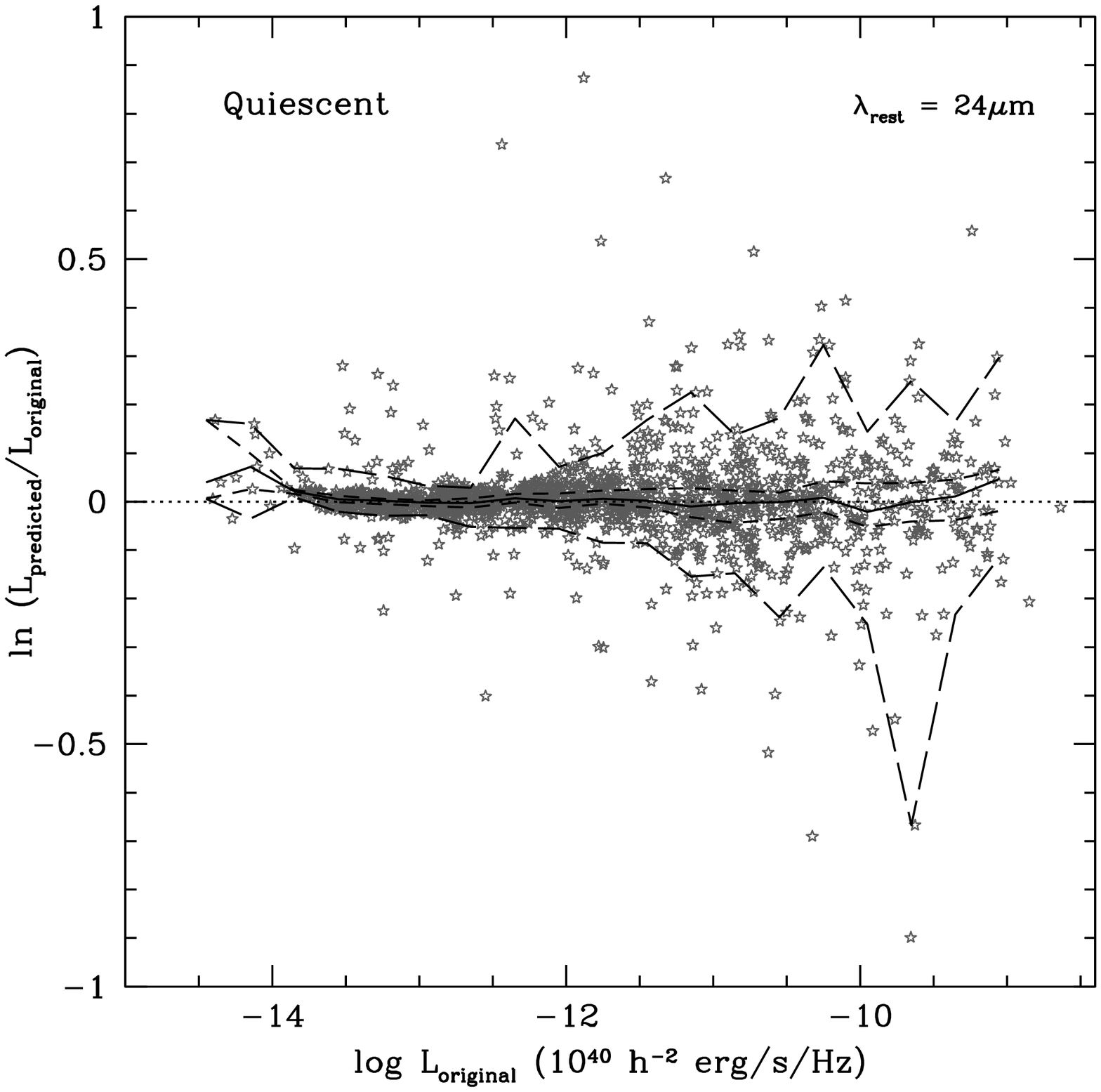}}
{\epsfxsize=8.truecm
\epsfbox[18 144 592 718]{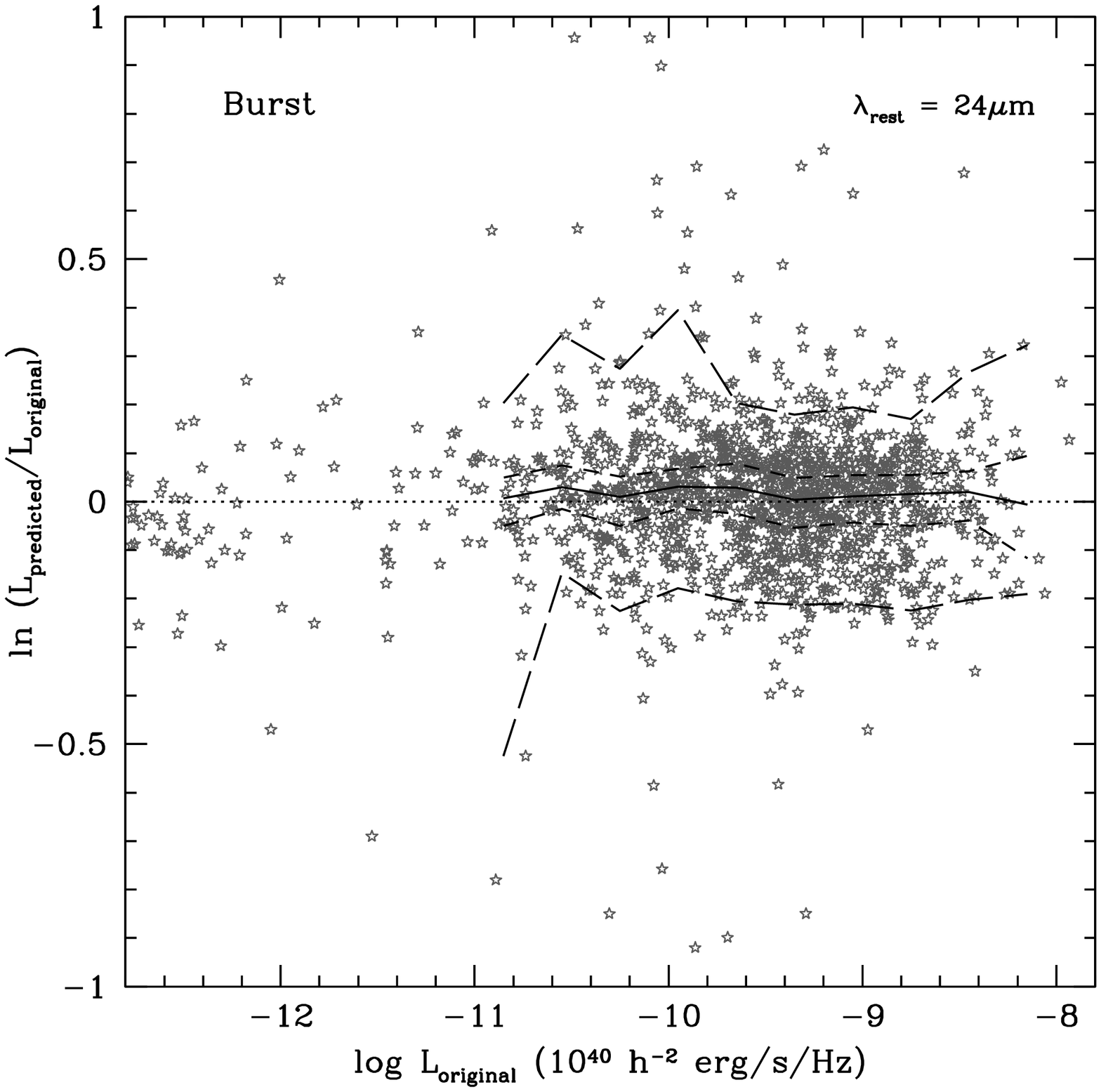}}
{\epsfxsize=8.truecm
\epsfbox[18 144 592 718]{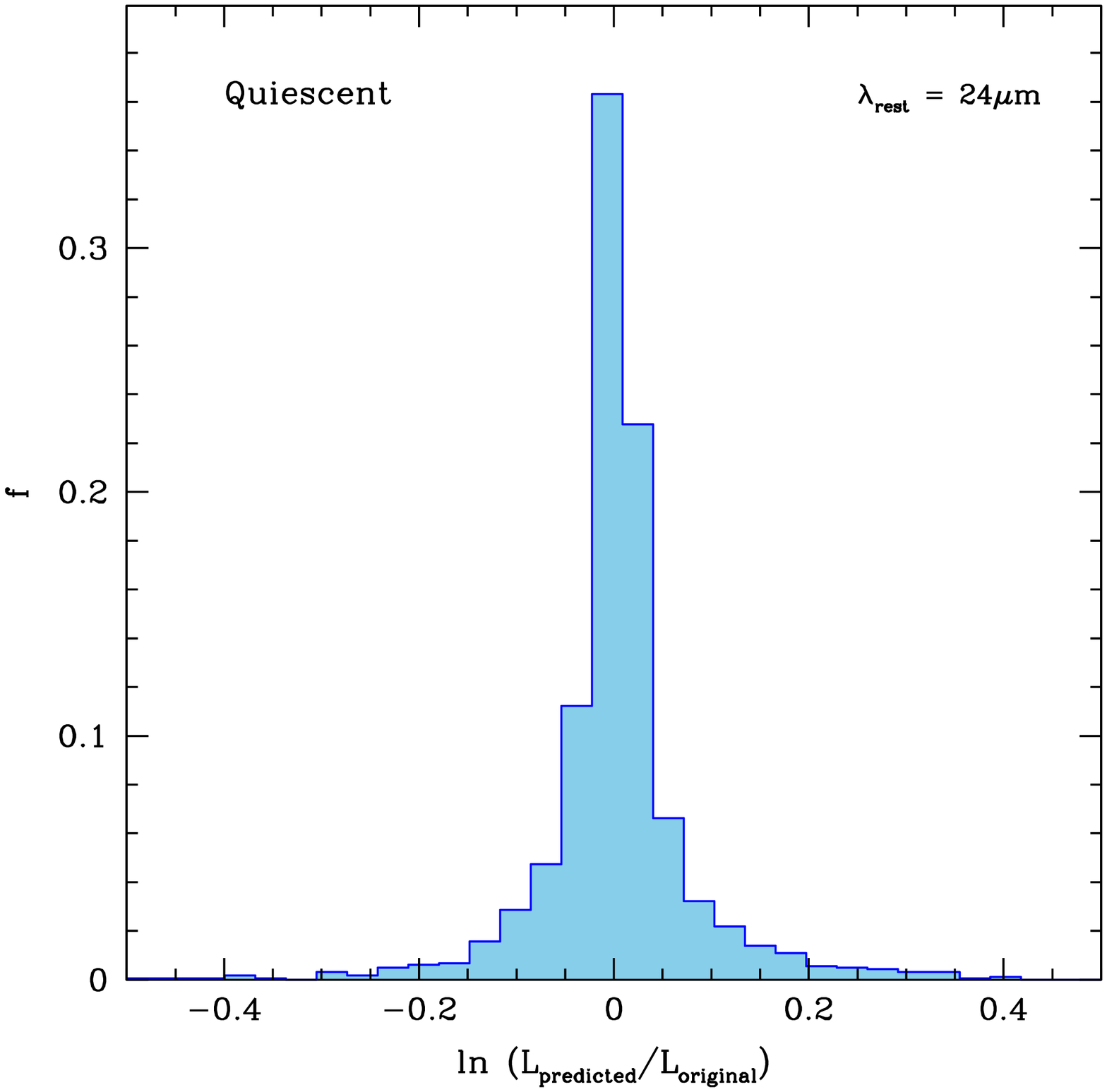}}
{\epsfxsize=8.truecm
\epsfbox[18 144 592 718]{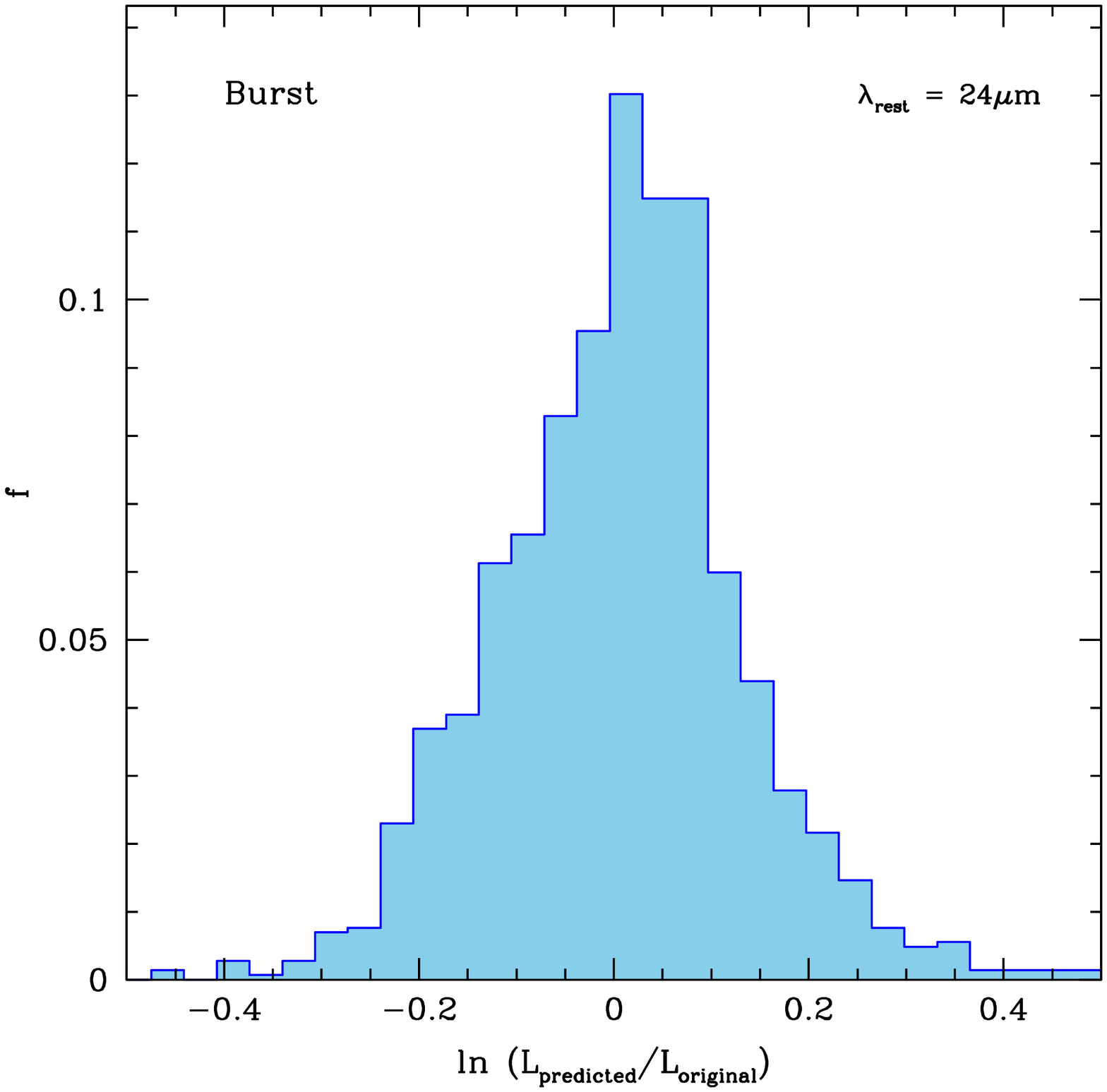}}
\caption{
The errors on the observer frame 24 $\mu$m luminosity predicted by the ANN 
for galaxies at $z=0.5$. The results for quiescent and burst galaxies 
are shown in the left and right panels respectively. 
The solid and dashed lines in the upper panels have the same meaning 
as in Fig.~\ref{fig:ann.compare.predicted.spectra}. 
The lower panels show the distribution of the logarithmic error. 
The range of the 
x-axes are set to the $1^{\rm st}$ and $99^{\rm th}$ percentiles 
of the distribution, and the histograms are normalized 
to give $\sum_i n_i = 1$.}
\label{fig:ann.24.errors}
\end{figure*}

\begin{table*}
\begin{center}
  \begin{tabular}{cccccccc}
  \hline
 Sample & $\varepsilon_L$ & $P_{|e|<10\%}$ & $\min$ & Q$_1$ & Q$_2$ & Q$_3$ & $\max$ \\
 \hline
 Quiescent	& 0.13     & 86.5 & -38.4    & -2.4 & -0.1 & 1.9 & 25.7 \\
 Bursts 	& 0.18     & 60.0 & -79.3    & -8.6 & -0.4 & 7.3 & 37.5 \\
  \hline
  \end{tabular}
   \caption{Summary statistics for the distribution of the error on the 
24 $\mu$m luminosities predicted by the ANN, for quiescent and burst 
galaxies at $z=0.5$. A description of the quantities is given in 
Table~\ref{tab:ann.predicted.spectra}.}
  \label{tab:ann.24.errors}
  \end{center}
 \end{table*}

Fig.~\ref{fig:ann.24.errors} shows the ratio of the predicted to true 
luminosity at rest-frame 24 $\mu$m, for quiescent and burst galaxies. 
The associated statistics are summarized in Table~\ref{tab:ann.24.errors}. 
As we found in the previous section, the ANN method performs better for 
quiescent galaxies; at rest-frame 24 $\mu$m, the root mean square 
logarithmic error is $\varepsilon_L=0.13$ and $P_{|e|<10\%} = 87\%$. For 
burst galaxies, the performance is not as good. The percentage of 
galaxies with errors within 10\% is 60\%. Different ANN architectures 
and input galaxy properties were tried without any improvement over 
these figures. As before, the error distribution is not correlated 
with luminosity. This suggests that the errors might not change the 
shape of the luminosity function. Also, we found no correlation between 
the error and several other galaxy properties.

\begin{figure}
{\epsfxsize=8.truecm
\epsfbox[18 144 592 718]{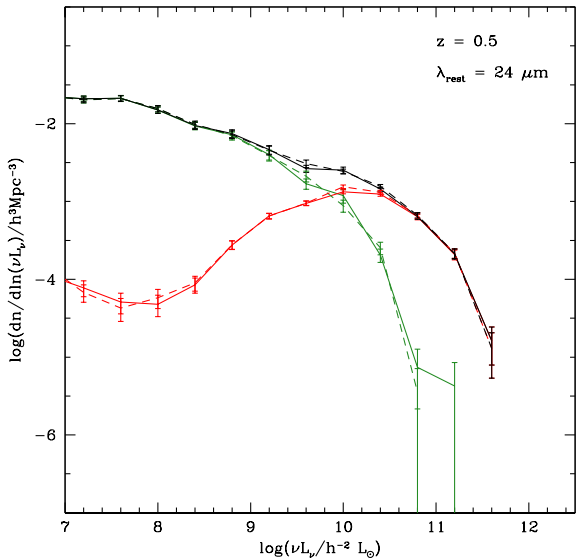}}
\caption{
The 24 $\mu$m luminosity function at $z=0.5$, given by original 
luminosities (solid lines) extracted from the \gr~spectra, and 
the luminosity function constructed from the luminosities 
predicted by the ANN (dashed lines). The black lines show the 
total luminosity function, with quiescent galaxies show by the 
green line and burst galaxies by the red line. The error bars  
indicate the Poisson uncertainties due to the number of 
galaxies in each luminosity bin.}
\label{fig:ann.lf.24}
\end{figure}

In Fig.~\ref{fig:ann.lf.24}, we compare the rest frame 24 $\mu$m 
luminosity functions at $z=0.5$, constructed from the luminosities 
predicted by the ANN and the original luminosities as given by \gr.
As with the UV comparison in the previous section, the luminosity function 
derived from the ANN predictions is in very good agreement with that 
obtained directly from \gr. Again, this success extends to the luminosity 
function of the 
burst sample, even though the errors are larger in this case.

\begin{figure*}
{\epsfxsize=8.truecm
\epsfbox[18 144 592 718]{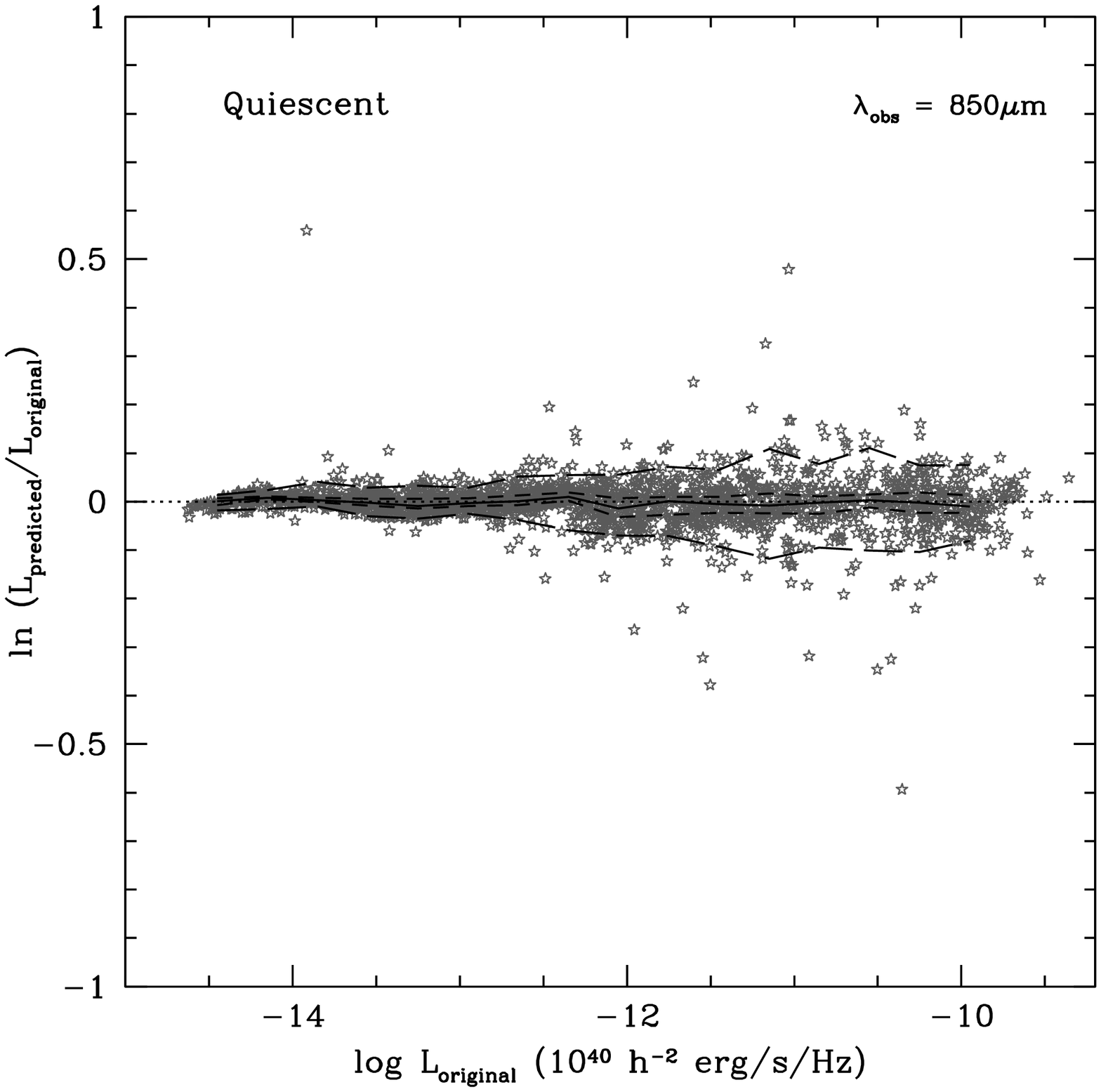}}
{\epsfxsize=8.truecm
\epsfbox[18 144 592 718]{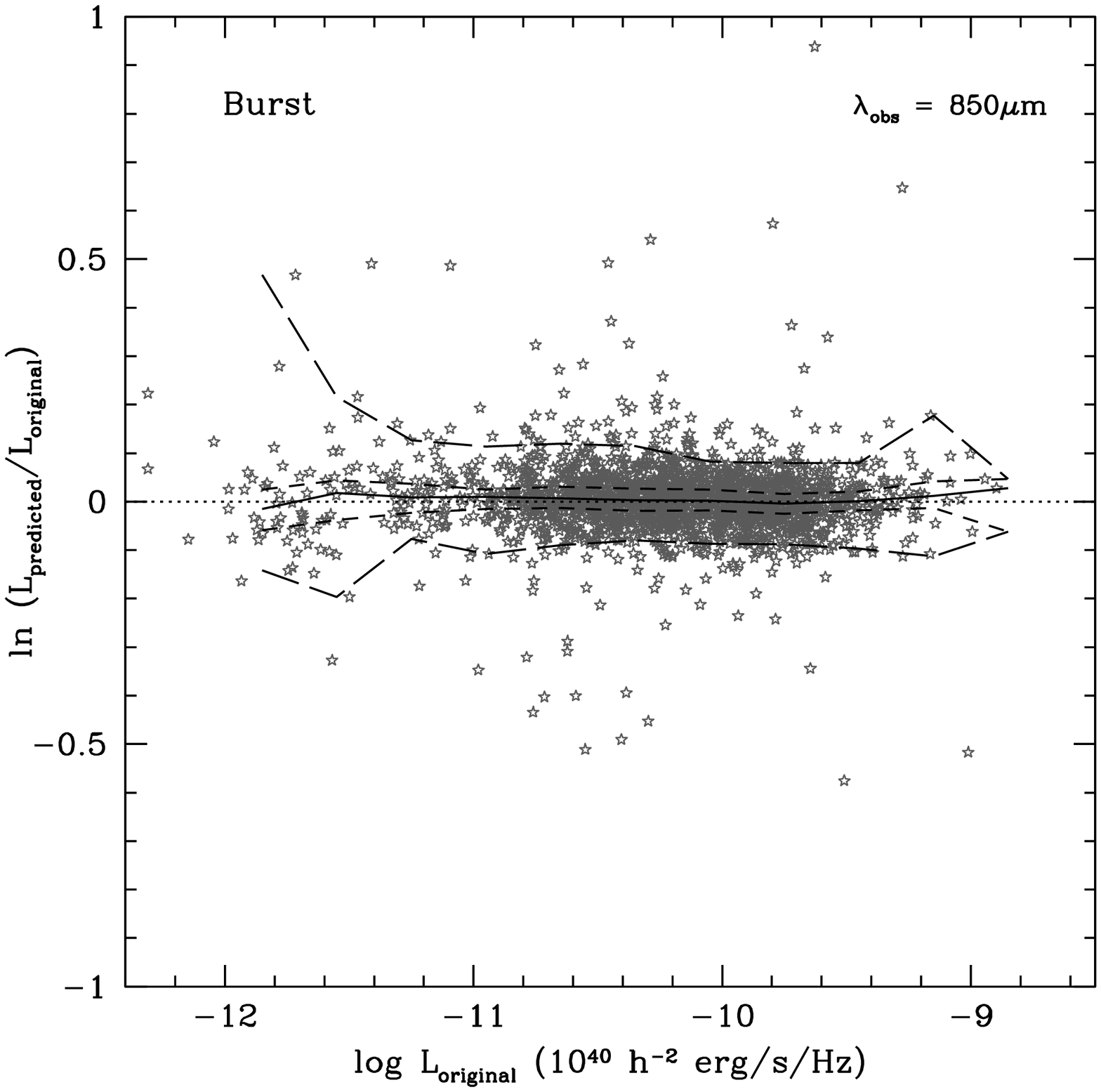}}
{\epsfxsize=8.truecm
\epsfbox[18 144 592 718]{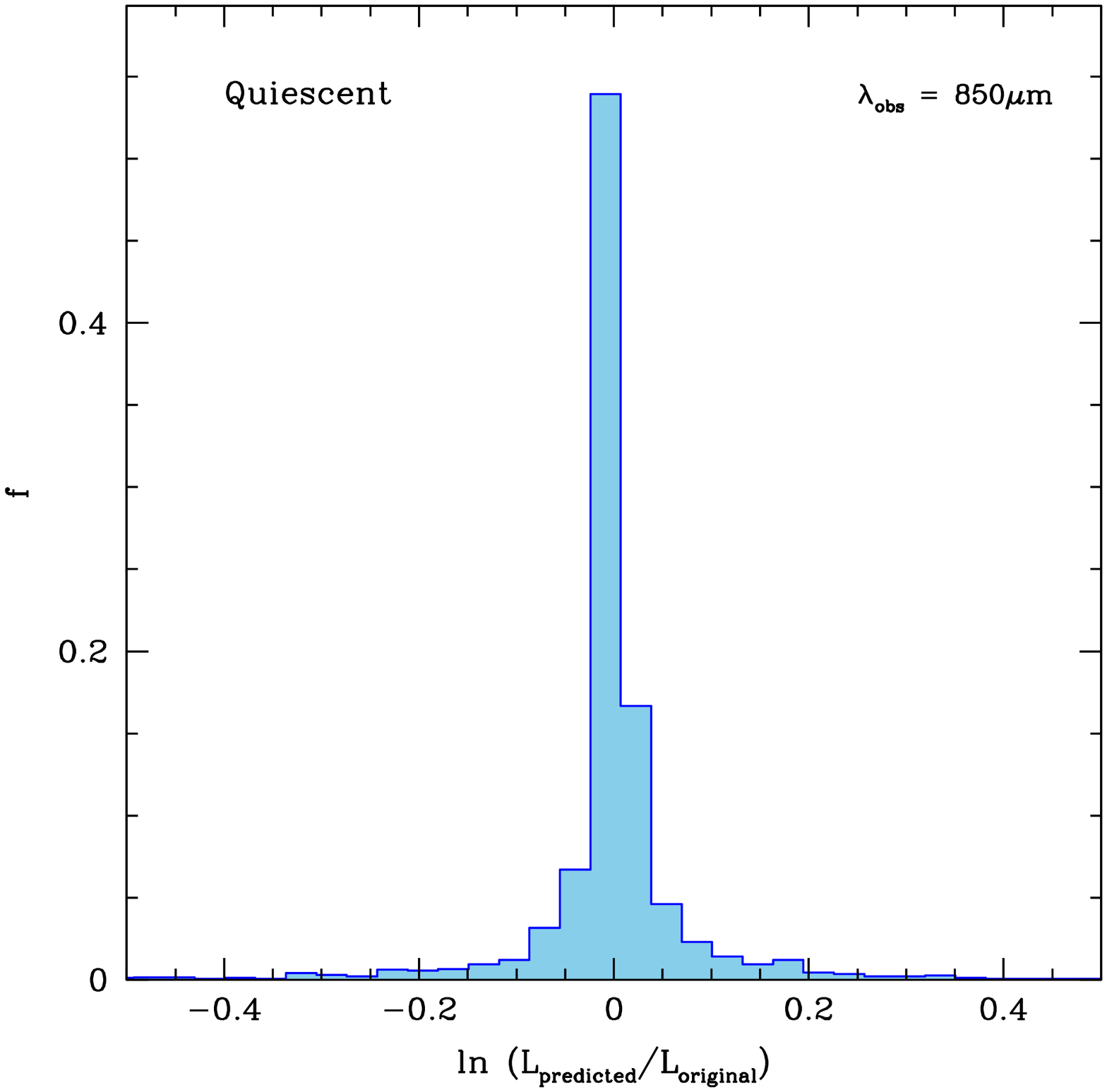}}
{\epsfxsize=8.truecm
\epsfbox[18 144 592 718]{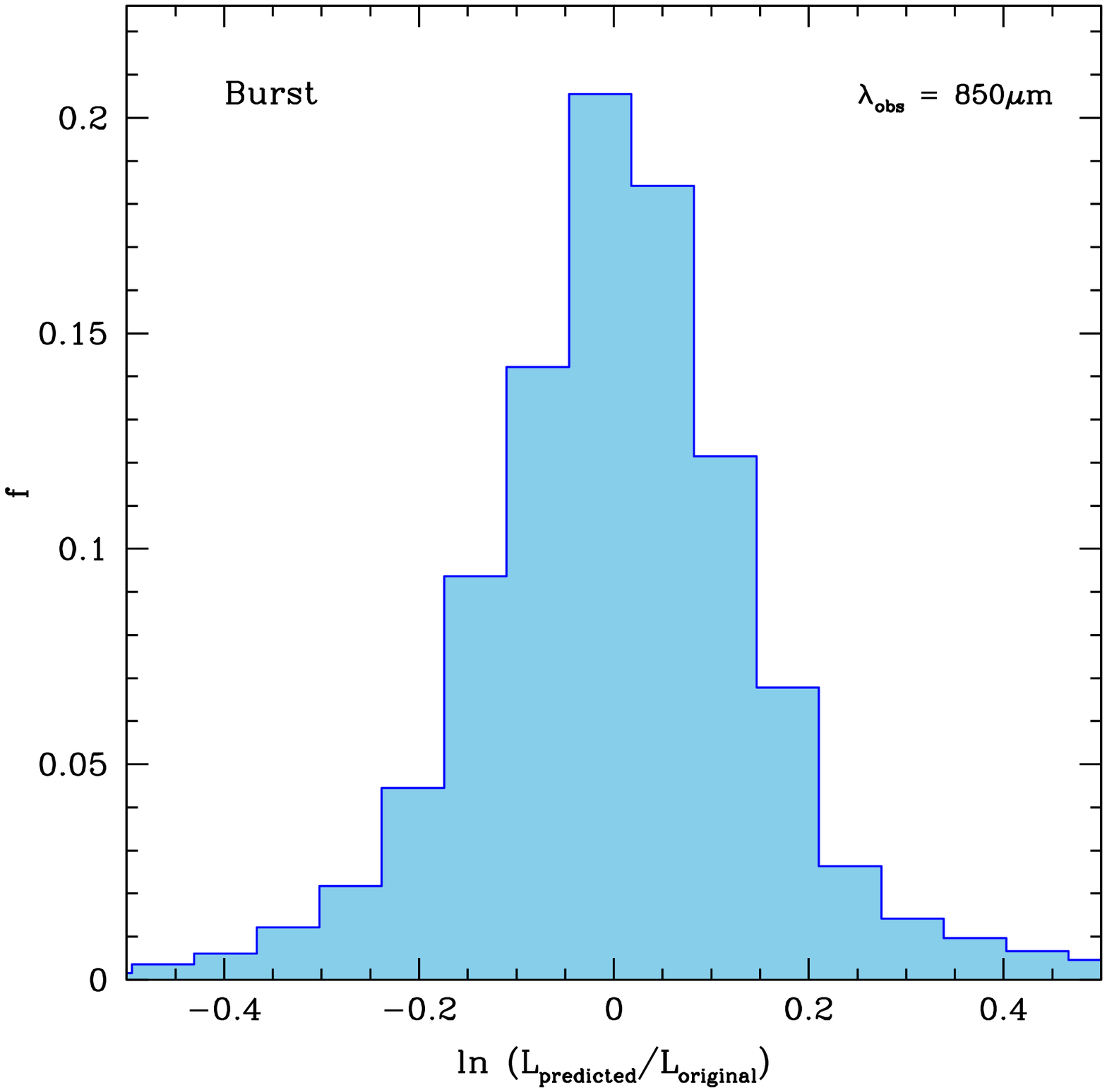}}
\caption{
The ratio of predicted to true luminosity at 850 $\mu$m 
in the observer frame. Quiescent galaxies are shown in 
the left-hand panels, while the results for the burst sample 
are plotted in the right-hand panels. 
The solid and dashed lines in the upper panels have the same 
meaning as in Fig.~\ref{fig:ann.compare.predicted.spectra}. 
The distribution of the logarithm of the ratio is presented 
in the lower panels. The range of the x-axes are set to 
the $1^{\rm st}$ and $99^{\rm th}$ percentiles of the 
distribution, and the histograms are normalized to 
give $\sum_i n_i = 1$.}
\label{fig:ann.850.errors}
\end{figure*}

\begin{table*}
\begin{center}
  \begin{tabular}{cccccccc}
  \hline
 Sample & $\varepsilon_L$ & $P_{|e|<10\%}$ & $\min$ & Q$_1$ & Q$_2$ & Q$_3$ & $\max$ \\
 \hline
 Quiescent	& 0.09     & 97.2 & -11.7    & -1.6 & 0.0 & 1.6 & 13.1 \\
 Bursts 	& 0.10     & 93.2 & -27.1    & -3.2 & -0.2 & 3.1 & 21.4 \\
  \hline
  \end{tabular}
   \caption{
Summary statistics for the predicted 850 $\mu$m luminosity 
error distribution, for quiescent and burst galaxies, at 
redshift $z=2$.  A description of these quantities can be found 
in Table~\ref{tab:ann.predicted.spectra}.}
  \label{tab:ann.850.errors}
  \end{center}
 \end{table*}

\subsection{Submillimetre Galaxies at $z=2$}

Submillimetre galaxies (SMGs) are thought to be predominantly dusty 
starbursts \citep{smail02}. The emission in the submillimetre region of 
the spectrum (around 850 $\mu$m) is due to the heating of dust when it 
absorbs the UV light emitted by young stars. It is also possible that 
some contribution to the flux at these wavelengths 
comes from the dust being heated by an AGN, although it is now believed 
that this process is secondary \citep{alexander, alexander2005}. 
In the standard picture, SMGs are galaxies with prodigious 
star formation rates $\sim 500-1000$ M$_{\odot}$ yr$^{-1}$. 
In the Baugh et~al. model, a top-heavy IMF is adopted in merger-driven 
starbursts. As a result, the model SMGs have more modest 
star formation rates. Nevertheless, the model SMGs are still the 
most massive galaxies in place with the highest star formation 
rates at the median redshift $z \sim 2$. 

Here we predict the observer frame 850 $\mu$m luminosities 
for galaxies at $z=2$, using the ANN. We compare our predictions with 
the correct values extracted from \gr~spectra, and evaluate the 
luminosity functions using both the predicted and original luminosities.

The predicted and original $850 \mu$m luminosities are compared in 
Fig.~\ref{fig:ann.850.errors}. Further information about the errors 
is presented in Table~\ref{tab:ann.850.errors}.
As shown in the previous section, the ANN predictions for the 
submillimetre are extremely good. We are able to reproduce 
the luminosity of more than 95\% of galaxies with an accuracy of 10\% 
or better, for both quiescent and burst galaxies. 
The success of the predictions is also reflected in the root mean 
square logarithmic error, which is 0.09 for quiescent galaxies 
and 0.10 for the burst sample. No clear correlation was found 
between the error and galaxy properties.

\begin{figure}
{\epsfxsize=8.truecm
\epsfbox[18 144 592 718]{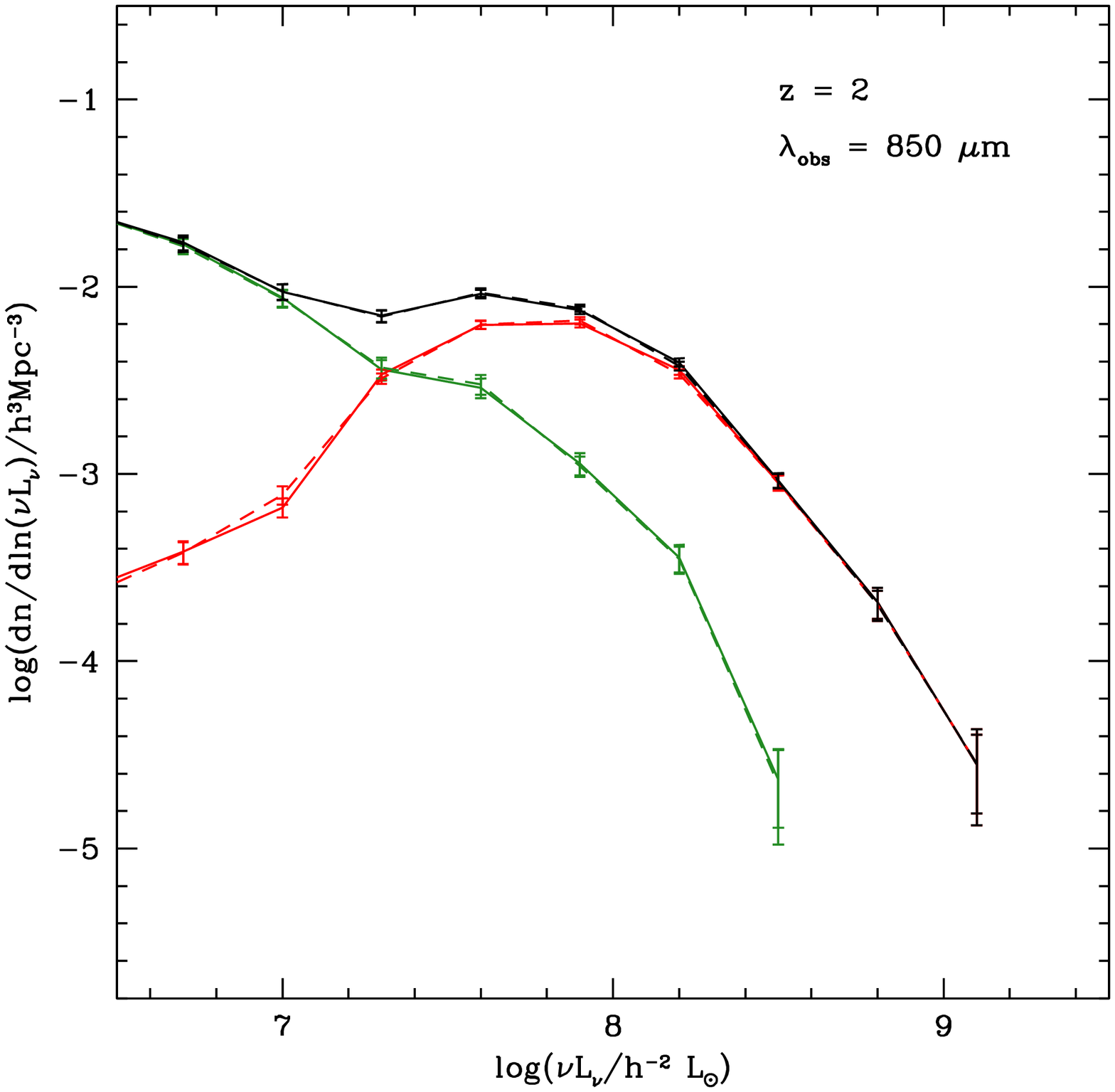}}
\caption{
The 850 $\mu$m observer frame luminosity function at $z=2$. 
The luminosity functions calculated from the original \gr~luminosities 
are shown by solid lines, while the ANN predicted values are plotted using 
dashed lines. The green and red lines show the luminosity function
 for the quiescent 
and burst galaxies, respectively, which are the components of the full 
sample, represented by the black line. The error bars indicate the 
Poisson uncertainties due to the number of galaxies simulated.}
\label{fig:ann.lf.850}
\end{figure}

The luminosity function in the observer frame 850 $\mu$m 
at $z=2$ is plotted in Fig.~\ref{fig:ann.lf.850}. Similar to what we 
found for the UV and mid-IR bands, the submillimetre luminosity functions 
calculated using the predicted luminosities are virtually indistinguishable 
from the functions constructed from the original 850 $\mu$m luminosities 
(extracted from \gr~spectra), for the whole luminosity range.

\section{Predicting galaxy colours}
\label{section:colours}
In this section we look at the performance of the ANN when predicting joint 
luminosity or colour distributions. In particular, we apply the ANN to the 
prediction of UV--submillimetre, mid-IR--submillimetre and UV--mid-IR colours, 
for a sample of \g~galaxies extracted at $z=0$. We define colour as 
L$_{\rm band\,1}$/L$_{\rm band\,2}$, where $L_{\rm band} \equiv
\langle L_{\nu} \rangle$, and the $\langle \rangle$ brackets denote an
average over the filter response of the passband.
We predict colours by training the ANN independently for each band. Hence, to 
predict a colour, two networks were trained, one for each luminosity. As we noted 
in Section~\ref{subsection:ann.predicted.lums}, this procedure produces better 
results than predicting both luminosities simultaneously using two output neurons. 

\begin{figure*}
{\epsfxsize=5.4truecm
\epsfbox[18 144 592 718]{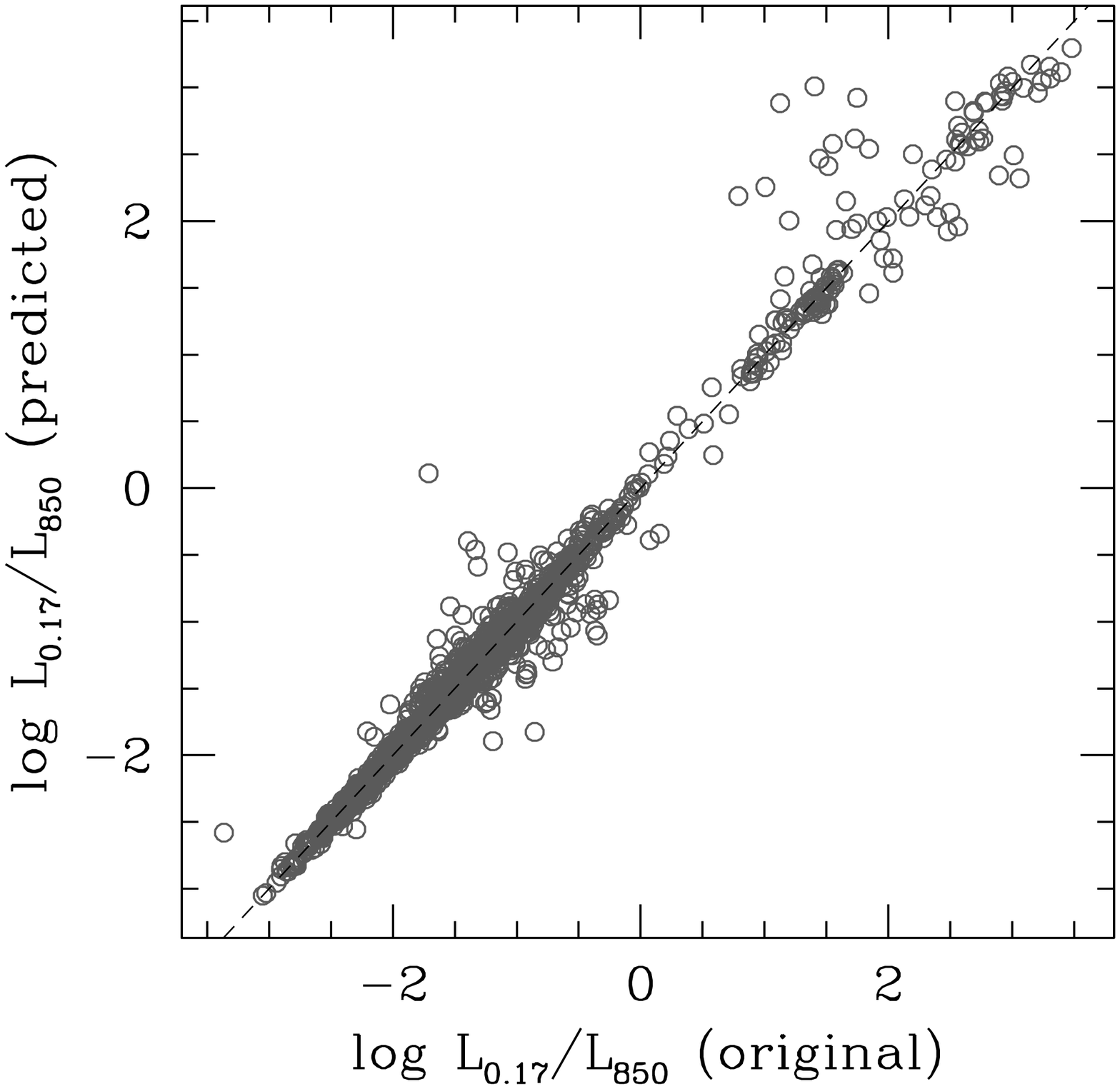}}
{\epsfxsize=5.4truecm
\epsfbox[18 144 592 718]{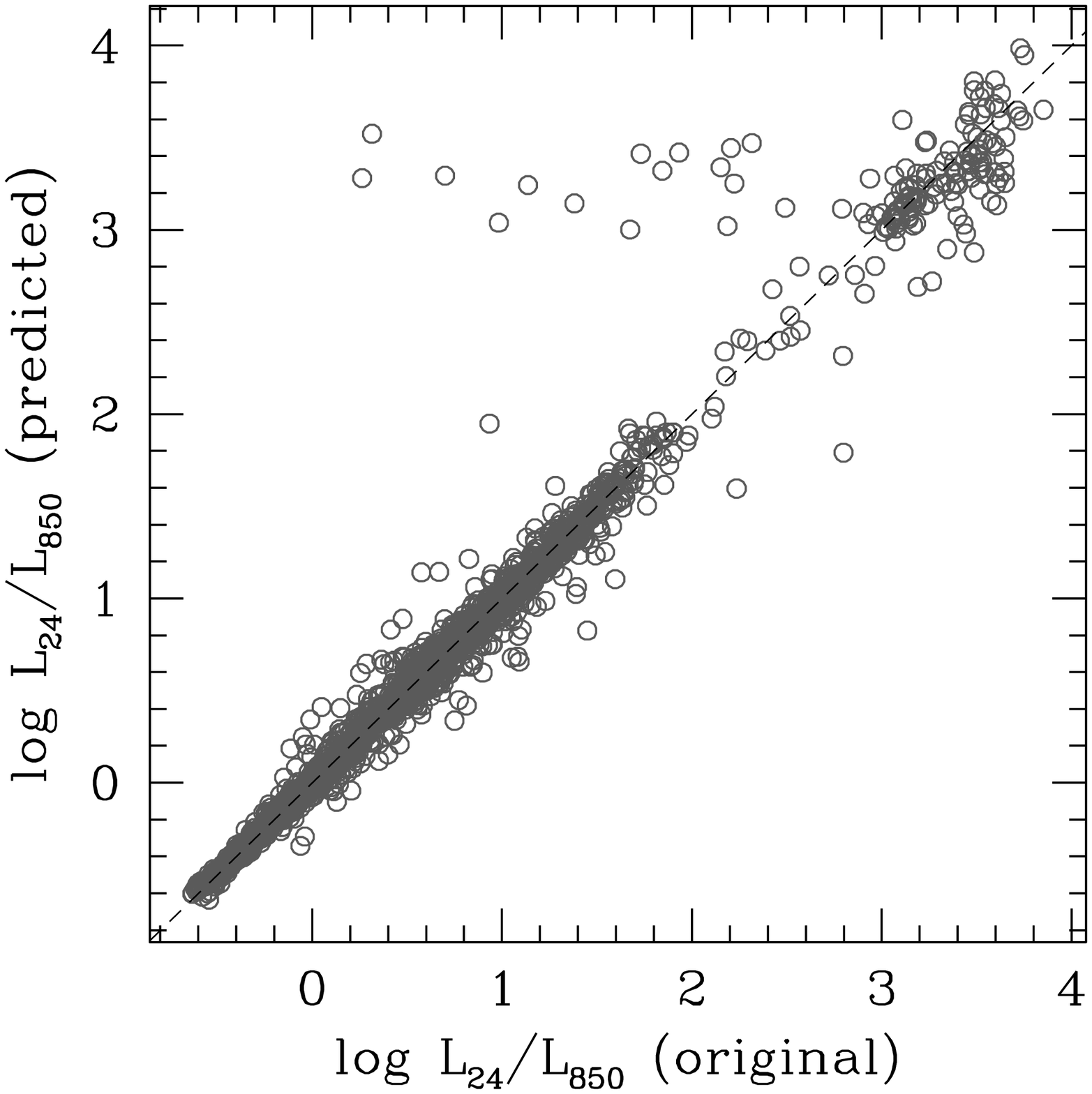}}
{\epsfxsize=5.4truecm
\epsfbox[18 144 592 718]{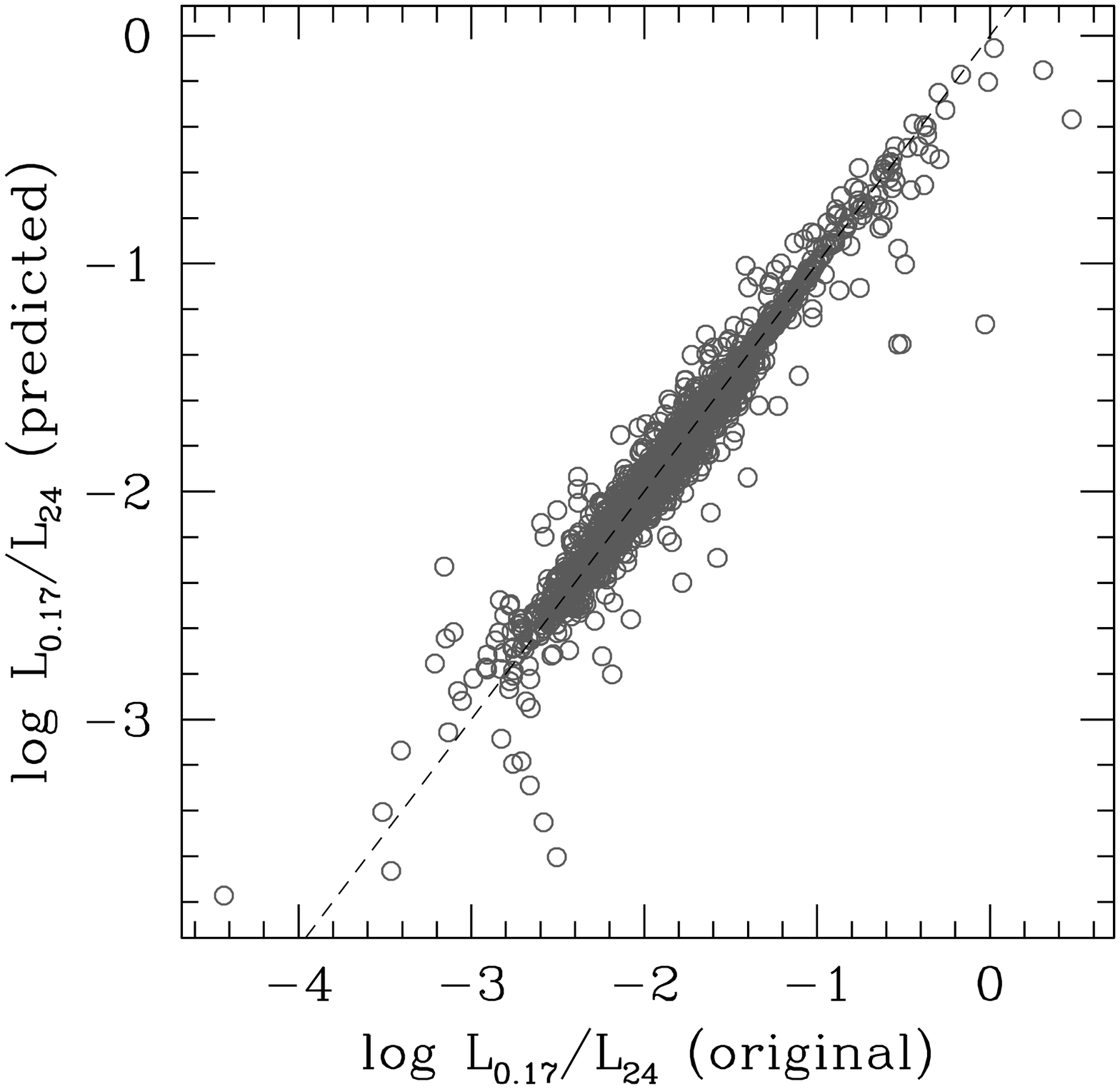}}
{\epsfxsize=5.4truecm
\epsfbox[18 144 592 718]{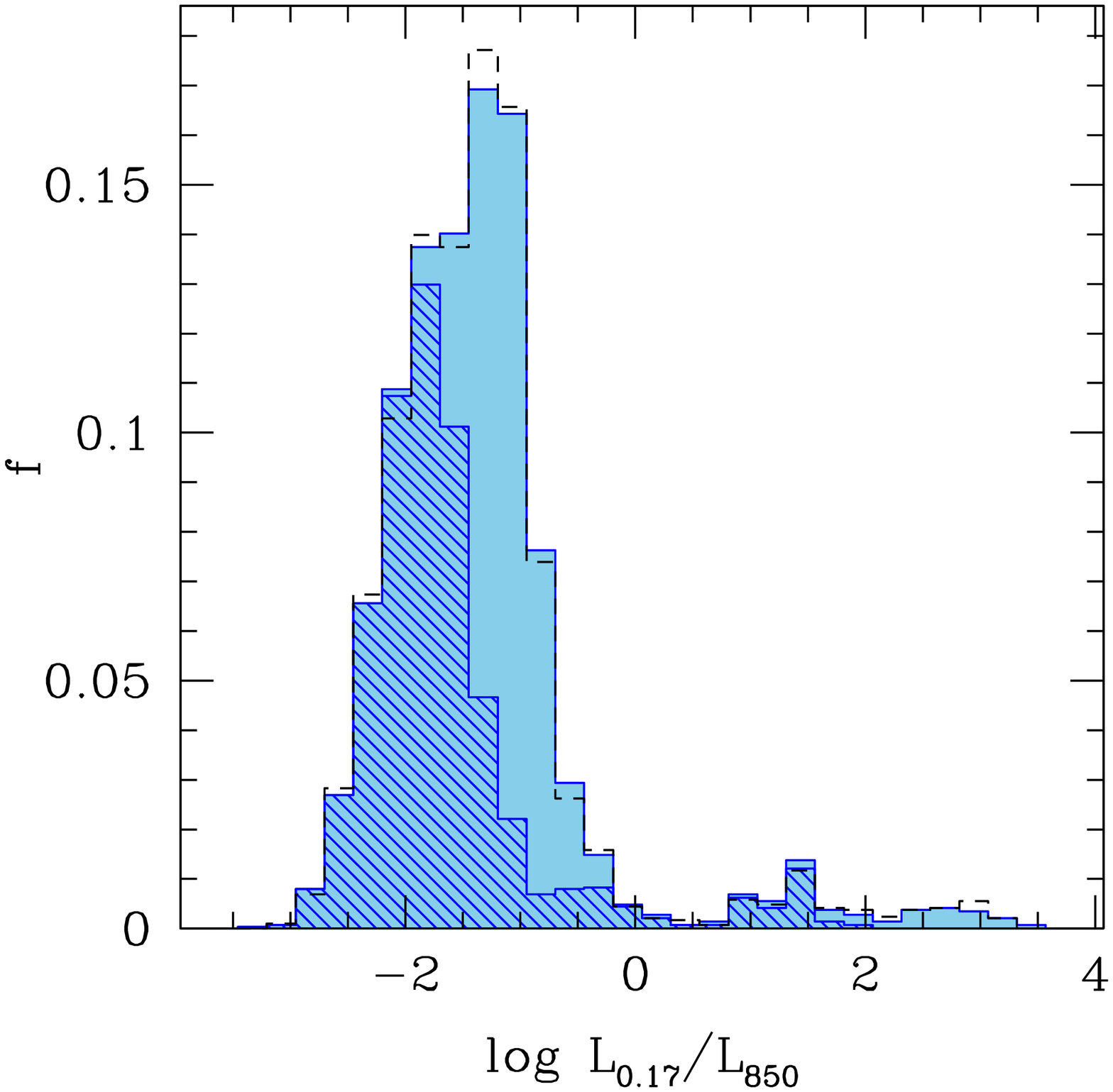}}
{\epsfxsize=5.4truecm
\epsfbox[18 144 592 718]{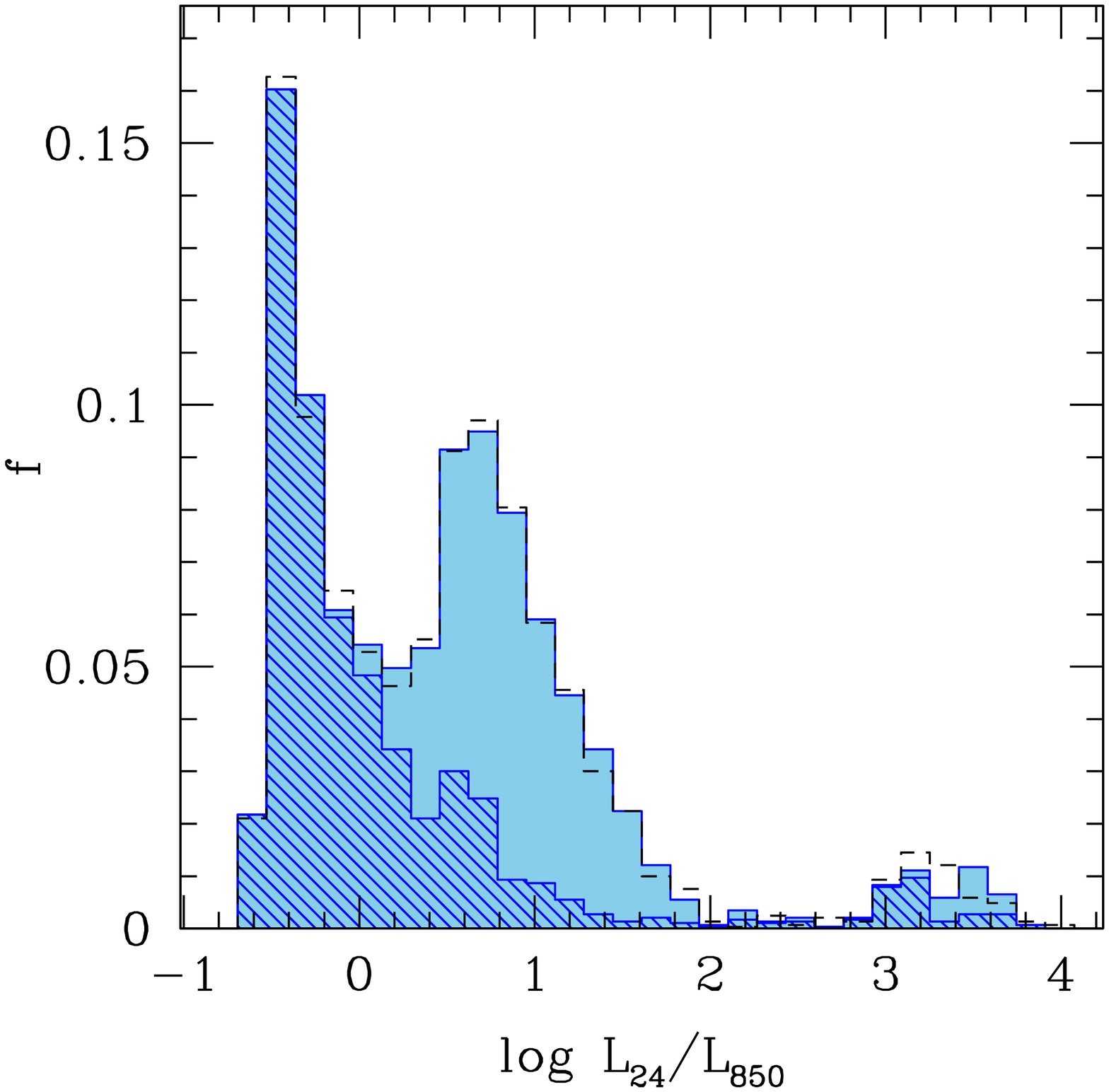}}
{\epsfxsize=5.4truecm
\epsfbox[18 144 592 718]{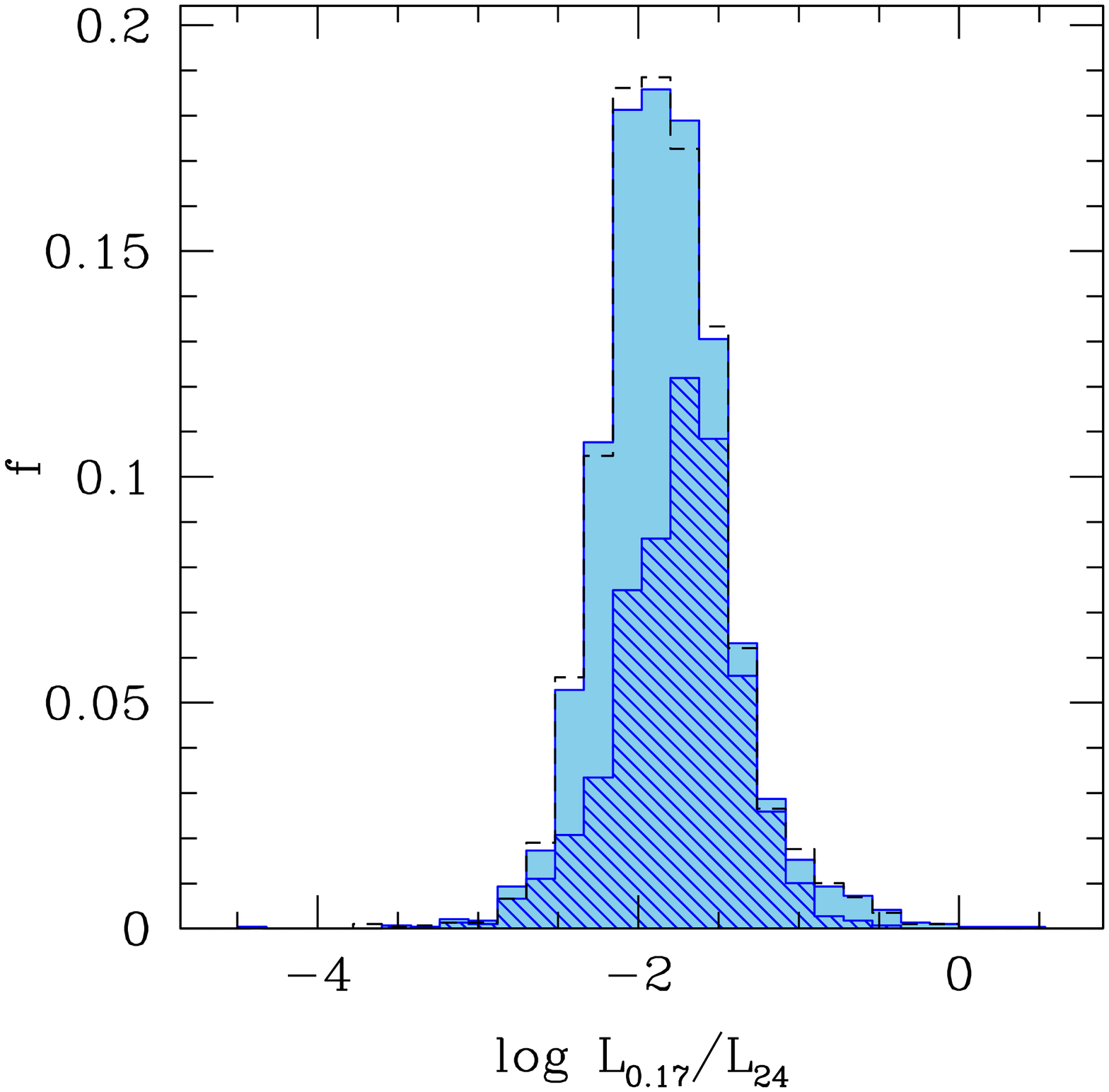}}
\caption{
The colour predicted by the ANN plotted against the true color for a 
sample of \g~galaxies extracted at redshift $z=0$. From left to right, we 
plot the $0.17\mu$m--$850\mu$m, $24\mu$m--$850\mu$m and $0.17\mu$m--$24\mu$m 
colours. In the lower panels, we show the distribution of colour. The blue 
solid-shaded histogram represents the true distribution, with the contribution 
from quiescent galaxies shown by the hatched histogram. The dashed line, 
unfilled histograms show the colours predicted by the ANN.}
\label{fig:colour1}
\end{figure*}

\begin{table*}
\begin{center}
  \begin{tabular}{ccccccccc}
  \hline
 Colour & Sample & $\varepsilon_L$ & $P_{|e|<10\%}$ & $p_1$ & Q$_1$ & Q$_2$ & Q$_3$ & $p_{99}$ \\
 \hline
 \multirow{2}{*}{$\frac{L_{0.17}}{L_{850}}$} 
 & Quiescent	& 0.14     & 78.9 & -42.1     & -3.7 & 0.0 & 3.8 & 34.4 \\
 & Bursts 	& 0.41     & 57.0 & -474.5   & -8.6 & 0.7 & 8.8 & 67.9 \\
\multirow{2}{*}{$\frac{L_{24}}{L_{850}}$} 
 & Quiescent	& 0.16     & 81.1 & -70.7     & -3.5 & 0.2 & 3.3 & 36.4 \\
 & Bursts 	& 0.52     & 58.7 & -733.1   & -7.8 & 0.8 & 8.2 & 58.0 \\
\multirow{2}{*}{$\frac{L_{0.17}}{L_{24}}$} 
 & Quiescent	& 0.16     & 78.3 & -40.3     & -3.2 & 0.2 & 4.2 & 33.8 \\
 & Bursts 	& 0.27     & 55.4 & -103.8   & -9.4 & 0.3 & 8.4 & 59.6 \\
  \hline
  \end{tabular}
   \caption{Statistics of the error distribution associated with the
prediction of colours at $z=0$, using ANN. A description of the
quantities is given in Table~\ref{tab:ann.predicted.spectra}.}
  \label{tab:ann.colour1}
  \end{center}
 \end{table*}

In the upper panels of Fig.~\ref{fig:colour1} we show the comparison 
between the predicted and true colours. The sample of galaxies used is 
defined in terms of stellar mass for quiescent galaxies and the stellar 
mass produced in the most recent burst for starbursts, as described above, 
and hence, as such, is not intended to match a particular observational 
selection.  
From left to right we plot 
the $0.17\mu$m--$850\mu$m, $24\mu$m--$850\mu$m and $0.17\mu$m--$24\mu$m 
colours. The error distributions are summarized in Table~\ref{tab:ann.colour1}.
This plot reveals that the ANN performs remarkably well when predicting 
colours. For quiescent galaxies, we find more than 78\% of the sample 
have colours within 10\% of the true colour. As noted in previous sections, 
the ANN is not capable of achieving such a performance for burst galaxies, 
with only $\sim 55\%$ of galaxies possessing predicted colours within 10\% 
of the expected values. Nevertheless, the distributions of the predicted 
colours, shown by the dashed, unshaded histograms in the lower panels of 
Fig.~\ref{fig:colour1}, are very similar to the true distributions 
(represented by the solid-shaded histograms). 

Fig.~\ref{fig:colour1} shows that at redshift $z=0$, \g~predicts that 
most of the galaxies have a $0.17\mu$m--$850\mu$m colour in the range 
0.001-1, with a median $\sim 0.04$. In this plot, the hatched histogram 
represents the contribution of the quiescent galaxies to the total colour 
distribution. We see that quiescent galaxies present redder UV--submillimetre 
colours (smaller luminosity ratios) than the burst population, by a factor 
of $\sim 8$. The distribution of the mid-IR--submillimetre colours shows 
a bi-modality, with one peak at $\sim 0.38$ and the second at $\sim 6.3$. 
This double-peaked distribution is a consequence of the different nature 
of quiescent and burst galaxies. The plot indicates that \g~modelled 
quiescent galaxies have distinctly redder $24\mu$m--$850\mu$m colours 
than burst galaxies. In the bottom-right panel of Fig.~\ref{fig:colour1} 
we plot the distribution of UV--mid-IR colour for galaxies at $z=0$. 
Model galaxies show colours between $\sim 0.001$--0.1, with a median 
around 0.01. Both quiescent and burst galaxies display similar 
distributions, with the former showing slightly bluer colours.

\section{The overlap between UV and submillimetre selected galaxies}
\label{section:clustering.colours}

The star formation history of the Universe has been probed at high
redshift using samples selected in the optical and at submillimetre
wavelengths \citep[e.g.][]{smail, steidel03}. Samples constructed in the
optical are sensitive to emission in the rest-frame UV at redshifts
$z>2$. The UV flux is very sensitive to dust extinction. Hence, to
estimate the true star formation density from such observations it is
necessary to apply a large extinction correction to the observed
flux. This problem does not apply to samples constructed at
submillimetre wavelengths. However, there are two different problems
to overcome in this case: How much of the dust heating is due to
extincted starlight and how much arises from AGN emission? What is the
conversion from submillimetre flux to total infrared luminosity? The
completeness of optically selected samples with regard to measuring
the star formation  density has been called into question. The
possibility has been put forward that heavily extincted star formation
could be completely missed in optically selected samples. To resolve
these issues it is important to establish the overlap between
optically and submillimetre selected samples \citep{adelberger}.

In this section, we shed some light on this problem by using the ANN
to predict the optical magnitudes and optical--sub-mm colour
distributions of galaxies selected from their sub-mm fluxes at redshift
$z=2$. Investigating in detail the overlap between UV-selected and
submillimetre-selected samples requires a large galaxy sample, which
can only be realistically generated using the ANN approach.  We choose
redshift $z=2$ for this comparison because it is the typical redshift
measured for galaxies in current faint sub-mm surveys
\citep{chapman2005}. One caveat to be borne in mind in this comparison
is that the current version of our model does not follow the heating
of the dust by AGN, so all the submillimetre radiation is emitted by
the reprocessing of UV starlight by dust.

\begin{figure}
{\epsfxsize=8.truecm
\epsfbox[18 144 592 718]{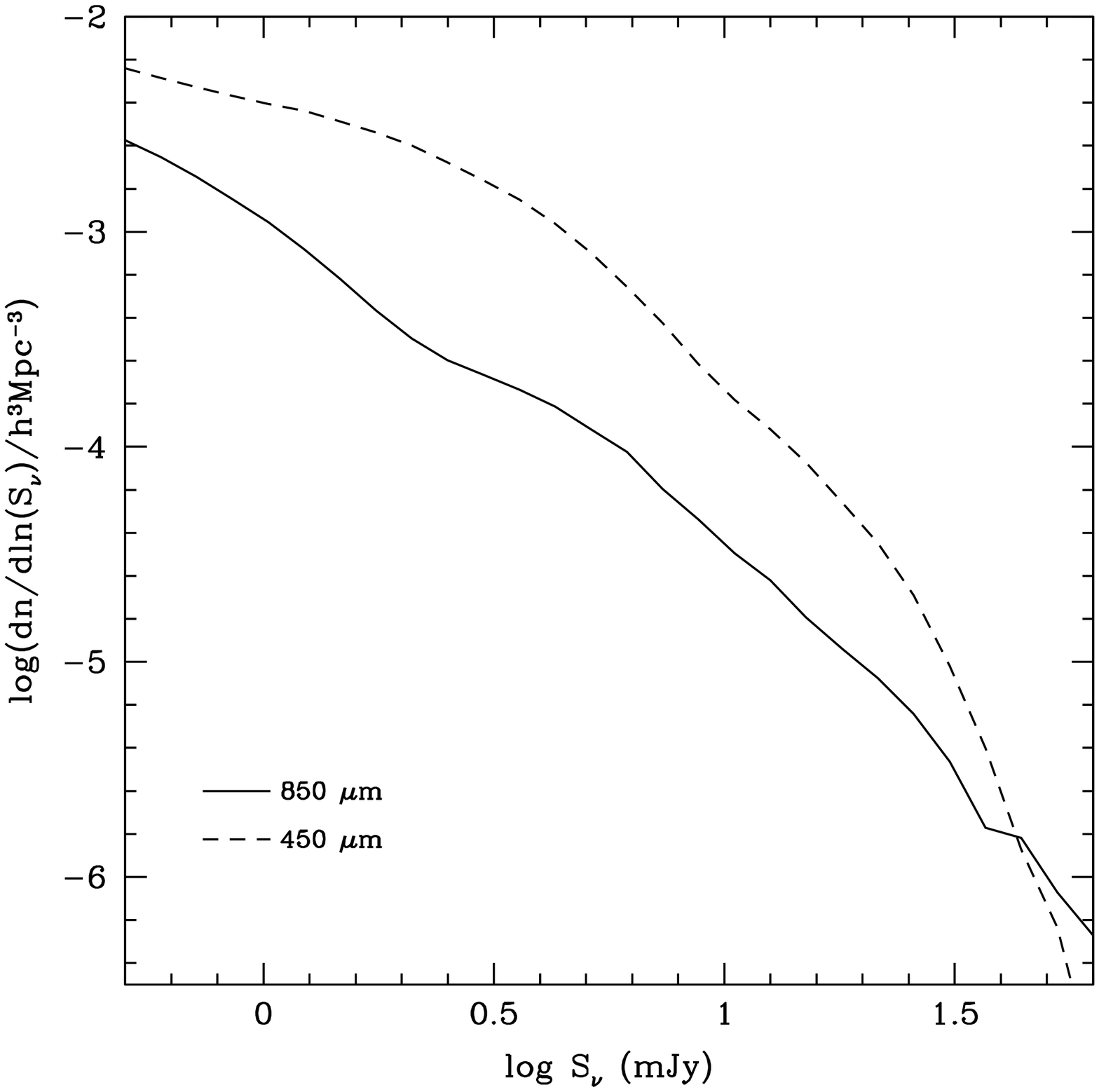}}
{\epsfxsize=8.truecm
\epsfbox[18 144 592 718]{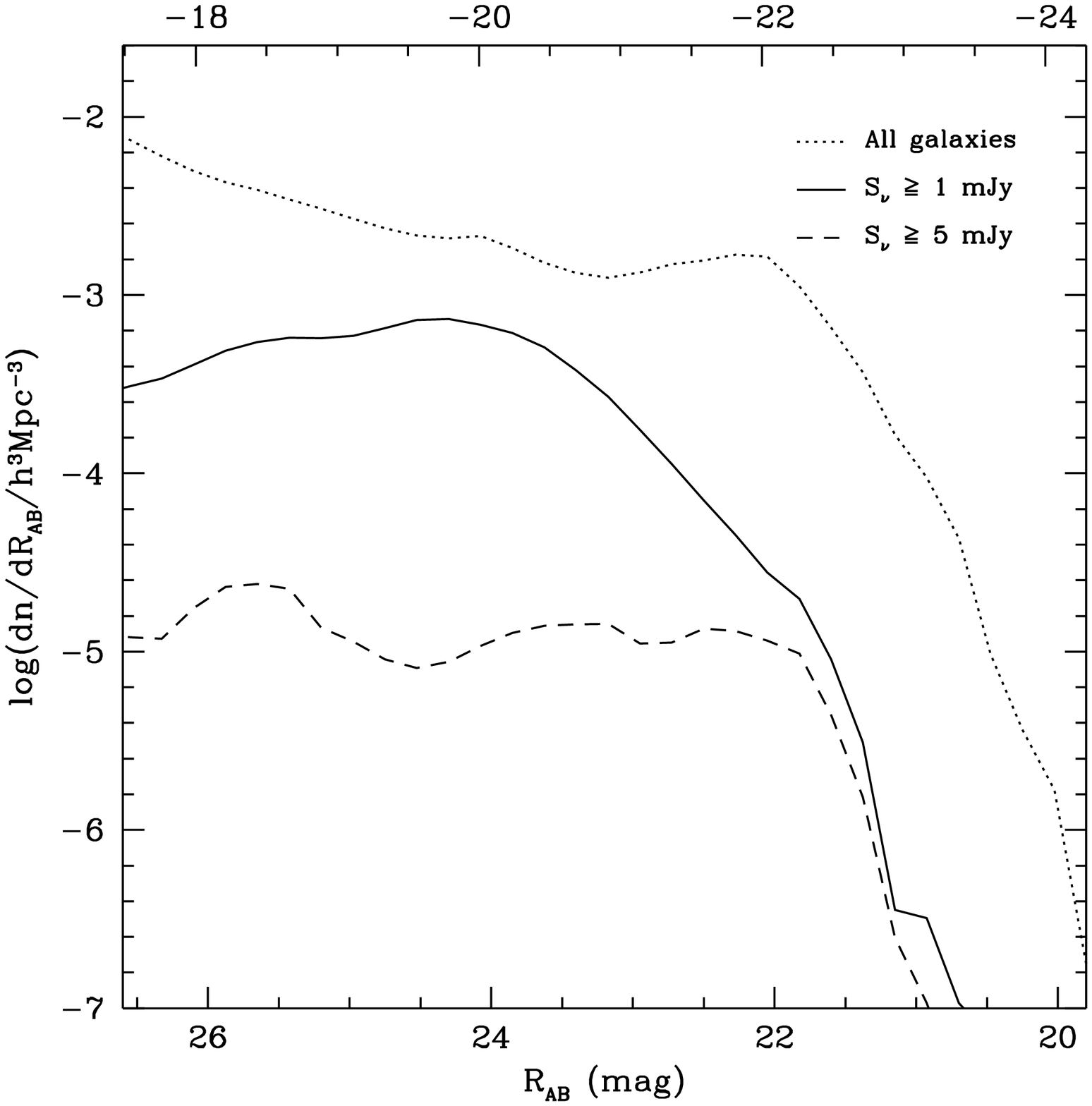}}
\caption{ Luminosity functions of sub-mm and rest-frame UV selected
galaxies at $z=2$.  Upper panel: luminosity functions at observed
wavelengths of 850 $\mu$m (solid line) and 450 $\mu$m (dashed
line). Lower panel: luminosity function in observed R-band,
corresponding to rest-frame UV, with apparent R-band magnitude shown
on bottom axis and rest-frame absolute AB magnitude shown on the top
axis. In the lower panel, we distinguish between the total R-band
luminosity function (dotted line), and that for the subsets of
galaxies with $850 \mu$m fluxes $S_{\nu}(850\, \mu{\rm m}) \geq 1$ mJy (solid line)
and $S_{\nu}(850\, \mu{\rm m})\geq 5$ mJy (dashed line). }
\label{fig:clustering.lfs}
\end{figure}

We first plot in Fig.~\ref{fig:clustering.lfs} the luminosity
functions at observed wavelengths of 450 $\mu$m and 850 $\mu$m (top
panel) and in the observed $R$-band, which at $z=2$ corresponds to the
rest-frame far-UV, $\lambda = 0.23\,\mu$m (bottom panel). The
integrated number density of sub-mm galaxies (SMGs) is listed in
Table~\ref{tab:clustering.density}.  As expected, the plot shows that
bright galaxies are rarer than fainter galaxies: submillimetre
galaxies with flux densities $S_{\nu}(850\, \mu{\rm m}) \geq 1$ mJy are $\sim 10$
times more abundant than galaxies with $S_{\nu}(850\, \mu{\rm m}) \geq 5$ mJy. For
these brighter galaxies, we calculate a space density of $7.9\times
10^{-5}$ Mpc$^{-3}$, which is of the same order as the value estimated
observationally by \citet{blain} of $2.7\times 10^{-5}$ Mpc$^{-3}$. At
a given flux, submillimetre selected galaxies at 450 $\mu$m are more
abundant than their $850\mu$m counterparts, with the exception perhaps
of galaxies brighter than 50 mJy. In our simulation, we find
approximately one submillimetre galaxy with a 450 $\mu$m flux density
greater than 1 mJy in every 220 Mpc$^3$.

The bottom panel of Fig.~\ref{fig:clustering.lfs} reveals that, in our
model, a large fraction of SMGs should be detectable in current deep
optical surveys. For example, around half of the SMGs with
$S_{\nu}(850\, \mu{\rm m}) \geq 5$ mJy are predicted to be brighter than
$R_{\rm AB}=25$, which is similar to the magnitude limit used by
\citet{steidel04} in their survey for star-forming galaxies at $z\sim
2$ using BX and BM colour selection on the rest-frame UV emission from
these galaxies.  (Note that for the adopted cosmology, $R_{\rm AB}=25$
corresponds to an absolute magnitude of $M_{\rm AB} - 5 \log\, h =
-19.1$ at this redshift.) We find a median magnitude $R_{\rm AB}=25.2$
for SMGs at $z=2$ with $S_{\nu}(850\, \mu{\rm m}) \geq 5$ mJy. This seems
quite consistent with the observational values from
\citet{chapman2005}, who found a median $R_{\rm AB}=25.4$ for a sample
of radio-detected SMGs with $S_{\nu}(850\, \mu{\rm m}) \gsim\, 5$ mJy.
This panel also reveals another important result: only $\approx 1\%$
of all the galaxies brighter than $R_{\rm AB}=25$, are predicted to
have 850$\mu{\rm m}$ flux densities brighter than 5 mJy.

\begin{table*}
\begin{center}
 \begin{tabular}{cccc}
 \hline
 Sample &   450 $\mu$m  & 850 $\mu$m & 850 $\mu$m and $R_{\rm AB}< 25$  mag \\
        & ($10^{-5}$ Mpc$^{-3}$) & ($10^{-5}$ Mpc$^{-3}$) & ($10^{-5}$ Mpc$^{-3}$)\\
 \hline
 Galaxies with $S_{\nu} \geq 0.5$ mJy & 773.9 & 211.2 & 124.6\\
 Galaxies with $S_{\nu} \geq 1$ mJy    & 456.3 & 83.5 & 53.8 \\
 Galaxies with $S_{\nu} \geq 5$ mJy    & 45.9 & 7.9 & 3.9 \\
 \hline
 \end{tabular}
  \caption{The space density of submillimetre galaxies in the
Millennium Simulation at $z=2$. We distinguish between galaxies with
$S_{\nu}(450\, \mu{\rm m})$ or $S_{\nu}(850\, \mu{\rm m}) \geq 0.5$
mJy, 1mJy and 5mJy respectively. In the third column, we further limit
our sample by only considering those galaxies brighter than $R_{\rm
AB}< 25$ mag. The number densities in the table are quoted in units of
$10^{-5}$ Mpc$^{-3}$.}
 \label{tab:clustering.density}
 \end{center}
\end{table*}

\begin{figure}
{\epsfxsize=8.truecm
\epsfbox[18 144 592 718]{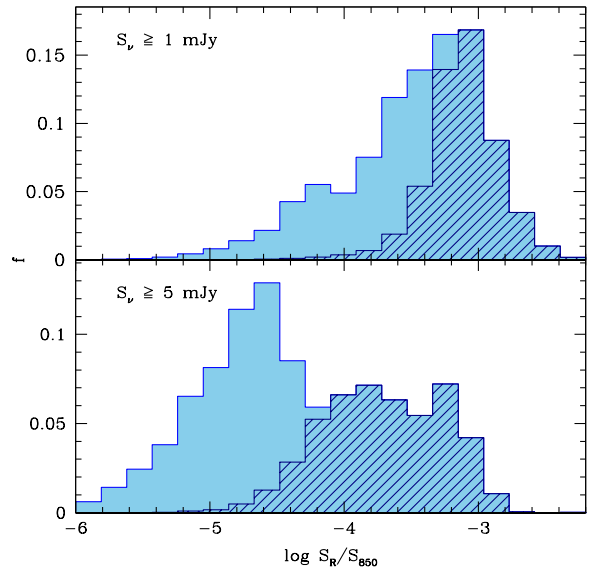}}
\caption{The distribution of observer-frame R-band -- 
submillimetre (850 $\mu$m) colours predicted by the ANN, for galaxies
in the Millennium Simulation at $z=2$. The colour is expressed as a
flux ratio $S_{\nu}(R)/ S_{\nu}(850\,\mu {\rm m})$.  The top and
bottom panels shows the distributions for galaxies with sub-mm flux
densities, $S_{\nu}(850\,\mu {\rm m})$, brighter than 1 and 5 mJy
respectively. In each panel, the filled histogram shows the full
distribution, while the hatched histogram shows the contribution to
this from galaxies which are also brighter than $R_{\rm AB}=25$ in the
optical.}
\label{fig:clustering.colours.Sv}
\end{figure}

In Fig.~\ref{fig:clustering.colours.Sv} we plot the distribution of
observer-frame R-band -- sub-mm (850 $\mu$m) colours predicted by the
ANN, for galaxies in the Millennium Simulation at $z=2$. We plot the
colour distributions for SMGs with flux densities
$S_{\nu}(850\,\mu{\rm m}) \geq 1$ and $\geq 5$ mJy in the two panels.
For galaxies with $S_{\nu}(850\,\mu{\rm m}) \geq 1$ mJy, we find a
median colour of $S_{\nu}(R)/ S_{\nu}(850\,\mu {\rm m}) \approx 4
\times 10^{-4}$. Brighter SMGs with $S_{\nu}(850\,\mu{\rm m}) \geq 5$
mJy display a colour distribution which is on average $\sim 10$ times
redder, with a median colour of $4 \times 10^{-5}$. The colour
distributions are also seen to be very broad, especially for the
brighter sub-mm flux limit, which covers a range $\sim 10^3$ in
colour. We also show (as hatched histograms) the colour distributions
which result for each sub-mm flux limit if we further select only
galaxies with optical magnitudes brighter than $R_{\rm AB}<25$. This
shows how we lose the redder part of the optical--sub-mm colour
distribution with this optical selection.



\section{Discussion and Conclusions}
\label{section:clustering.conclusions}

In this paper we have introduced a new method to rapidly predict
accurate spectral energy distributions over a wide wavelength range
from a small number of galaxy properties, using artificial neural
networks (ANN). Granato et~al. (2000) combined the {\tt GALFORM}
semi-analytical galaxy formation code with the spectro-photometric
code {\tt GRASIL}. The use of {\tt GRASIL} allows a more comprehensive
and accurate treatment of the effect of dust on the SED of the galaxy,
predicting the dust emission in the mid- and far-IR regions, as well
as improving the accuracy of the predicted spectra in the
UV. Unfortunately, {\tt GRASIL} takes several minutes to run for each
galaxy, which prohibits the direct application of this code to
populate large dark matter simulation volumes with galaxies.  The ANN
provides a fast, simple and flexible means to calculate accurate
galaxy spectra based on \gr. Here, we have carried out the first tests
of the method and present applications to galaxy luminosities and
colours.

The ANN is trained using a sample of galaxies for which \gr~has been
run to compute spectra. We found that the ANN approach performs well
when predicting galaxy spectra from galaxy properties. The best
performing ANN architecture we found is a simple supervised,
feed-forward net, composed of 12 input galaxy properties, two hidden
layers with 30 neurons each, and one output neuron. The ANN works best
when predicting the luminosity at one wavelength at a time, rather
than the whole spectrum.  Due to the inherent variety in the spectra
of galaxies which are undergoing a burst of star formation, or which
recently underwent a burst, we found it best to train the ANN
separately for samples of quiescent and bursting galaxies. The ANN
needs to be trained at each redshift of interest and for each set of
\g~plus \gr~parameters.  The luminosities predicted by the ANN agree
remarkably well with those computed directly using \gr. In the
observer frame $850\mu$m at $z=2$, over 90\% of the ANN predicted
luminosities lie within 10\% of the true luminosities calculated
directly from \gr. The ANN works somewhat less well in the UV and
mid-IR. Nevertheless, at all the wavelengths considered we find that
the luminosity functions predicted by the ANN are in excellent
agreement with those computed directly with \gr.

The ANN also performs well when predicting the colours of galaxies.
In this case, the ANN is trained for each band individually.  Given
this success, we applied the ANN to investigate the overlap between
samples of rest-frame UV and sub-mm selected galaxies at $z= 2$.  This
problem is ideally suited to the ANN approach, as it requires a large
sample of galaxies covering a wide range of luminosity. Although we
predict that 50\% of bright submillimetre sources (850$\mu$m flux
greater than 5 mJy) should have optical magnitudes brighter than
$R_{\rm AB}<25$, these SMGs make up only a small fraction of an
optically selected sample at the same magnitude limit.  In an
optically selected sample of galaxies at $z=2$ brighter than $R_{\rm
AB}<25$, 10\% are predicted to have an 850 $\mu$m flux brighter than 1
mJy and 1\% are expected to be brighter than 5 mJy. These predictions
seem consistent with recent observational constraints
\citep[e.g.][]{chapman2005}.

The success of our new ANN approach in generating accurate predictions
of the spectral energy distributions of large samples of galaxies
means that we can now produce mock catalogues of galaxies for
forthcoming surveys such as the Herschel ATLAS survey, which will
cover 600 square degrees in five far-infrared bands, and the SCUBA-2
Cosmology Legacy Survey, which will cover around 40 square degrees to
a fainter flux limit in the sub-mm.  In a companion paper we apply the
ANN technique to populate the Millennium Simulation with galaxies with
accurate sub-mm fluxes to make predictions for the clustering of dusty
galaxies.

\subsection*{ACKNOWLEDGEMENTS}

CA gratefully acknowledges support in the form of a scholarship
from the Science and Technology Foundation (FCT), Portugal. This work
was supported in part by the Science and Technology Facilities Council
and by the Royal Society. CSF is a Royal Society Wolfson Research Merit 
Award holder.

\end{document}